\documentclass[12pt]{article}
\pdfoutput=1
\usepackage{jheppub}
\usepackage{epsf}
\usepackage{epsfig}
\usepackage{amsmath}
\usepackage{amsfonts}
\usepackage{amssymb}
\usepackage{graphicx}
\usepackage{subcaption}
\usepackage{color}
\usepackage{psfrag}
\usepackage{hyperref}
\usepackage{tikz}
\usepackage[export]{adjustbox}
\newcommand{\bmat}{\left(\begin{array}}
\newcommand{\emat}{\end{array}\right)}

\def\yzero{\smash{\hbox{$y\kern-4pt\raise1pt\hbox{${}^\circ$}$}}}

\def\beq{\begin{equation}}
\def\eeq{\end{equation}}
\def\beqa{\begin{eqnarray}}
\def\eeqa{\end{eqnarray}}

\def\-{\hphantom{-}}

\def\s2{\frac{1}{\sqrt2}}

\def\beq{\begin{equation}}
\def\eeq{\end{equation}}
\def\beqa{\begin{eqnarray}}
\def\eeqa{\end{eqnarray}}

\def\Tr{{\rm Tr \,}}

\def\IF{\relax{\rm I\kern-.18em F}}
\def\II{\relax{\rm I\kern-.18em I}}

\def\Dsl{\,\raise.15ex\hbox{/}\mkern-13.5mu D} %this one can be subscripted
\def\IC{{\bf C}}
\def\IS{{\bf S}}
\def\IR{{\bf R}}
\def\IZ{{\bf Z}}
\def\IX{{\bf X}}

\def\IT{{\bf T}}

\def\NN{{\cal N}}

\def\C{{\bf C}}

\newcommand{\drawsquare}[2]{\hbox{%
\rule{#2pt}{#1pt}\hskip-#2pt%  left vertical
\rule{#1pt}{#2pt}\hskip-#1pt%  lower horizontal
\rule[#1pt]{#1pt}{#2pt}}\rule[#1pt]{#2pt}{#2pt}\hskip-#2pt%  upper horizontal
\rule{#2pt}{#1pt}}% right vertical

\newcommand{\fund}{~\raisebox{-.5pt}{\drawsquare{6.5}{0.4}}~}
\newcommand{\antifund}{~\overline{\raisebox{-.5pt}{\drawsquare{6.5}{0.4}}}~}

\catcode`\@=11   
\newdimen\@rotdimen
\newbox\@rotbox  

\def\@vspec#1{\special{ps:#1}}%  passes #1 verbatim to the output
\def\@rotstart#1{\@vspec{gsave currentpoint currentpoint translate
   #1 neg exch neg exch translate}}% #1 can be any origin-fixing transformation
\def\@rotfinish{\@vspec{currentpoint grestore moveto}}% gets back in synch 
\def\@rotr#1{\@rotdimen=\ht#1\advance\@rotdimen by\dp#1%
   \hbox to\@rotdimen{\hskip\ht#1\vbox to\wd#1{\@rotstart{90 rotate}%
   \box#1\vss}\hss}\@rotfinish}
\def\@rotl#1{\@rotdimen=\ht#1\advance\@rotdimen by\dp#1%
   \hbox to\@rotdimen{\vbox to\wd#1{\vskip\wd#1\@rotstart{270 rotate}%
   \box#1\vss}\hss}\@rotfinish}%
\def\@rotu#1{\@rotdimen=\ht#1\advance\@rotdimen by\dp#1%
   \hbox to\wd#1{\hskip\wd#1\vbox to\@rotdimen{\vskip\@rotdimen
   \@rotstart{-1 dup scale}\box#1\vss}\hss}\@rotfinish}%
\def\@rotf#1{\hbox to\wd#1{\hskip\wd#1\@rotstart{-1 1 scale}%
   \box#1\hss}\@rotfinish}%
\def\rotate{\@ifnextchar[{\@rotate}{\@rotate[l]}}
\def\@rotate[#1]#2{\setbox\@rotbox=\hbox{#2}\@nameuse{@rot#1}\@rotbox}

\catcode`\@=12
\DeclareMathOperator{\U}{U}
\DeclareMathOperator{\SU}{SU}

\newcommand{\coma}{\text{ , }}
\newcommand{\fstop}{\text{ .}}

\newcommand{\e}{\text{ and }}

\begin{document}

%----------------------------------------------------------------------%
%  numbering equations with section number
%----------------------------------------------------------------------%
\makeatletter
\@addtoreset{equation}{section}
\makeatother
\renewcommand{\theequation}{\thesection.\arabic{equation}}
%----------------------------------------------------------------------%
%  title page
%----------------------------------------------------------------------%
\pagestyle{empty}
%\vspace*{1.0in}

\rightline{ IFT-UAM/CSIC-19-96}
\vspace{3cm}

\begin{center}
\LARGE{\bf Discrete Symmetries in Dimer Diagrams 
%\\ and the Swampland Distance Conjectures \\[12mm] 
}
\large{Eduardo Garc\'{\i}a-Valdecasas, Alessandro Mininno,  Angel M. Uranga\\[4mm]}
\footnotesize{Instituto de F\'{\i}sica Te\'orica IFT-UAM/CSIC,\\[-0.3em] 
C/ Nicol\'as Cabrera 13-15, 
Campus de Cantoblanco, 28049 Madrid, Spain} 
\footnotesize{\href{mailto:eduardo.garcia.valdecasas@gmail.com}{eduardo.garcia.valdecasas@gmail.com}, \href{mailto:alessandro.mininno@uam.es}{alessandro.mininno@uam.es},  \href{mailto:angel.uranga@csic.es}{angel.uranga@csic.es}}

\vspace*{20mm}

\small{\bf Abstract} \\%[5mm]
\end{center}
\begin{center}
\begin{minipage}[h]{17.0cm}
{\small We apply dimer diagram techniques to uncover discrete global symmetries in the fields theories on D3-branes at singularities given by general orbifolds of general toric Calabi-Yau threefold singularities. The discrete symmetries are discrete Heisenberg groups, with two $\IZ_N$ generators $A$, $B$ with commutation $AB=CBA$, with $C$ a central element. This fully generalizes earlier observations in particular orbifolds of $\IC^3$, the conifold and $Y_{p,q}$. The solution for any orbifold of a given parent theory follows from a universal structure in the infinite dimer in $\IR^2$ giving the covering space of the unit cell of the parent theory before orbifolding. The generator $A$ is realized as a shift in the dimer diagram, associated to the orbifold quantum symmetry; the action of $B$ is determined by equations describing a 1-form in the dimer graph in the unit cell of the parent theory with twisted boundary conditions; finally, $C$ is an element of the (mesonic and baryonic) non-anomalous $\U(1)$ symmetries, determined by geometric identities involving the elements of the dimer graph of the parent theory. These discrete global symmetries of the quiver gauge theories are holographically dual to discrete gauge symmetries from torsion cycles in the horizon, as we also briefly discuss. Our findings allow to easily construct the discrete symmetries for infinite classes of orbifolds. We provide explicit examples by constructing the discrete symmetries for the infinite classes of general orbifolds of $\IC^3$, conifold, and complex cones over the toric del Pezzo surfaces, $dP_1$, $dP_2$ and $dP_3$.}
\end{minipage}
\end{center}
\newpage
%----------------------------------------------------------------------%
%  Resetting of counters
%----------------------------------------------------------------------%
\setcounter{page}{1}
\pagestyle{plain}
\renewcommand{\thefootnote}{\arabic{footnote}}
\setcounter{footnote}{0}
%----------------------------------------------------------------------%
%  Paper begins
%----------------------------------------------------------------------%

\tableofcontents

\vspace*{1cm}

\section{Introduction and conclusions}

Discrete symmetries are key to our understanding of quantum field theory and the Standard Model, and it is an interesting question to address their realization in fundamental theories like string theory. In particular, the general arguments about absence of global symmetries in theories of quantum gravity (see \cite{Banks:1988yz,Abbott:1989jw,Coleman:1989zu} for early viewpoints, and e.g.\cite{Kallosh:1995hi,Banks:2010zn} and references therein, for more recent discussions) suggest that discrete symmetries should have a gauge nature in such theories \cite{Alford:1988sj,Krauss:1988zc,Alford:1989ch,Preskill:1990bm,Alford:1990mk,Alford:1990pt,
Alford:1991vr,Alford:1992yx} (see \cite{Harlow:2018jwu,Harlow:2018tng,Craig:2018yvw} for recent discussions in the swampland \cite{Vafa:2005ui,Brennan:2017rbf,Palti:2019pca} context).

Discrete gauge symmetries have been studied in string theory from different perspectives. Abelian gauge symmetries, and their application to MSSM-like models have been explored in D-brane models in \cite{Camara:2011jg,BerasaluceGonzalez:2011wy,Ibanez:2012wg,BerasaluceGonzalez:2012zn,Berasaluce-Gonzalez:2013sna}.
Non-abelian discrete gauge symmetries in 4d string compactifications were systematically studied in \cite{BerasaluceGonzalez:2012vb}. In fact, the first appearance of non-abelian discrete gauge symmetries in string theory arose in \cite{Gukov:1998kn} in the gravity dual of the quiver gauge theory on D3-branes at the $\IC^3/\IZ_3$ singularity. This was subsequently generalized to other particular orbifolds of $\IC^3$, the conifold and $Y_{p,q}$ in \cite{Burrington:2006uu,Burrington:2007mj}. The symmetries were constructed as global discrete symmetries of the quiver theory, by laboriously solving the conditions of invariance of the superpotential and cancellation of discrete gauge anomalies. The symmetries correspond to discrete Heisenberg groups, with $\IZ_N$ generators $A$, $B$ anticommuting to a central element $C$, namely $AB=CBA$. In the gravity dual, the discrete symmetries arise from torsion homology cycles, and the non-abelian nature is encoded in brane creation effects among the $\IZ_N$ charged objects \cite{Gukov:1998kn},  or alternatively in the KK reduction of Chern-Simons terms for torsion forms with non-trivial relations \cite{BerasaluceGonzalez:2012vb}.

In this paper we apply the powerful description of D3-branes at toric CY threefold singularities in terms of dimer diagrams \cite{Hanany:2005ve,Franco:2005rj,Kennaway:2007tq} to unravel the underlying structure of discrete symmetries in general orbifolds of general toric singularities. We find discrete Heisenberg groups for the whole class of theories, generalizing earlier results for particular examples. We show that the discrete symmetry structure for any orbifold of a given parent theory follows from a universal structure in the infinite dimer in $\IR^2$ giving the covering space of the unit cell of the parent theory. The general structure is as follows. The generator $A$ is realized as a shift in the dimer diagram, associated to the orbifold quantum symmetry; the action of $B$ is determined by equations describing a 1-form in the dimer graph in the unit cell of the parent theory with twisted boundary conditions. The element $C$ is a discrete subgroup of the non-anomalous $\U(1)$ symmetries (mesonic, and baryonic, if present), determined by a simple set of equations related to geometric identities among the elements of the dimer graph in the parent theory.

Our findings allow to easily construct the discrete symmetries for infinite classes of orbifolds. To illustrate the power of our methods, we provide explicit examples by constructing the discrete symmetries for general orbifolds of $\IC^3$, conifold, and complex cones over the toric del Pezzo surfaces, $dP_1$, $dP_2$ and $dP_3$.

These discrete global symmetries of the quiver gauge theories are holographically dual to discrete gauge symmetries from torsion cycles in the horizon, as we also briefly discuss. Our techniques thus provide the largest ensemble of discrete gauge symmetries in string theory models, in this case in AdS. They thus provide a natural setup to explore the properties of discrete symmetries in AdS quantum gravity, with interesting interplay with holography and hopefully with the swampland constraints for AdS vacua \cite{Ooguri:2016pdq,Harlow:2018jwu,Harlow:2018tng,Lust:2019zwm}.

\medskip

The paper is organized as follows. In Section~\ref{sec:intro} we revisit the dimer diagram description of quiver gauge theories on D3-branes at toric CY threefold singularities. We review the ingredients of their dimer diagrams and periodic quivers in Section~\ref{sec:dimer-quiver}, and describe their continuous $\U(1)$ symmetries in Section~\ref{sec:cont-symm}. In Section~\ref{sec:geomident} we re-derive the latter from a new ingredient, which we dub Geometric Identities for the dimer regarded as a graph. In Section~\ref{sec:discrete-appetizer} we give a first pass discussion of discrete symmetries for orbifolds of toric geometries. In Section~\ref{sec:general-orbifolds-general-toric} we describe general orbifolds of general toric singularities and the corresponding gauge theories. In Section~\ref{sec:general-heisenberg} we describe the general structure of the discrete Heisenberg groups, and in Section~\ref{sec:discrete-covering} we uncover their origin from an underlying structure of a 1-form defined on the infinite dimer in $\IR^2$ of the parent theory. We exploit this understanding to solve by inspection the discrete symmetries for general orbifolds of $\IC^3$, in Section~\ref{sec:example-general-c3}, and of the conifold, in Section~\ref{sec:orbifold-conifold-app}. In Section~\ref{sec:general-solution} we provide the systematic procedure to construct the explicit solution for the discrete symmetries of a general orbifold of a general toric singularity, in terms of equations for 1-forms on the graph of the parent theory dimer/quiver in its unit cell (with twisted boundary conditions). Section~\ref{sec:examples} is devoted to the explicit construction of the discrete symmetries in infinite families of orbifolds. In Sections~\ref{sec:general-solution-c3} and \ref{sec:general-solution-conifold} we recover the discrete symmetries for orbifolds of $\IC^3$ and the conifold, and in Section~\ref{sec:general-solution-dp1} we construct the discrete symmetries for the infinite class of general orbifolds of the $dP_1$ theory. Further examples are postponed to Appendix \ref{sec:more-examples}, in particular infinite classes of orbifolds of the $dP_2$ theory (in appendix \ref{sec:general-solution-dp2}) and of the $dP_3$ theory (in appendix \ref{sec:general-solution-dp3}). Finally, Section~\ref{sec:gravity} contains a sketch of the realization of these symmetries in the gravity dual, in terms of torsion classes in the 5d horizon geometry. Appendix \ref{sec:cohomology} introduces some topological concepts for the dimer/quiver graphs, useful for the discussion in the main text.

\medskip

\section{Dimer diagrams and quiver gauge theories}
\label{sec:intro}

\subsection{Dimer diagram and periodic quiver}
\label{sec:dimer-quiver}

The gauge theories on D3-branes at toric CY threefold singularities are efficiently encoded in dimer diagrams \cite{Hanany:2005ve,Franco:2005rj,Kennaway:2007tq}. These are bipartite tilings of $\IT^2$. The bipartite property means that  vertices can be colored black and white, with edges joining vertices of different color; it endows edges with an orientation e.g. from black to white nodes, and an orientation around vertices e.g. clockwise (resp. counter-clockwise) for black (resp. white) vertices.

\begin{figure}[!htp]
	\centering
	\begin{subfigure}[l]{0.4\textwidth}
		\begin{center}
			\includegraphics[width=\textwidth]{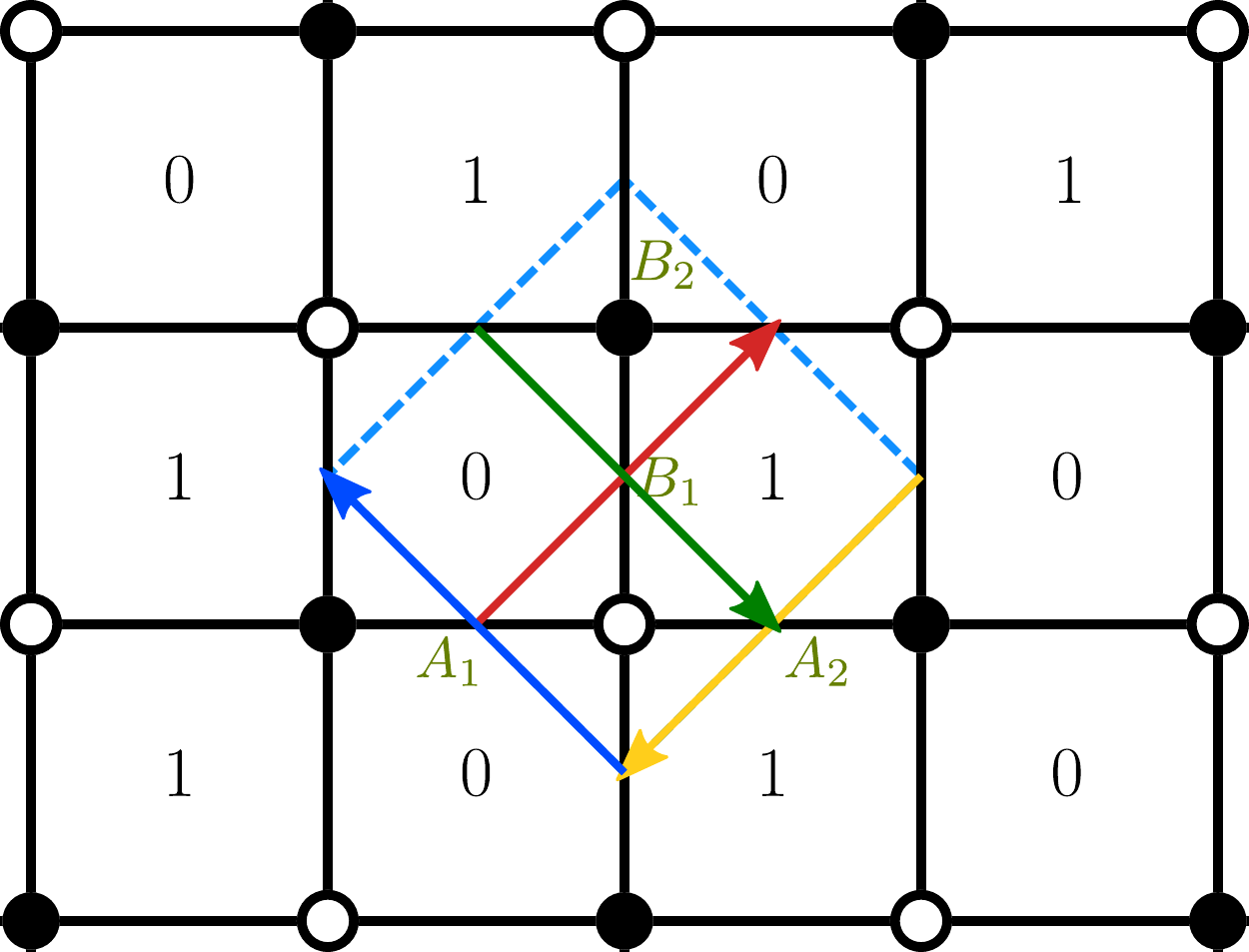}
			\caption{}
			\label{dimer-conifold}
		\end{center}
	\end{subfigure}\hspace{10mm}
	\begin{subfigure}[r]{0.5\textwidth}
		\begin{center}
			\includegraphics[width=\textwidth]{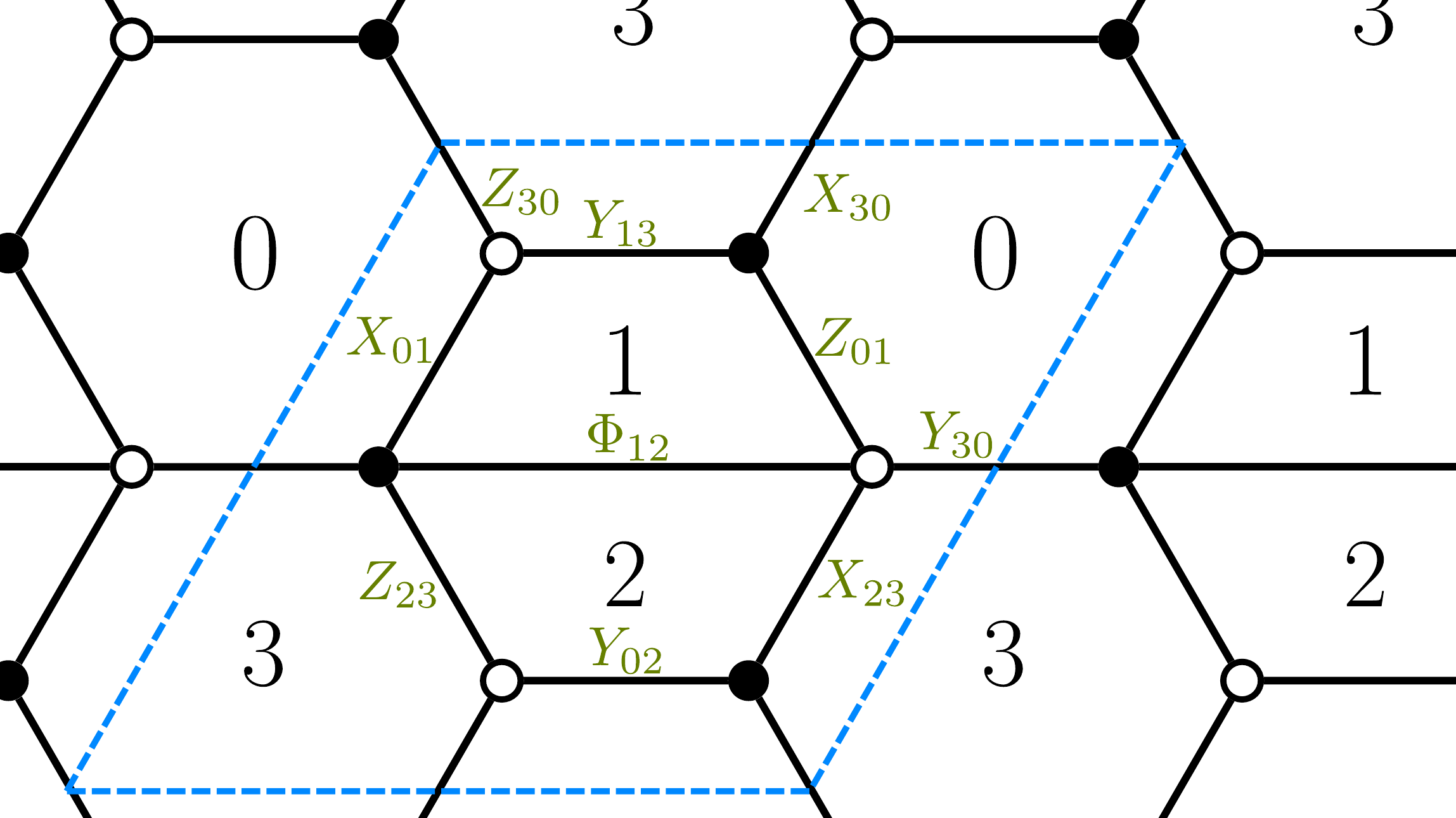}
			\caption{}
			\label{dimer-dp1}
		\end{center}
	\end{subfigure}
	\caption{Dimer diagram for the conifold with its zig-zag paths (\ref{dimer-conifold}) and the dimer diagram for $dP_1$ (\ref{dimer-dp1}).}
	\label{dimer-conifold-dp1} 
\end{figure}

The correspondence with the gauge theory is such that each face $F_a$ corresponds to a gauge factor $\SU(N_a)$. Actually, there is a $\U(N_a)$ symmetry group, but the $\U(1)_a$ factor is generically massive due to $BF$ couplings with closed string modes \cite{Ibanez:1998qp} (often also involved in a Green-Schwarz mechanism to cancel mixed anomalies). These $\U(1)$'s remain as (perturbatively exact) global symmetries, and will play a prominent role in this paper.

The correspondence also sets that each edge $E_i$, which separates faces $F_a$ and $F_b$, corresponds to a bifundamental chiral multiplet $(\fund_a,\antifund_b)$ if one crosses $E_i$ with positive orientation in going from $F_a$ to $F_b$. Finally, black and white vertices, denoted by $V_{\alpha}$ or $V'_{\alpha}$, respectively, introduce superpotential terms $\pm\Tr (\Phi_{E_1}\ldots \Phi_{E_n})$, with $\{ E_1,\ldots,E_n\}$ is the ordered set of edges surrounding the vertex, and the sign is positive or negative for black and white nodes, respectively.

The dimer diagram for the conifold and the $dP_1$ theory are shown in Figure \ref{dimer-conifold-dp1}. 

\begin{figure}[!htp]
	\centering
	\begin{subfigure}[l]{0.4\textwidth}
		\begin{center}
			\includegraphics[width=\textwidth]{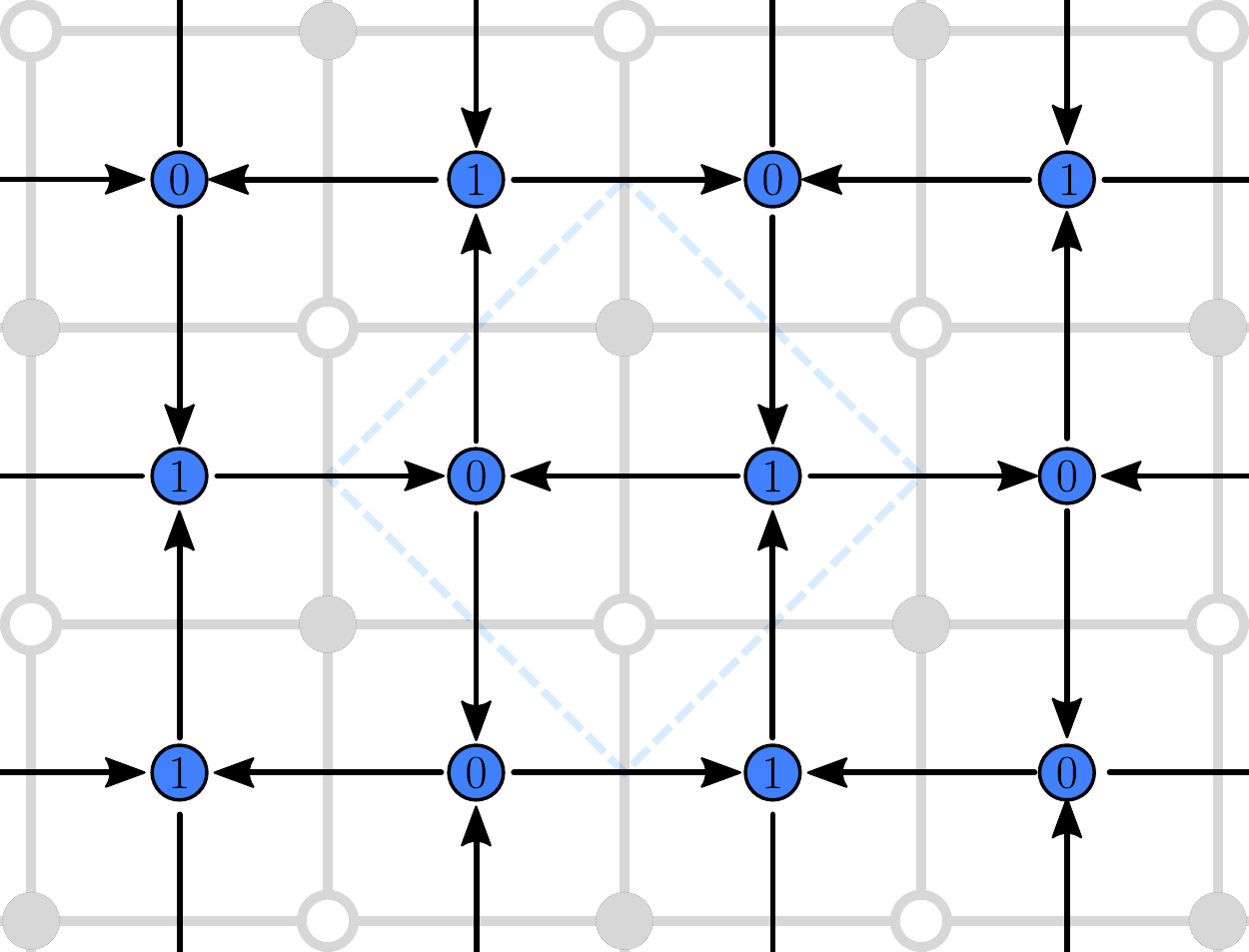}
			\caption{}
			\label{periodicquiver-conifold}
		\end{center}
	\end{subfigure}\hspace{10mm}
	\begin{subfigure}[r]{0.5\textwidth}
		\begin{center}
			\includegraphics[width=\textwidth]{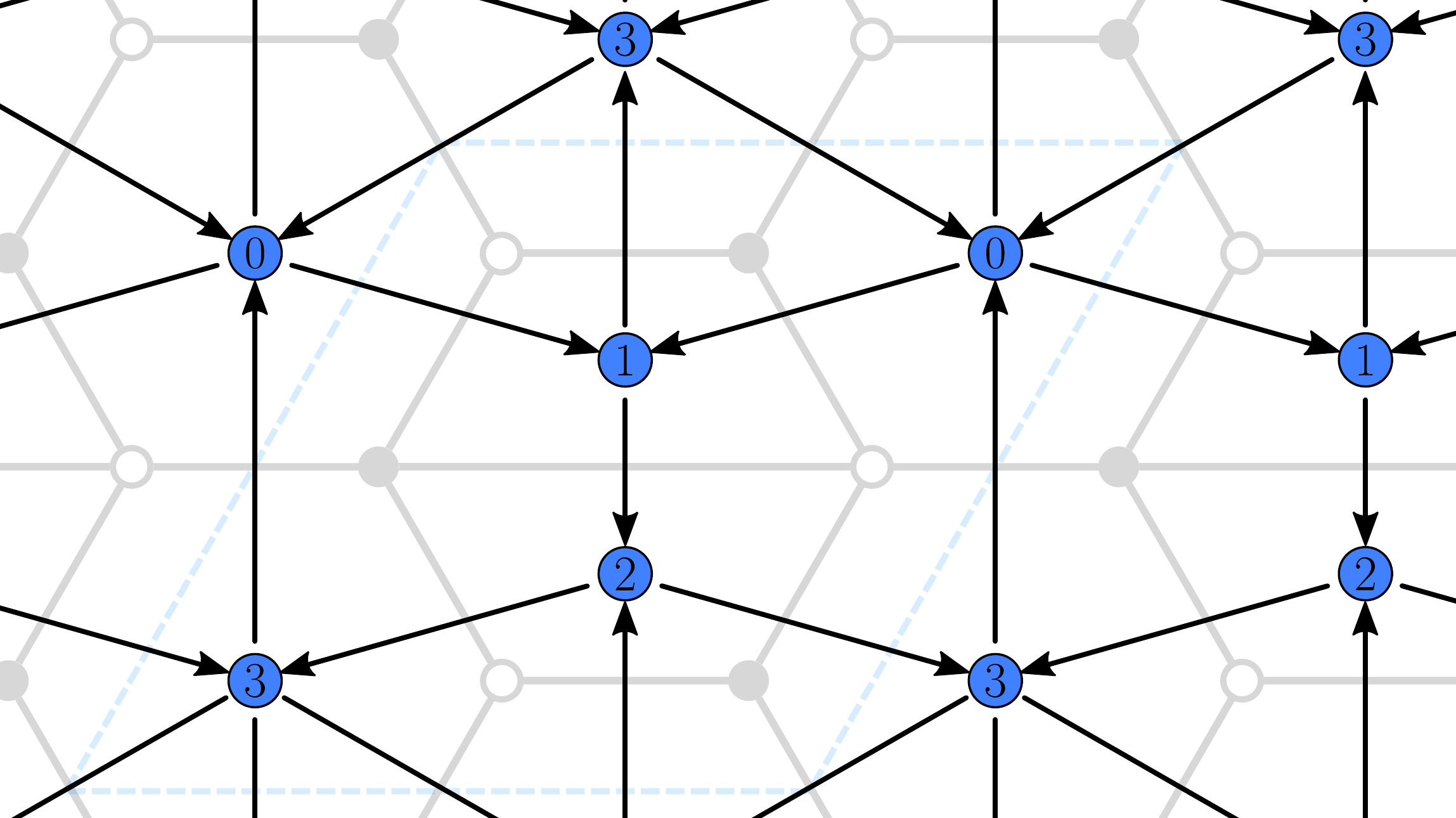}
			\caption{}
			\label{periodicquiver-dp1}
		\end{center}
	\end{subfigure}
	\caption{Periodic quiver for the conifold (\ref{periodicquiver-conifold}) and $dP_1$ (\ref{periodicquiver-dp1}).}
	\label{periodicquiver-conifold-dp1} 
\end{figure}

It will be convenient to introduce the periodic quiver as the dual of the dimer diagram. Namely, each face is replaced by a node, which we continue to denote by $F_a$; each edge $E_i$ separating face $F_a$ from face $F_b$ is replaced by an oriented arrow (again denoted by $E_i$) between nodes $F_a$ and $F_b$; and vertices $V_{\alpha}$, $V'_{\alpha}$ now correspond to plaquettes of arrows with clockwise or counter-clockwise orientation, respectively. Namely, in the periodic quiver, nodes correspond to gauge factors, arrows to bifundamental matter, and plaquettes to superpotential couplings. The periodic quiver is similar to the standard quivers used to describe gauge theories, with the extra periodic structure providing also the information about the superpotential. The periodic quivers for the conifold and $dP_1$ theories are given in Figure \ref{periodicquiver-conifold-dp1}.

It will be useful to have in mind that these ingredients in the dimer and the periodic quiver allow to define a (co)homology in the corresponding diagrams, ultimately related to the (co)homology in the underlying 2-torus. We have collected this description in Appendix \ref{sec:cohomology}.

\medskip

In the above description we have considered general ranks $N_a$. In general, these are constrained by cancellation of non-abelian anomalies. Denoting the (net) number of bifundamentals $(\fund_a,\antifund_b)$ by $I_{ab}$ (defined as an anti-symmetric matrix, with negative entries indicating matter in the conjugate representation), the conditions are
\beqa
\sum_a N_a I_{ab}\, =\, 0 \quad \forall b \fstop
\label{nonab-anomaly-cancellation}
\eeqa
In a bipartite dimer, any face has an even number of edges, so if we choose $N_a=N$ for all $a$ there are cancellations among consecutive edges and the anomaly-cancellation constraints are satisfied. This corresponds to the so-called regular or dynamical D3-branes (which can move off the singular point and explore the geometry). Choices of non-equal ranks include the so-called fractional branes, which can be regarded as higher-dimensional branes wrapped on the cycles collapsed at the singular point. In the following we will focus on regular D-branes, and mostly have the case $N=1$ in mind.

\medskip

\begin{figure}[!htp]
	\centering
	\begin{subfigure}[l]{0.45\textwidth}
		\centering
		\includegraphics[scale=0.5]{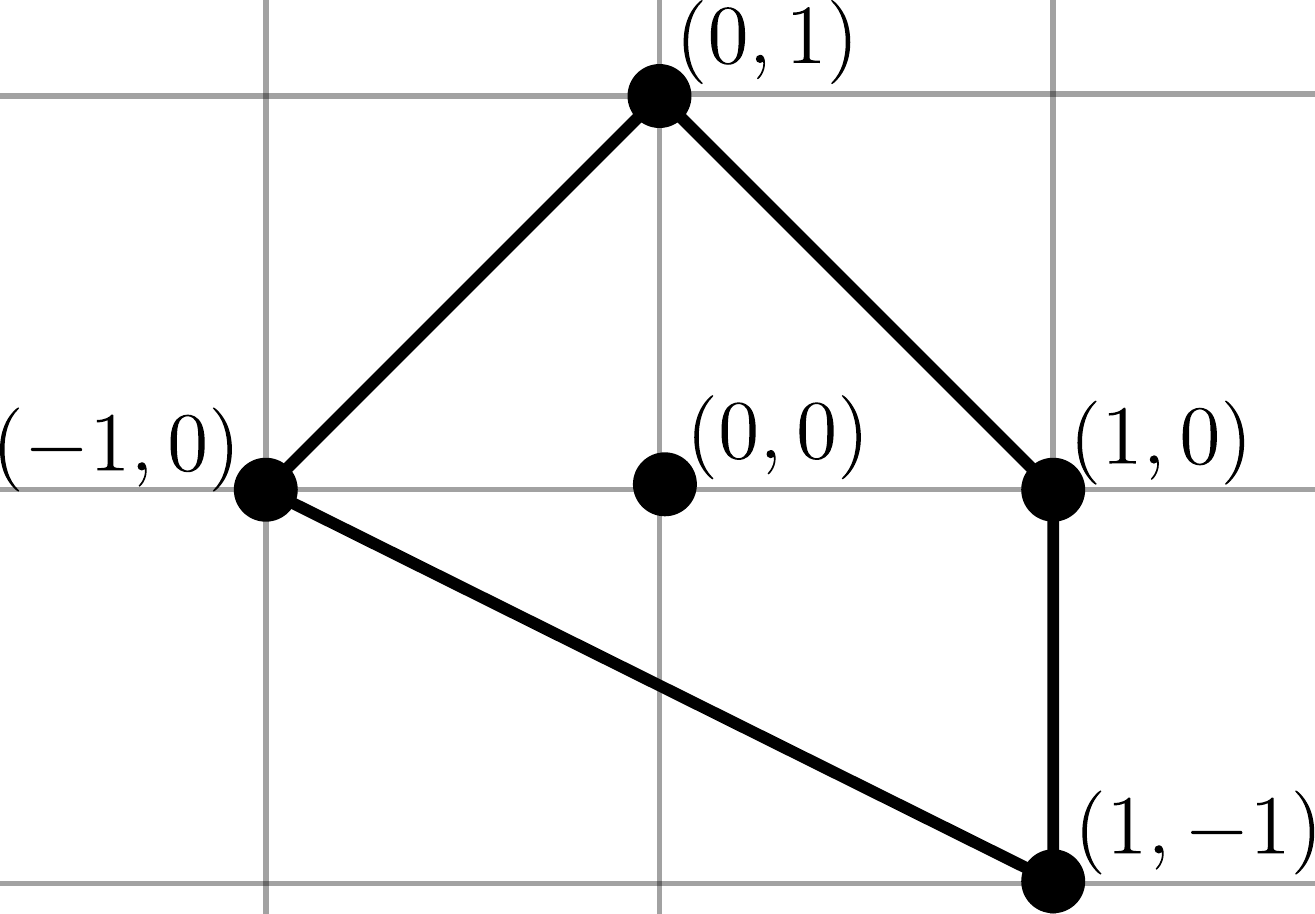}
		\caption{}
		\label{toricdiagram-dp1}
	\end{subfigure}
	\begin{subfigure}[r]{0.45\textwidth}
		\centering
		\includegraphics[scale=0.4]{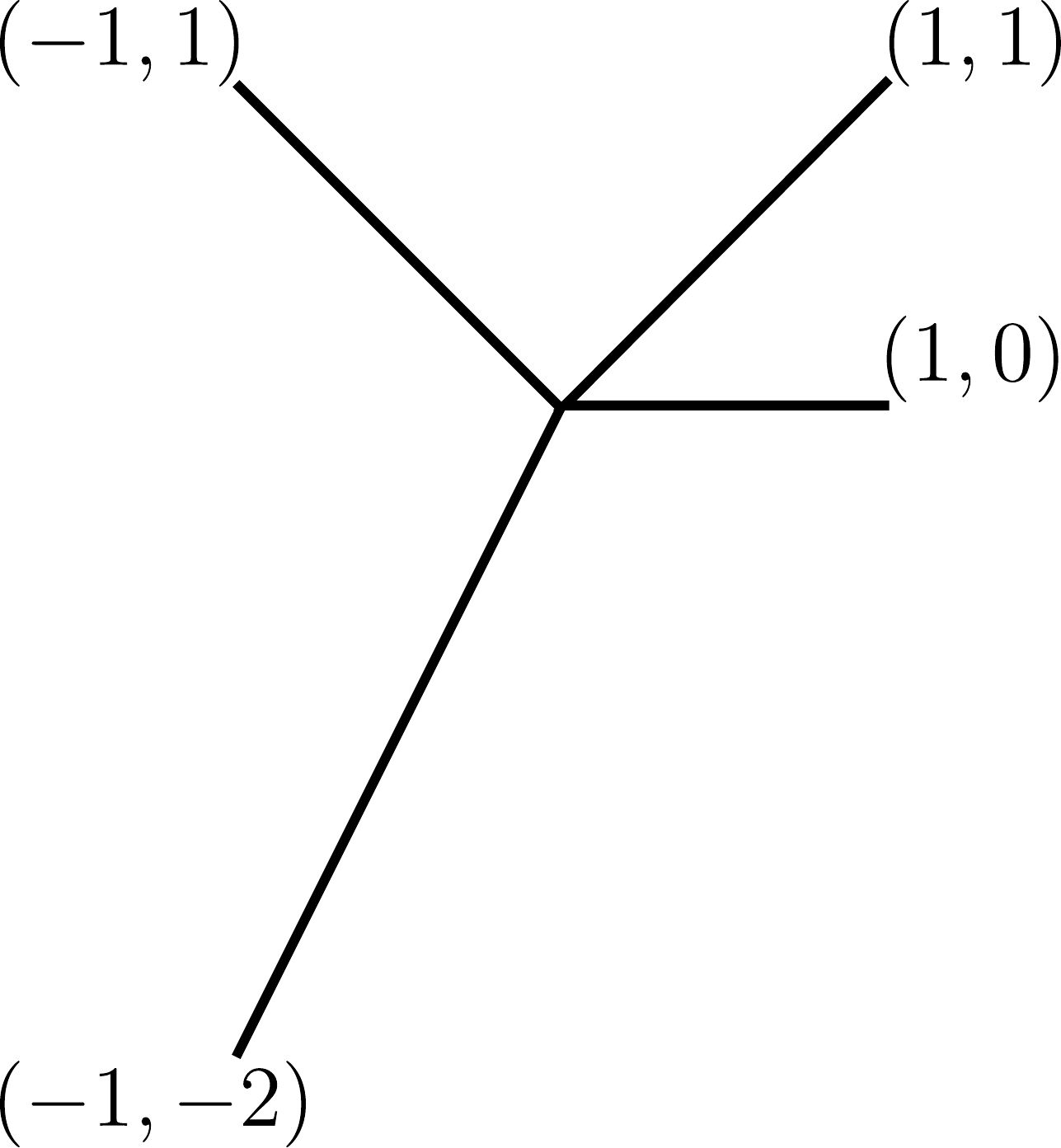}
		\caption{}
		\label{pqweb-dp1}
	\end{subfigure}
	\caption{Toric diagram (\ref{toricdiagram-dp1}) and (p,q)-web diagram (\ref{pqweb-dp1}) for $dP_1$ theory.}
	\label{toricdiagram-pqweb-dp1}
\end{figure}

The toric geometry associated to a given dimer diagram can be recovered in several (equivalent) ways. A very direct method is to introduce zig-zag paths. A zig-zag path in the dimer is a consecutive sequence of edges such that the path turns maximally to the left at e.g. black vertices and maximally to the right at white nodes. It can be depicted as a oriented path following edges and forced to cross them in the middle, see Figure \ref{dimer-conifold}. Each edge is thus crossed by two oppositely oriented zig-zag paths. Zig-zag paths cannot self-intersect, and define $(p,q)$ 1-cycles in the dimer 2-torus. Each such path corresponds to an  external leg in the web diagram\footnote{This correspondence is by defining a height function, defined as an integer-valued stepwise function increasing by one unit as one crosses the path (with positive orientation). The labels of the external leg in the web diagram are obtained as the jumps of the height function along the two basic 1-cycles in the 2-torus. In practice, this is equivalent (up to some relabeling) to just taking the $(p,q)$ labels of the zig-zag path to be those of the external leg of the web diagram.} corresponding to the toric threefold singularity, namely the diagram dual to the toric data, see Figure \ref{pqweb-dp1}. 

Intuitively, this follows because the threefold geometry can be obtained as the mesonic moduli space of the gauge theory, and the zig-zag paths correspond to mesons of the gauge theory (obtained as the trace of the product of bifundamentals corresponding to the sequence of edges); notice that the F-term relations imply that mesons are only defined by the homology classes of the paths.

\begin{figure}[!htp]
	\begin{center}
		\includegraphics[scale=0.2]{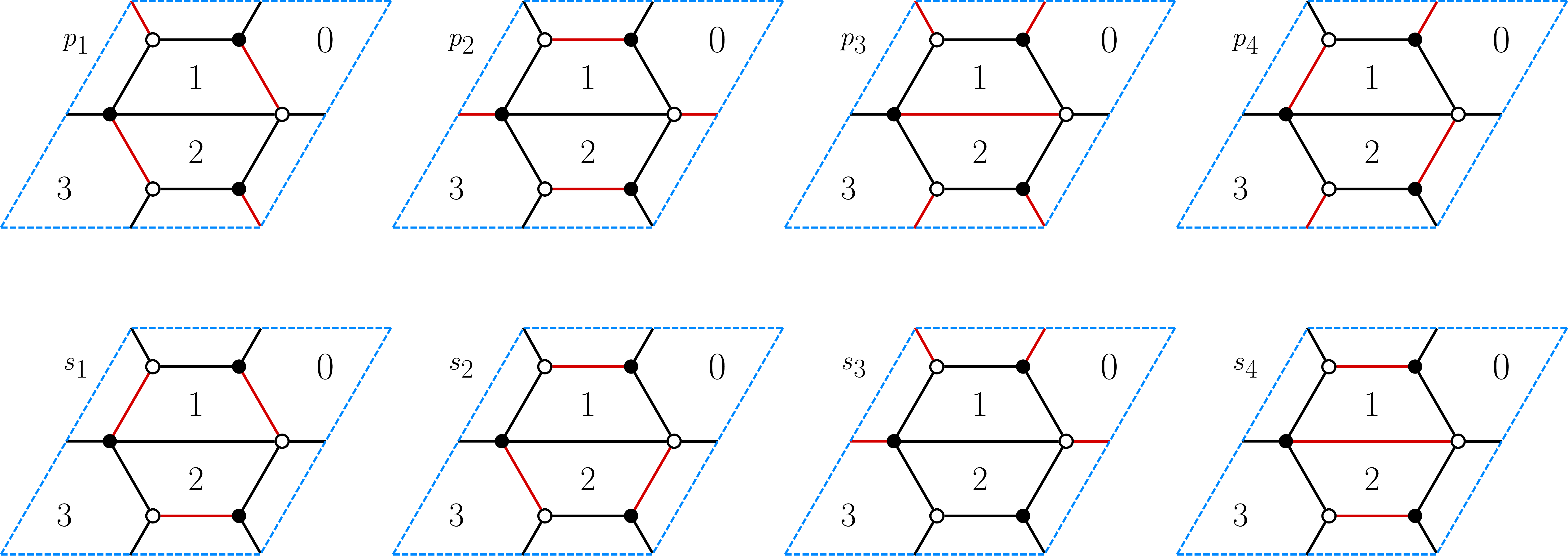}
	\end{center}
	\caption{Set of perfect matchings for $dP_1$ theory.} 
	\label{perfectmatchings-dp1} 
\end{figure}

A more detailed method to obtain the threefold geometry is  by introducing perfect matchings. A perfect matching in the dimer diagram is a set of edges such that each vertex in the dimer belongs to just one edge in the set, see Figure \ref{perfectmatchings-dp1} for an example. 
There is no closed formula for the number of perfect matchings in a given dimer, but it is easily determined in most examples.
At this point, one can recover the toric diagram as follows: regarding the perfect matchings $p_A$ as 1-chains (with orientation of edges from black to white nodes), one can fix a reference matching $p_0$ and obtain a set of 1-cycles in the dimer given by $p_A-p_0$. The $(p,q)$ labels of these 1-cycles correspond to the coordinates of points in the toric diagram of the threefold singularity, see Figure \ref{toricdiagram-dp1}. 
Note that this description is related to the previous paragraph because zig-zag paths can be obtained as differences of perfect matchings at consecutive external points in the toric diagram, namely segments in the toric diagram dual to precisely external legs in the web diagram.

An even more detailed relation with the toric description is by noticing that perfect matchings correspond to coordinates in the linear sigma model (or holomorphic quotient) description of the mesonic moduli space of the gauge theory. Let us review it, as it is useful to describe $\U(1)$ symmetries. In general, the bifundamentals are not useful coordinates to describe the moduli space, because they are constrained by the F-term conditions; perfect matchings are an efficient ingredient to solve these constraints automatically. The key idea is to define the bifundamentals in terms of the perfect matchings by the following relation
\beqa
\Phi_{E_i}\, =\, \prod_A \, p_A^{k_{i,A}} \coma
\label{bifund-pm}
\eeqa
where $k_{i,A}=1$ if $E_i$ belongs to the perfect matching $p_A$, and is zero otherwise. With this relation, all F-term constraints for the bifundamentals are solved automatically, with no restriction on the $p_A$. This is related to the fact that, from the very definition of perfect matchings, any term in the superpotential is given by the product of all perfect matchings
\beqa
W\sim \prod_A p_A \fstop
\label{supo-pm}
\eeqa
On the other hand, the above relation (\ref{bifund-pm}) introduces a redundancy, as $\IC^*$ transformations of the $p_A$ which leave the $\Phi_{E_i}$ invariant. These are defined by a set of charges $q_{A,r}$ satisfying
\beqa
\sum_A k_{i,A}q_{A,r}\, =\, 0 \quad \forall i \fstop
\label{eq:pms-Fterm}
\eeqa
The moduli space of F-flat directions is thus generated by the complex coordinates $p_A$ modulo these $\IC^*$ actions.
In addition, to obtain the mesonic moduli space we have to quotient by the $\U(1)$'s associated to the faces in the dimer. A bifundamental $\Phi_{ab}$ in the $(\fund_a,\antifund_b)$ carries charges $(+1,-1)$ under $\U(1)_a\times \U(1)_b$, and we need to translate these to charges for the coordinates $p_A$. Denoting by $q_{i,a}$ the charge of the bifundamental $\Phi_{E_i}$ under $\U(1)_a$, we introduce a matrix  of charges $q_{A,a}$ satisfying
\beqa
q_{i,a}\, =\, \sum_A k_{i,A} q_{A,a}\quad \forall i \fstop
\label{eq:pms-Dterms}
\eeqa
The charges $q_{A,a}$ define $\IC^*$ actions on the $p_A$ which implement the $\U(1)_a$ actions at the level of perfect matchings. The mesonic moduli space is thus obtained as the quotient of the complex coordinates $p_A$ by the $\IC^*$ actions generated by the charges $q_{A,r}$ and $q_{A,a}$. 

For illustration, consider one explicit example. In Figure~\ref{perfectmatchings-dp1} we see the perfect matchings for $dP_1$ theory. The moduli space of $F-$flat directions is found solving Equation~\eqref{eq:pms-Fterm}, which in our case, for instance, gives the following two $\IC^*$ actions on the perfect matchings:

\begin{equation}
	\left(
	\begin{array}{cccccccc}
		p_1 & p_2 & p_3 & p_4 & s_1 & s_2 & s_3 & s_4\\
		\hline
		1 & 0 & -1 & 1 & -1 & -1 & 0 & 1 \\
		-1 & -1 & 0 & -1 & 1 & 1 & 1 & 0 
	\end{array}
	\right) \fstop
\end{equation}
The other $\C^*$ actions can be found using Equation~\eqref{eq:pms-Dterms}, obtaining 
\begin{equation}
	\left(
	\begin{array}{cccccccc}
		p_1 & p_2 & p_3 & p_4 & s_1 & s_2 & s_3 & s_4\\
		\hline
		0 & 0 & 0 & 0 & 0 & 0 & 1 & -1 \\
		0 & 0 & 0 & 0 & 1 & 0 & 0 & -1 \\
		0 & 0 & 0 & 0 & 0 & 1 & 0 & -1 
	\end{array}
	\right)\fstop
\end{equation}
Combining the two matrices you get the complete set of $\IC^*$ actions on the perfect matchings which defines the mesonic moduli space. The kernel of the matrix is

\begin{equation}
	\left(
	\begin{array}{cccccccc}
		0 & -1 & 1 & 1 & 0 & 0 & 0 & 0 \\
		1 & 0 & 0 & -1 & 0 & 0 & 0 & 0 \\
		-1 & 1 & -1 & 0 & 0 & 0 & 0 & 0 \\
	\end{array}
	\right)\fstop
\end{equation}
The columns of this matrix are coordinates of points in 3d. All the points are on the plane defined by the equation $y=-x-z$ and, on that plane, the toric diagram of Figure~\ref{toricdiagram-dp1} is reproduced.

\subsection{Continuous $\U(1)$ symmetries}
\label{sec:cont-symm}

In the above discussion, it is implicit that the number of perfect matchings minus the number of $\IC^*$ actions is equal to 3, so that the symplectic quotient defines a threefold. An important implication is that the resulting geometry enjoys a $\U(1)^3$ symmetry (namely, the toric action making it a {\em toric} geometry). Namely, $\IC^*$ actions on the $p_A$ which are orthogonal to those we are quotienting by. Labeling them with $m=0,1,2$, their charges $q_{m,A}$ are given by the kernel of the combined matrix $(q_{A,a}|q_{A,r})$, namely satisfy:
\beqa
\sum_A q_{m,A} \, q_{A,a}=0 \quad \forall a\quad , \quad  \sum_A q_{m,A}\, q_{A,r}=0\quad \forall r \fstop
\eeqa
One of these $\U(1)$'s (which we label with $m=0$) is an R-symmetry\footnote{In the superconformal case, the actual R-symmetry is in general a combination determined by $a$-maximization, see \cite{Bertolini:2004xf,Benvenuti:2004dy}.}, whereas two linear combinations satisfying $\sum_A q_{m,A}=0$, $m=1,2$, leave $W$ in (\ref{supo-pm}) invariant, and correspond to $\U(1)^2$ mesonic symmetries.
In addition, there are $\U(1)_a$ symmetries associated to the faces. These correspond to baryonic symmetries, most of which are in fact anomalous. The mixed $\U(1)_a$-$\SU(N_b)^2$ anomaly is given by
\beqa
A_{ab}\propto N_a \, I_{ab} \quad {\rm (no}\; {\rm sum)} \coma
\eeqa
where $N_a$ arises as a normalization of the $\U(1)_a$ generator and $I_{ab}$ is defined as around (\ref{nonab-anomaly-cancellation}). It is thus clear that, denoting by $Q_a$ the generator of $\U(1)_a$, a general linear combination 
\beqa
Q_{\bf B}\, =\,\sum_a n_a Q_a
\label{baryonic-u1-combination}
\eeqa
defines an anomaly free baryonic $\U(1)$ when the $n_a$ satisfy the anomaly cancellation condition (\ref{nonab-anomaly-cancellation}), namely
\beqa
\sum_a n_a I_{ab}\,=\,0 \fstop
\label{baryonic-u1-anomaly-freedom}
\eeqa

That is, there is an anomaly-free baryonic $\U(1)$ for coefficients $n_a$ such that they could define a fractional brane. Let us emphasize, however, that for the anomaly-free baryonic $\U(1)$ to exist, it is not necessary that the fractional brane is present; hence our use of lowercase $n_a$ instead of $N_a$ in (\ref{baryonic-u1-combination}).

\subsection{A new toolkit: $\U(1)$ global symmetries from Geometric Identities}
\label{sec:geomident}

In this section we introduce a new ingredient, which to our knowledge has not appeared in the literature. 

Let us discuss global $\U(1)$ symmetries in the gauge theory from a somewhat more abstract perspective, using the topological intuitions in the dimer/quiver diagrams introduced in Appendix \ref{sec:cohomology}.

A $\U(1)$ symmetry is an assignment of charges to the edges $E_i$ (or arrows $E_i$) in the dimer (resp. quiver) diagram of the gauge theory. We may regard this as a 1-form $\gamma$ on the quiver, namely a map that to each arrow $E_i$ assigns a number (the charge) $\gamma(E_i)$. Regarding the arrow as a 1-chain, this is also the integral of the 1-form over the 1-chain. One may also regard it as a 1-form in the dimer, which we also denote $\gamma$.

These charge assignments are constrained by demanding invariance of the superpotential. This means that for each plaquette $V_\alpha$ (or $V_\alpha'$) in the quiver, with boundary given by a concatenation of arrows $\{E_1,E_2,\ldots, E_n\}$, the 1-form $\gamma$ satisfies
\beqa
\partial V_{\alpha}, \partial V_{\alpha}'\quad \to \quad\gamma(E_1)\, +\,\gamma(E_2)\,+\cdots +\, \gamma(E_n)\, =\, 0 \fstop
\label{closed-form-invsupo}
\eeqa
Recalling from Appendix \ref{sec:cohomology} the definition of exterior derivative and using Stokes' theorem over the plaquette, we have
\beqa
d\gamma=0 \fstop
\eeqa
Regarding $\gamma$ as realized in the dimer, these correspond to the so-called harmonic maps in the math literature. We stick to the nomenclature suggested by the notation, and refer to these as closed 1-forms in the periodic quiver. 

As expected, closed 1-forms in finite graphs without torus periodicities must be exact, namely there exists a 0-form $f$ in the quiver (namely, a map assigning a number $f(F_a)$ to each quiver node $F_a$) such that
\beqa
\gamma=df \fstop
\eeqa
Namely, if we denote by $h(E_i)$, $t(E_i)$ the head and tail of the arrow $E_i$, then
\beqa
\gamma(E_i)\, =\, f(t(E_i))\, -\, h(t(E_i)) \fstop
\eeqa
Physically, if we denote $n_a\equiv f(F_a)$, this means that the charge assignment for edges given by $\gamma$ is just inherited from the $\U(1)_a$ charges via a linear combination 
\beqa
Q\, =\,\sum_a n_a Q_a \fstop
%\label{baryonic-u1-combination}
\eeqa
Namely just like (\ref{baryonic-u1-combination}), with the only difference that we are not yet demanding cancellation of anomalies. As explained above, such linear combinations in the toroidal graph correspond to (still possibly anomalous) baryonic $\U(1)$ symmetries. Hence, mesonic $\U(1)$ symmetries are defined as closed 1-forms (i.e. symmetries of the superpotential) which are not exact (i.e. are not baryonic), and correspond to combinations of the two independent homology classes of 1-forms in the 2-torus. Hence we recover the $\U(1)^2$ mesonic symmetry.

%As explained above, the existence of exact (hence closed) 1-forms in the 2-torus is possible when the theory admits fractional branes. Let us revisit the result as follows.

Let us discuss the anomaly cancellation conditions more explicitly, as follows. For each face $F_a$ in the dimer (resp. node in the quiver), surrounded by a concatenation of edges (resp. arrows) $\{E_1,\ldots,E_m\}$, the mixed $\SU(N_a)^2$
anomaly cancellation conditions read
\beqa
{\tilde \partial} F_a\quad \to \quad \gamma(E_1)\, +\cdots +\, \gamma(E_m)\, =\, 0 \fstop
\label{form-anomcancel}
\eeqa
Note that since the natural orientation of edges does not allow to write this equation as over the boundary of $F_a$ in the dimer, hence we use the notation ${\tilde \partial}$ for this `signed' boundary.

In this language, an anomaly free $\U(1)$ symmetry is a charge assignment satisfying the conditions (\ref{closed-form-invsupo}) and (\ref{form-anomcancel}). These form an homogeneous linear system of equations, with the number of unknowns given by the number $E$ of edges in the dimer, and with the number of equations given by the number $V$ of vertices plus the number $F$ of faces. Since the dimer is a tiling of the 2-torus, it satisfies
\beqa
F+V=E\fstop
\eeqa
Hence, the only non-trivial solution defining $\U(1)$ charges must require the existence of linear relations among the equations. Indeed, a general dimer always has two such relations, which we may write
\begin{align}
\begin{split}
\sum_\alpha\, \partial V_\alpha\, -\, \sum_\alpha \partial V_{\alpha}'\,=\, 0 \coma\\
\sum_a {\tilde \partial}F_a\, - \, \sum_\alpha\, \partial V_\alpha\,-\, \sum_\alpha \partial V_{\alpha}'\,=\, 0\fstop
\label{geometric-identities}
\end{split}
\end{align}
These can be regarded as {\em geometric identities} which the elements of the dimer/quiver diagrams satisfy. The two anomaly free solutions which exist for any general dimer due to these universal geometric identities correspond to the $\U(1)^2$ mesonic symmetries. 

In addition, we know that theories admitting fractional branes, have additional anomaly-free baryonic $\U(1)$'s. These correspond to linear combinations (\ref{baryonic-u1-combination}) satisfying (\ref{baryonic-u1-anomaly-freedom}). This requires that the above linear system of equations admits further geometric identities for theories admitting fractional branes. Indeed, in such cases it is possible to show that $\sum_a n_a {\tilde \partial}F_a$ can be recast as a combination of $\partial V_\alpha$ and $\partial V_{\alpha}'$. 
We will find explicit examples in later sections.

\medskip

Incidentally, we would like to mention that, if we interpret $\gamma(E_i)$ not as charges, but as the exact anomalous dimensions for the bifundamental chiral multiplets, the above analysis is very closely related to the Leigh-Strassler characterization of marginal couplings in $\NN=1$ SCFTs \cite{Leigh:1995ep} (see \cite{Imamura:2007dc} for a discussion in the present context). Namely, the conditions of vanishing of the exact beta functions for the superpotential couplings (at $V_\alpha$, $V_{\alpha}'$) and for the gauge couplings (at $F_a$) form an inhomogeneous linear system of equations for $\gamma(E_i)$, whose associated homogeneous linear system is precisely given by the above, (\ref{closed-form-invsupo}) and (\ref{form-anomcancel}). Moreover, the inhomogeneous system of equations satisfies relations given precisely by the universal (\ref{geometric-identities}), allowing for the existence of a marginal coupling corresponding to the complexified coupling constant of the diagonal gauge group on the recombined regular brane. Additional geometric identities imply additional marginal couplings, associated to the gauge couplings of the corresponding fractional branes. 

\medskip

Let us conclude by mentioning the realization of these $\U(1)$ symmetries and marginal couplings in the holographic dual. For systems of D3-branes at toric singularities, there is a generic $\U(1)^3$ isometry in the horizon, which includes the R-symmetry and the $\U(1)^2$ mesonic symmetry. There is also a universal massless scalar, given by the axio-dilaton, dual to the marginal coupling. If the theory admits additional anomaly-free rank assignments (fractional branes), the gravity dual internal space $\IX_5$ contains homology 3-cycles, supporting additional $\U(1)$'s arising from integrating the RR 4-form over them; also, their dual 2-cycles produce additional massless scalars from integrating the NSNS and RR 2-forms, which are duals to the additional marginal couplings of the holographic dual gauge theory.

The purpose of this paper is to  extend this matching to the discrete symmetries. The natural arena are orbifolds.

\section{Discrete Symmetries in Orbifolds of Toric Geometries: An appetizer}
\label{sec:discrete-appetizer}

The purpose of this paper is to uncover the discrete symmetries in gauge/gravity duals corresponding to orbifolds of toric geometries. We thus start with a general description of orbifolds of general toric CY threefold singularities.

\subsection{General Orbifolds of General Toric Theories}
\label{sec:general-orbifolds-general-toric}

Consider a general toric gauge theory, with a dimer (resp. quiver) diagram with unit cell ${\cal C}$, with faces $F_a$, edges $E_i$ and vertices $V_\alpha,V_{\alpha}'$. There is a general procedure to construct general abelian $\IZ_N$ orbifolds of this theory \cite{Schmaltz:1998bg,Uranga:1998vf} as follows. Denote by $Q_1$, $Q_2$ the two mesonic $\U(1)$'s, normalized to have minimal charge $\pm 1$, and consider the linear combination 
\beqa
Q_\theta \,=\, p_1 Q_1 \,+\, p_2 Q_2 \quad p_1,p_2\in\IZ\; ,\; {\rm \gcd}(p_1,p_2)=1\fstop
\label{Qtheta}
\eeqa
Let us denote by $k_{E_i}$ the charge under $Q_\theta$ of the bifundamental associated to the edge $E_i$. We consider the action of the generator $\theta$ of the orbifold group $\IZ_N$ to be given by
\beqa
\theta\; : \quad \Phi_{E_i}\; \rightarrow \exp \Big(\, 2\pi i\, \frac{k_{E_i}}N\, \Big) \, \Phi_{E_i}\fstop
\label{theta-on-fields}
\eeqa
In addition, there is an action of $\theta$ on the gauge degrees of freedom, inspired by the action of Chan-Paton indices of D-branes. Namely, such that an object in the fundamental representation of $\U(N_a)$ transforms with an order $N$ matrix
\beqa
\gamma_{\theta,a}\,=\, {\rm diag}\; \big({\bf 1}_{n_{a,0}},\, e^{2\pi i/N}\, {\bf 1}_{n_{a,1}},\,\ldots, e^{2\pi i (N-1)/N}\, {\bf 1}_{n_{a,N-1}} \; \big) 
\eeqa
and with its inverse on anti-fundamental representations.

The orbifold theory is obtained by removing fields of the parent theory which are not invariant under the combined action of the mesonic and gauge action. In particular, gauge bosons are singlets under the mesonic action, so demanding invariance of the generators $\lambda_a$ of $\U(N_a)$ under the gauge action of $\gamma_{\theta,a}$
\beqa
\lambda_a\,=\, \gamma_{\theta,a}\, \lambda_a\, \gamma_{\theta,a}^{\; -1}
\eeqa
breaks the group as follows
\beqa
\bigotimes_a U(N_a)\; \rightarrow \; \bigotimes_a\, \bigotimes_{r=0}^{N-1} \, U(n_{a,r}) \fstop
\eeqa
Here the treatment of the $\U(1)$'s is as discussed above, namely they are realized just as (potentially anomalous) global symmetries.

For an edge $E_i$ separating two faces $F_a$, $F_b$ in the dimer (respectively, an arrow with $t(E_i)=F_a$, $h(E_i)=F_b$ in the quiver), and with charge $q_{E_i}$ under (\ref{Qtheta}), the invariance of the bifundamental field $\Phi_{E_i}$, regarded as a matrix is
\beqa
\Phi_{E_i} \, =\, e^{2\pi i\, k_{E_i}/N} \, \gamma_{\theta,a}\,\Phi_{E_i}\, \gamma_{\theta,b}^{\; -1} \coma
\eeqa
leading to a projection pattern of the bifundamental $E_i$ into a set of bifundamentals $E_{i,r}$ as follows
\beqa
(\fund_a,\antifund_b)\; \rightarrow \; \bigoplus_{r=0}^{N-1}\, (\fund_{a,r},\antifund_{b,r+k_{E_i}} )\fstop
\eeqa
Finally, the superpotential of the orbifold theory is obtained by simply replacing the surviving fields in the superpotential terms of the parent theory. It is easy to see that a superpotential term in a vertex $V_\alpha$ (or $V_{\alpha}'$), describing the interaction of a concatenated set of fields 
$\{E_{1},\ldots,E_{n}\}$, 
leads to a set of superpotential terms $V_{\alpha,r}$ (resp. $V_{\alpha,r}'$) describing the interaction of the set of fields $\{E_{1,r},\ldots,E_{n,r}\}$.

\begin{figure}[!htp]
	\begin{center}
		\includegraphics[width=0.6\textwidth]{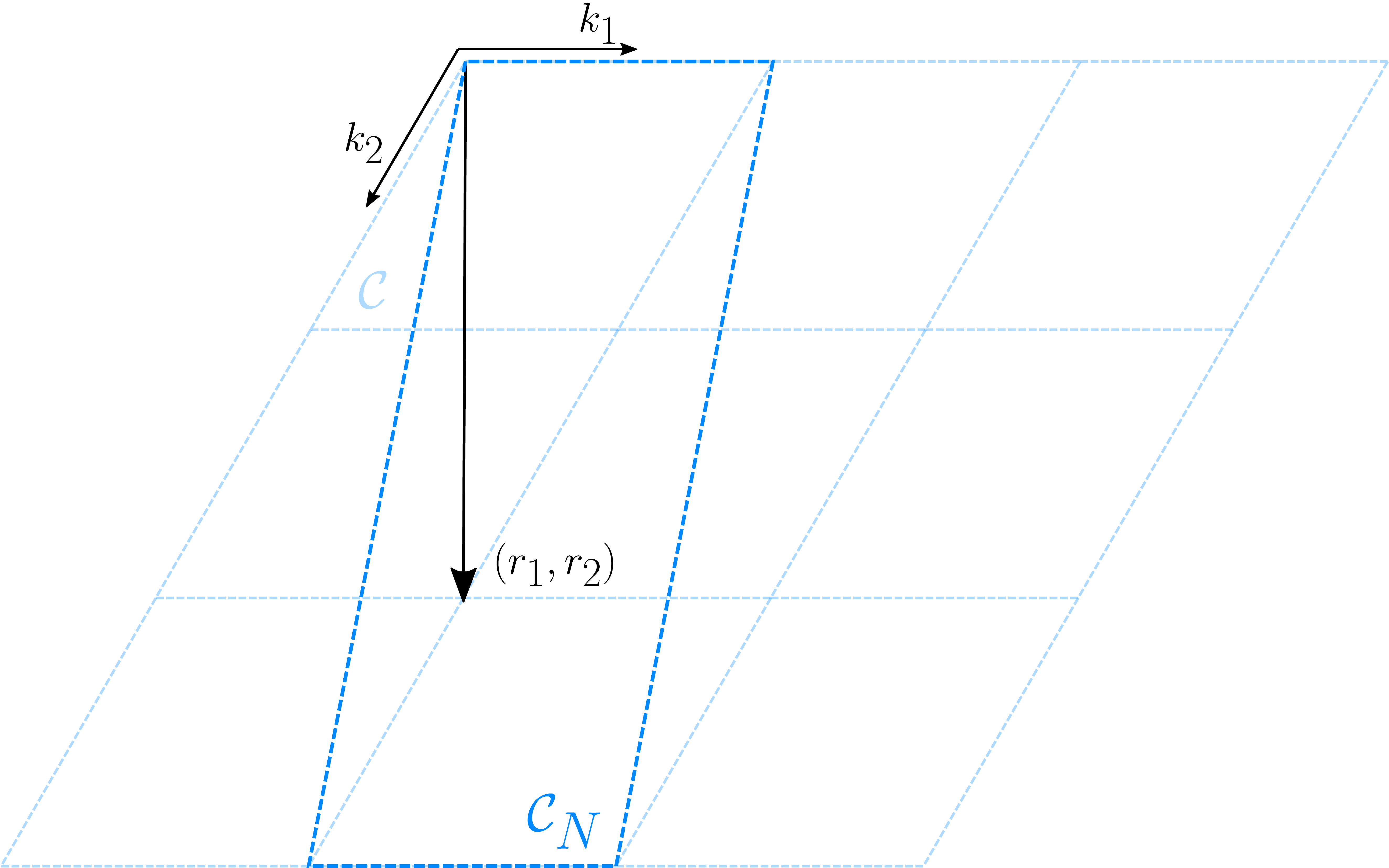}
	\end{center}
	\caption{General orbifolding in a dimer using a periodic array of unit cells (in light blue). The final unit cell $\mathcal{C}_N$ is shown in blue.}
	\label{GeneralOrbifold} 
\end{figure}

The orbifold theory is described by a dimer/quiver diagram whose unit cell ${\cal C}_N$ is obtained by taking $N$ copies of the unit cell ${\cal C}$ of the parent theory. Hence, each ingredient of the parent theory has $N$ descendant copies in the orbifold theory. We can be more explicit about how the different copies of ${\cal C}$ are adjoined to form ${\cal C}_N$, as follows, see Figure \ref{GeneralOrbifold}. Consider the infinite periodic array in $\IR^2$ corresponding to the parent theory, and choose a unit cell ${\cal C}$, and two basis 1-cycles. The latter correspond to vectors in $\IR^2$ defining the periodicities. The infinite copies of ${\cal C}$ can be labeled by two indices $(r_1,r_2)$ according to their position in the direction of the basic 1-cycles. Consider now mesons of the theory in ${\cal C}$, with winding numbers $(1,0)$ and $(0,1)$, and denote by $k_1$, $k_2$ the charges of these mesons under $Q_\theta$ (\ref{Qtheta}), namely just the sums of the $Q_\theta$ charges $k_{E_i}$ of the edges/arrows $E_i$ involved in the corresponding meson. The mesons can be regarded as open paths in the infinite array, starting from a face in ${\cal C}$ to its copy face/node in the copies of ${\cal C}$ at the two independent adjacent positions. Now regarding the infinite array as the covering space for the orbifold theory, the open path joins a starting face/node $F_{a,r}$ with the faces/nodes $F_{a,r+k_1}$, $F_{a,r+k_2}$ in the two adjacent copies. In general, the copy of the face/node $F_{a,r}$ in the copy of ${\cal C}$ at the position $(r_1,r_2)$ in the infinite array corresponds to the face/node $F_{a,r+r_1k_1+r_2k_2}$. 

%\begin{figure}[!htp]
%		\begin{center}
%			\includegraphics[scale=0.4]{dualpol-mesons-dp1}
%		\end{center}
%	\caption{Dual polygon for the $dP_1$ theory, whose toric diagram is in Figure~\ref{toricdiagram-dp1}. The coordinates represent the charges of the mesons under the two mesonic symmetries.}
%	\label{dualpol-mesons-dp1} 
%\end{figure}

The integers $k_1$, $k_2$ determine the action of the orbifold on the mesons, namely on the coordinates of the toric geometry. In fact, they are related to the construction of orbifolds in terms of toric data. Basically, the $\IZ_N$ orbifold of any toric geometry is obtained by refining the two-dimensional lattice by an order $N$ vector, which in our present context is $(k_1,k_2)/N$. The action on the mesons is inherited from this by the standard relation between mesons and toric data as explained in Section~\ref{sec:dimer-quiver}.  See Figure \ref{dP1torZN} for an example.

Note that in general, if e.g. ${\rm \gcd}(k_1,N)=1$, we may take the unit cell ${\cal C}_N$ of the orbifold theory as the $N$ copies of ${\cal C}$ in the direction of the $(1,0)$ 1-cycle in ${\cal C}$. However, we prefer to work in the infinite array, and work for general $k_1$, $k_2$ with no special relation with $N$.
On the other hand, notice that since all $N$ copies of the unit cells arise in the infinite array, we may choose ${\rm \gcd}(k_1,k_2)=1$.

\subsection{Structure of the Discrete Heisenberg group}
\label{sec:general-heisenberg}

In the above construction, there is a manifest global discrete symmetry, with generator $A$ acting as $r\to r+1$ on the labels of the $N$ copies of the fields of the parent theory, namely
\begin{align}
\begin{split}
A:\, & F_{a,r}\;\to \; F_{a,r+1} \quad \Rightarrow \quad  \SU(n_{a,r})\;\to \SU(n_{a,r+1} ) \\
& E_{i,r}\;\to E_{i,r+1} \quad \Rightarrow \quad \Phi_{E_{i,r}}\;\to\; \Phi_{E_{i,r+1}} \\
& V_{\alpha,r} \; \to \; V_{\alpha,r+1} \quad {\rm (similar}\;{\rm for} \; V_{\alpha,r}'\; {\rm )}\fstop
\end{split}
\end{align}

This is just a $\IZ_N$ rotation of the theory, which in the context of orbifolds of $\IC^3$ is often referred to as quantum symmetry (as it is a symmetry of the quantum worldsheet theory, in the sense of the $\alpha'$ expansion).

This transformation corresponds to a shift of the unit cell of the parent theory ${\cal C}$ in the ${\cal C}_N$, which in fact is most easily discussed in the infinite periodic array in $\IR^2$. The shifts of ${\cal C}$ to the adjacent unit cells in the two independent directions correspond to the operations $A^{k_1}$ and $A^{k_2}$, respectively. Since ${\rm \gcd}(k_1,k_2)=1$, by Bezout's theorem there exist integers $r_1,r_2$ such that $r_1k_2+r_2k_2=1$, hence $A$ corresponds to the shift of the unit cell ${\cal C}$ to its copy in the position $(r_1,r_2)$.

\medskip

As pioneered in \cite{Gukov:1998kn}, and further explored in \cite{Burrington:2006uu} (see also \cite{Burrington:2006aw,Burrington:2006pu,Burrington:2007mj}) there are several examples or orbifolds of simple geometries, this discrete group can be accompanied by further symmetry generator $B$, defined as phase rotations of the bifundamentals, such that the symmetry is enhanced to a discrete Heisenberg group. 

\begin{figure}[!htp]
	\begin{center}
		\includegraphics[scale=0.35]{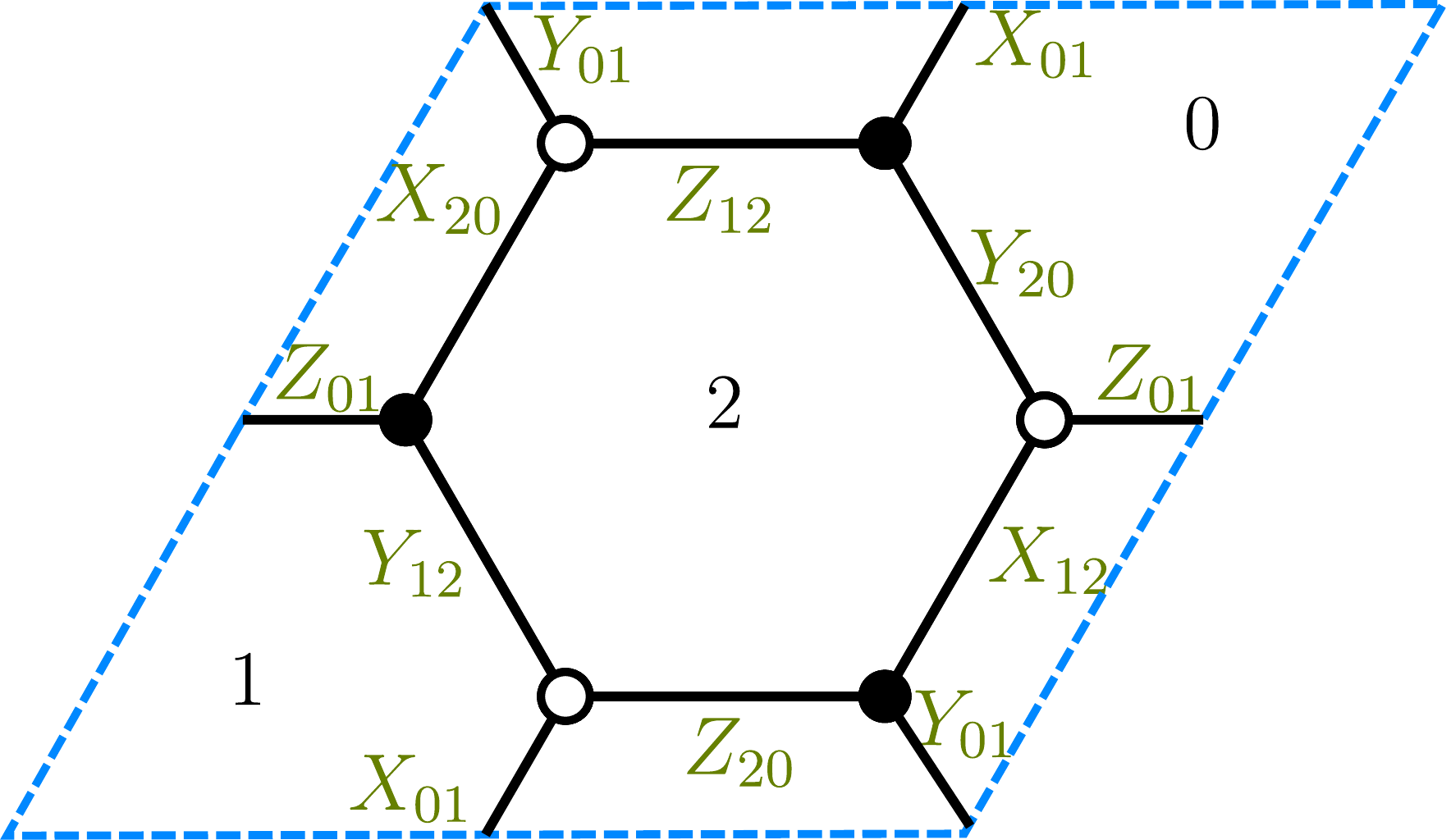}
	\end{center}
	\caption{Dimer diagram for the orbifold $\IC^3/\IZ_3$.} 
	\label{dimer_c3z3} 
\end{figure}

Consider for instance the orbifold $\IC^3/\IZ_3$ studied in \cite{Gukov:1998kn}. There are 3 gauge factors $\SU(N_r)$ with bifundamentals $X_{r,r+1}$, $Y_{r,r+1}$, $Z_{r,r-2}$ and a superpotential as follows from the dimer in Figure \ref{dimer_c3z3}. We can define $\omega=\exp[2\pi i/(3N)]$, so that on top of the global $\SU(3)$ symmetry, acting on the fields associated to the three complex planes, there is a global symmetry $B$ under which the fields transform, for instance, as
\begin{equation}
	B\,\left(\begin{array}{c}
		X_{01}\\ Y_{01} \\ Z_{01} 
	\end{array}\right)=\omega  \left(\begin{array}{c}
		X_{01}\\ Y_{01} \\ Z_{01} 
	\end{array}\right) \coma B\, \left(\begin{array}{c}
		X_{12}\\ Y_{12} \\ Z_{12} 
	\end{array}\right)=\omega^{-1} \left(\begin{array}{c}
		X_{12}\\ Y_{12} \\ Z_{12} 
	\end{array} \right)\e B\, \left(\begin{array}{c}
		X_{20}\\ Y_{20} \\ Z_{20} 
	\end{array}\right) = \left(\begin{array}{c}
		X_{20}\\ Y_{20} \\ Z_{20} 
	\end{array}\right) \fstop
\end{equation}
The actions $A$ and $B$ satisfy the commutation relation
\beqa
AB\,=\, CBA\coma
\eeqa
where the action of $C$ is
\begin{equation}
	C\,\left(\begin{array}{c}
		X_{01}\\ Y_{01} \\ Z_{01} 
	\end{array}\right)=\omega^{-2}  \left(\begin{array}{c}
		X_{01}\\ Y_{01} \\ Z_{01} 
	\end{array} \right)\coma C\, \left(\begin{array}{c}
		X_{12}\\ Y_{12} \\ Z_{12} 
	\end{array}\right)=\omega \left(\begin{array}{c}
		X_{12}\\ Y_{12} \\ Z_{12} 
	\end{array}\right) \e C\, \left(\begin{array}{c}
		X_{20}\\ Y_{20} \\ Z_{20} 
	\end{array}\right) = \omega\left(\begin{array}{c}
		X_{20}\\ Y_{20} \\ Z_{20} 
	\end{array}\right)\,,
\end{equation}
and commutes with both $A$ and $B$ so, it is central. Hence we recover a discrete Heisenberg group $H_3$, also known as $\Delta_{27}$.

There are examples of other orbifolds of $\IC^3$ studied for instance in \cite{Burrington:2006uu}. In the case of $\IC^3/\IZ_5$, we can call $W_{r,r+1}=(X_{r,r+1},Y_{r,r+1})$, while we keep $Z_{r,r-2}$ as it is. Let us also define $\omega=\exp[2\pi i/(5N)]$, so that the Heisenberg algebra is given by
	
\begin{align}
	\begin{split}
	B &\, \left( W_{01}\coma W_{12}\coma W_{23}\coma W_{34} \coma W_{40} \right)  = \left( W_{01}\coma \omega W_{12}\coma \omega^2 W_{23}\coma \omega^3 W_{34} \coma \omega^{-6}W_{40} \right)\coma \\
	B &\,  \left( Z_{03}\coma Z_{14}\coma Z_{20}\coma Z_{31} \coma Z_{42} \right)= \left( \omega^{3} Z_{03}\coma \omega^{6}Z_{14}\coma \omega^{-1}Z_{20}\coma \omega^{-3}Z_{31} \coma \omega^{-5}Z_{42} \right) \coma\\
	C &\, \left( W_{01}\coma W_{12}\coma W_{23}\coma W_{34} \coma W_{40} \right)  =\left(\omega W_{01}\coma \omega W_{12}\coma\omega W_{23}\coma\omega W_{34} \coma\omega^{-4} W_{40} \right)\coma\\
	C &\, \left( Z_{03}\coma Z_{14}\coma Z_{20}\coma Z_{31} \coma Z_{42} \right) = \left( \omega^{3}Z_{03}\coma \omega^{3}Z_{14}\coma \omega^{-2}Z_{20}\coma \omega^{-2}Z_{31} \coma \omega^{-2}Z_{42} \right) \fstop
	\end{split}
\end{align}
	
\medskip

We are looking for this structure on general orbifolds of general toric geometries. As explained above, the symmetry $A$ corresponds to the shift $r\to r+1$, which implements an order $N$ cyclic permutation among the gauge factors, acting correspondingly on the bifundamentals and superpotential terms. In addition, we look for an action $B$ under which the different bifundamental fields $E_{i,r}$ will transform with charges $b_{E_{i,r}}$, which in general depend on $r$. The actions $A$ and $B$ should anticommute to an action $C$, under which the bifundamentals $E_{i,r}$ transform with charges $c_{E_{i}}$, which, in order for $C$ to be central (and in particular commute with $A$), must be independent of $r$.

The procedure to construct the solution for these charges in general orbifolds of general toric theories is explained in Section~\ref{sec:general-solution}. Before entering this discussion, it is useful to introduce an important viewpoint.

\subsection{Discrete symmetries from the covering space}
\label{sec:discrete-covering}

Consider a given parent geometry, and the quotients defined by a $\IZ_N$ group with generator $\theta$ defined by an action (\ref{Qtheta}) associated to the two integers $(p_1,p_2)$. As discussed, there is a $\Gamma=\IZ_N$ quantum symmetry whose generator we denote by $A_N$, to emphasize its order. We also have a $\IZ_N$ generated by $B_N$, under which bifundamentals have charges $b_{E_i,r}$ defined modulo $N$.

In this section we are going to uncover the existence of a most important structure for the $B$ symmetry for the family of orbifolds of a given parent theory,  for fixed $(k_1,k_2)$, but different orders $N$.

The first observation is that it is useful to regard the charge assignments for the $B_N$ symmetry in the infinite periodic array in $\IR^2$, with periodicities (for fixed $k_1,k_2$) depending on $N$. This is motivated by the following argument. We regard the set of $B_N$ charges $b_{E_{i,r}}$ as defined for arbitrary $r\in\IZ$, hence on the infinite periodic dimer/quiver, but satisfying the periodicity $b_{E_{i,r}}=b_{E_{i,r+N}}$ mod $N$. We also have to impose the condition that $B_N$ leaves the superpotential invariant, and that it has no mixed anomaly with the gauge factors. These can be written
\begin{align}
\begin{split}
& \sum_{\partial V_{\alpha,r}} b_{E_{i,r}}\, =\, 0 \coma \\
&\sum_{\tilde\partial F_{a,r}} b_{E_{i,r}}\, =\, 0\fstop
\label{periodic-constraints}
\end{split}
\end{align}
Here the equations have to be satisfied modulo $N$. However, we now show that they actually must be satisfied as equations for integers, without resorting to the mod $N$ condition, as follows.

It is a familiar fact that quotienting the orbifold theory by the quantum symmetry $\Gamma$, one recovers the parent theory back. Similarly, if we consider some non-prime order $N=pN'$ with $p,N'\in \IZ$, and consider the element $(A_N)^{N'}$, it generates a $\IZ_{p}$ subgroup $\Gamma'\subset \Gamma$. Quotienting the theory by $\Gamma'$ should result in a theory which is a $\IZ_{N'}$ quotient of the parent, with the same pair $(k_1,k_2)$. This merely corresponds to considering the unit cell ${\cal C}_N$ of the $\IZ_N$ orbifold, and imposing the identification $r\to r+p$ on all elements to achieve a unit cell ${\cal C}_{N'}$ for the $\IZ_{N'}$ orbifold theory. Going back to the infinite periodic version, we have an initial set of charges $b_{E_{i,r}}$ (defined mod $N$) and we are changing from a periodicity set by $N$ to a periodicity set by $N'$. The requirement that the initial charges are compatible with the symmetry $B_{N'}$ of the discrete Heisenberg group of the $\IZ_{N'}$ theory implies that $b_{E_{i,r}}=b_{E_{i,r+p}}$ mod $N'$. Transferring this to the set of constraints (\ref{periodic-constraints}), and we find that they must be obeyed {\em modulo} $N'$. Considering $N$'s large enough, or rather, with large enough number of divisors, it is easy to convince oneself that the equations (\ref{periodic-constraints}) have to be obeyed in $\IZ$, without use of the mod $N$ conditions. 

In other words, the family of $\IZ_N$ theories, for fixed $(k_1,k_2)$ and varying $N$, has $B$ charges inherited from a universal assignment of integer charges $b_{E_{i,r}}\in\IZ$,  in the infinite periodic quiver/dimer i.e. $r\in \IZ$. The charges for the theory of a given $N$ are obtained by restricting the integer charges modulo $N$. The fact that this is compatible with the periodicities $b_{E_{i,r}}=b_{E_{i,r+N}}$ mod $N$, for any $N$, implies that  (\ref{periodic-constraints}) have to be obeyed directly, not modulo $N$.

In the following we ignore the sub-index $N$ in the discrete symmetry generators like $B$, $C$, and mostly work in whole families of $\IZ_N$ orbifolds, for fixed $(k_1,k_2)$, but varying $N$. As anticipated, this is most efficiently done by working on the infinite periodic array, with $B$ charges realized as integer charges therein.

\medskip

The fact that the constraints (\ref{periodic-constraints}) are defined without using the modulo $N$ condition has an interesting implication. In the language of appendix \ref{sec:cohomology}, the set of charges can be regarded as defining a 1-form $\gamma$, namely $\gamma(E_{i,r})=b_{E_{i,r}}$. Then the invariance of the superpotential requires the 1-form to be closed
\beqa
d\gamma=0\fstop
\eeqa
As explained above, solutions in the dimer 2-torus correspond to continuous $\U(1)$ mesonic (for non-exact $\gamma\neq df$) or baryonic (for exact $\gamma=df$) symmetries. This, together with the above considerations, suggests to look for symmetries defined by 1-forms $\gamma$ defined on the covering infinite periodic array. In $\IR^2$, any closed form must be exact $\gamma=df$; hence, we introduce a 0-form $f$ on the infinite periodic array. More concretely, using a label $r$ for the infinite set of faces/nodes, we assign an integer $n_a\equiv f(F_r)$ to each face of the (infinite) dimer (resp. node of the infinite quiver). 
This amounts to choosing a formal infinite linear combination of the $\U(1)_{a,r}$ generators $Q_{a,r}$
\beqa
Q_{B}\,=\, \sum_{a,r} \, n_{a,r} \, Q_{a,r}\fstop
\label{q-b}
\eeqa
So that a bifundamental associated to an edge $E$ separating faces $F_r$ and $F_s$ (resp. an arrow from node $F_{a,r}$ to node $F_{b,s}$) has an associated $B$ charge 
\beqa
b_E\,=\,\gamma(E)\,= \, n_{a,r}\, -\, n_{b,s} \quad , \quad {\rm for}\; \; t(E)=F_{a,r}\;  ,\; h(E)=F_{b,s}\fstop
\label{q-b-charges}
\eeqa

The values of $n_r$ are further constrained by cancellation of anomalies. In the following we use the description in term of the linear combination $Q_B$ to construct the discrete symmetries in several infinite classes of models, by solving the anomaly cancellation conditions by inspection. A systematic procedure to solve general orbifolds of general geometries is given in Section~\ref{sec:general-solution}.

\subsection{Example 1: The infinite class of general orbifolds of $\IC^3$}
\label{sec:example-general-c3}

Consider the infinite class of general orbifolds of $\IC^3$,defined by a generator $\theta$ acting as
\beqa
x \to e^{2\pi i \, k_1/N} \,x \quad , \quad  y \to e^{2\pi i \, k_2/N} \,y  \quad , \quad  z \to e^{2\pi i \, k_3/N} \,z\coma
\label{orbifold-action-c3}
\eeqa
with $k_1+k_2+k_3=0$, so we take the twist vector $(k_1,k_2,-k_1-k_2)/N$. The notation $k_1$, $k_2$ is chosen with hindsight to agree with their meaning in Section~\ref{sec:general-orbifolds-general-toric}.

The parent theory of D3-branes in flat space $\IC^3$ has three adjoints, $X$, $Y$, $Z$.  They are the basic mesons parametrizing $\IC^3$, so the orbifold action on them is inherited from (\ref{orbifold-action-c3}). They  carry charges $(1,0)$, $(0,1)$ and $(-1,-1)$ under the mesonic $\U(1)^2$, so this action corresponds to the $Q_\theta$ combination (\ref{Qtheta}) with $p_1=k_1$, $p_2=k_2$.

\begin{figure}[!htp]
	\begin{center}
		\includegraphics[width=0.5\textwidth]{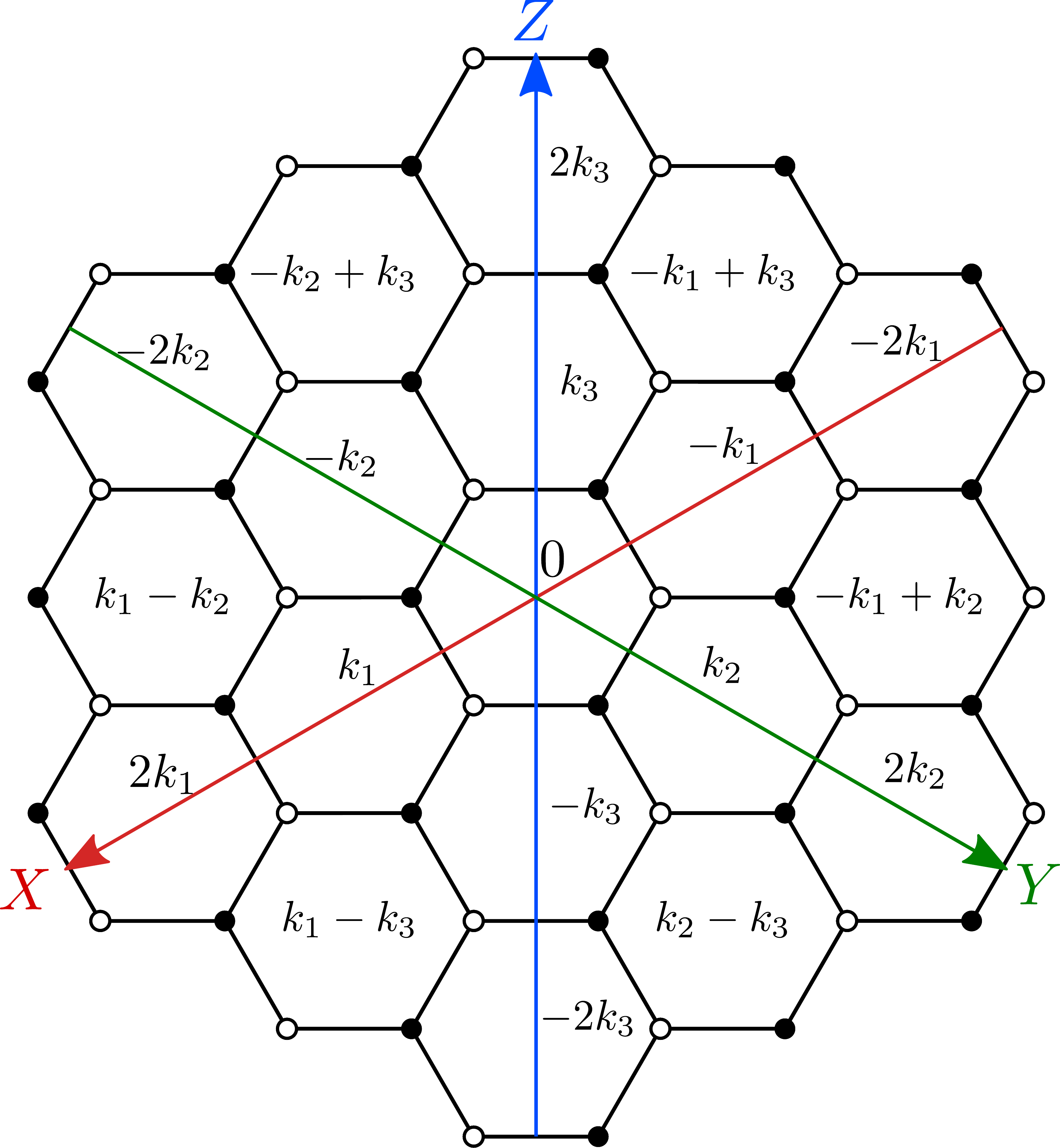}
		\caption{Dimer diagram for a general $\IC^3/\IZ_N$ orbifold.}
		\label{dimer-orbifold-general}
	\end{center}
\end{figure}

In the orbifold, the gauge group is a product of unitary factors $\U(n)_r$, with $r=0,\ldots, N-1$, and there are bifundamental fields
\beqa
X_{r,r+k_1}\quad , \quad Y_{r,r+k_2} \quad , \quad Z_{r,r+k_3}\coma
\eeqa
where the sub-indices denote the bifundamental representation, i.e. $\Phi_{rs}$ transforms in the $(\fund_r,\antifund_s)$. All the information, including the cubic superpotential, is encoded in a honeycomb dimer,  where now the unit cell contains $N$ different faces, labeled by $r=0,\ldots, N-1$, and where the index of the faces changes by $k_1$ and $k_2$ between neighboring faces, in the two independent directions (and hence, by $-k_1-k_2$ in the third, not linearly independent, direction). As explained, we prefer to consider the general class of orbifolds for arbitrary $N$, by considering the infinite periodic array in $\IR^2$, as shown in Figure \ref{dimer-orbifold-general}.

\begin{figure}[!htp]
	\begin{center}
		\includegraphics[width=0.4\textwidth]{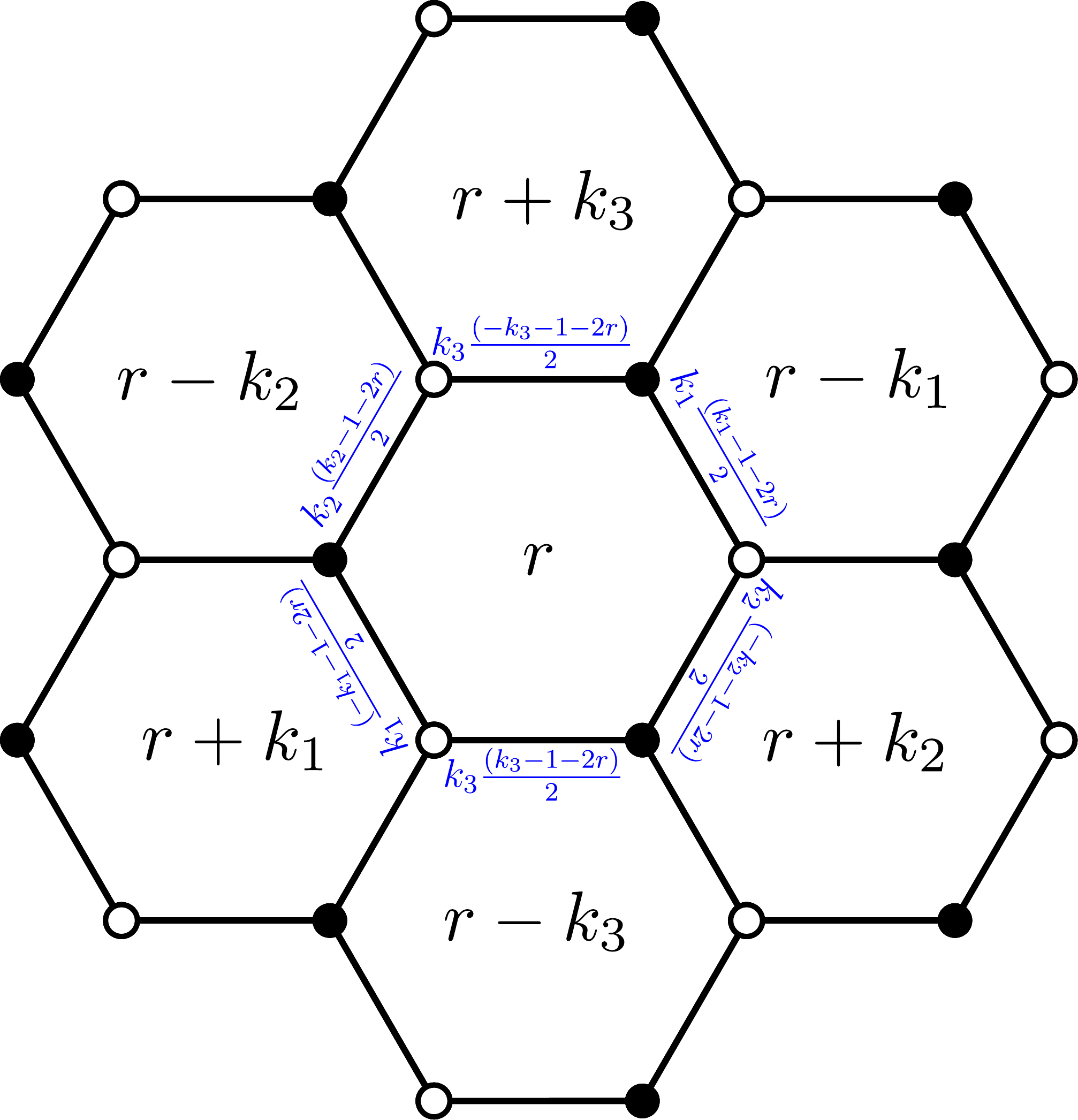}
		\caption{General face and B-charges (in blue) for bifundamentals  for a general $\IC^3/\IZ_N$ orbifold.
		}
		\label{dimer-orbifold-charges} 
	\end{center}
\end{figure}

As explained above, the generator $A$ of the Heisenberg group is realized as the action $r\to r+1$. On the other hand, the generator $B$ can be obtained from the combination (\ref{q-b}), dropping the index $a$ since there is only one face in the parent theory. The anomaly cancellation conditions for the coefficients $n_r$ thus reads
\beqa
n_{r+k_1}\, -\, n_{r-k_1}\, +\, n_{r+k_2}\,-\, n_{r-k_2}\,+\,n_{r+k_3}\,-\,n_{r-k_3}\,=\,0\fstop
\eeqa
It is not difficult to use known examples to try and infer a viable solution to the anomaly conditions, given by
\beqa
n_r\, =\, \frac {r(r+1)}2\fstop
\eeqa
This will be re-derived in Section~\ref{sec:general-solution-c3} from a general procedure, but for the moment we take it at face value. Using the charges for bifundamentals (\ref{q-b-charges}), see Figure \ref{dimer-orbifold-charges}, it is straightforward to check that the anomalies for an arbitrary face cancel.

We can now extract the charges under the $C$ symmetry. From the commutation relation $AB=CBA$ and the fact that $C$ is central, we have $AB^{k_1}=C^{k_1}B^{k_1} A$, which implies that the charges of the different bifundamentals under the $C$ symmetry can be obtained from the difference of charges of two copies of the bifundamental related by $r\to r+k_1$. The result is that for bifundamentals of $X$, $Y$ or $Z$ kind, the $C$-charge is given by 
\beqa
Q_C(E)= k_i\quad {\rm with}\quad i=1,2,3 	\quad {\rm for}\; {\rm bifundamentals}\; {\rm of} \; X, Y, Z\; {\rm kind} \fstop
\eeqa
It is straightforward to check that the anomalies cancel, and that periodicities are satisfied. In fact, this follows directly from the fact that the $C$-charges must be equal for all copies of a given bifundamental. This means that the $C$-charge can be defined on the dimer of the 2-torus {\em of the parent theory}. In other words, it is part of the mesonic $\U(1)^2$ symmetry of the $\IC^3$ theory (note that there are no baryonic $\U(1)$'s in this case), as is moreover clear from the above explicit charges.

\medskip

Given this universal solution, we can now find the discrete symmetry for any $\IZ_N$ orbifold by interpreting the labels $r$ mod $N$, and thus recovering the unit cell ${\cal C}_N$ of the orbifold theory as dictated by the corresponding identifications of faces in the infinite dimer. It is easy to check that the set of $B$ charges for the bifundamentals respects the corresponding periodicities, as follows.
Consider moving in the direction of $r\to r+k_1$, until we hit $r$ again (mod $N$). If we denote ${\rm \gcd}(k_1,N)=p$, this will happen after $N/p$ steps, so we have an identification $r\sim r+k_1N/p$.  The charges of all the bifundamentals charged under $\U(1)_r$ shifts by an amount $k_i k_1 N/p$ (with $i=1,2,3$ for fields of the $X$, $Y$ or $Z$ kind, respectively), which is 0 mod $N$ in all cases. Clearly, a similar result is obtained for the identifications in the directions $r\to r+k_2$ or $t\to r+k_3$.

Hence we have explicitly constructed the discrete Heisenberg groups $H_N$ for all orbifolds $\IC^3/\IZ_N$ with twist vector $(k_1,k_2,-k_1-k_2)/N$. We invite the reader to check that this general solution reproduces all known examples of discrete symmetries in orbifolds of $\IC^3$, in particular those of Section~\ref{sec:general-heisenberg}.

\medskip

\subsection{Example 2: The infinite class of general orbifolds of the conifold}
\label{sec:orbifold-conifold-app}

We now consider the infinite class of general orbifolds of the conifold. As show in Figure~\ref{dimer-conifold}, the conifold theory is described by two factors $\SU(N)_0 \times \SU(N)_1$, and bifundamentals $A_1$, $A_2$ in the $(\fund,\antifund)$, and $B_1$, $B_2$ in the $(\antifund,\fund)$. We define the orbifold by the action of its generator $\theta$ on these fields
\begin{align}
\begin{split}
\theta: & A_1 \; \rightarrow\; e^{2\pi i \, \frac{p_1}N} A_1 \quad ,\quad A_2 \; \rightarrow\; e^{-2\pi i \, \frac{p_1}N} A_2 \\
& B_1 \; \rightarrow\; e^{2\pi i \, \frac{p_2}N} B_1 \quad ,\quad B_2 \; \rightarrow\; e^{-2\pi i \, \frac{p_2}N} B_2 \fstop
\label{orbifold-action-conifold}
\end{split}
\end{align}
This agrees with the notation (\ref{Qtheta}), by noticing that the charges of $A_1$, $A_2$, $B_1$, $B_2$ under the mesonic $\U(1)^2$ symmetries \footnote{In this case, they are part of a larger $\SU(2)^2$ global symmetry.} are $(1,0)$, $(-1,0)$, $(0,1)$, $(0,-1)$.

Introducing the mesons
\beqa
x\,=\, A_1B_1\; , \; y\,=\, A_2B_2\; ,\; z\,=\,A_1B_2\; ,\; w\,=\, A_2B_1\coma
\eeqa
which satisfy $xy=zw$, the orbifold action is 
\beqa
\theta: \;  x\to e^{2\pi i\, \frac{p_1+p_2}N}\, x \; ,\; y \to e^{2\pi i\, \frac{-p_1-p_2}N}\, y 
\; ,\; z \to e^{2\pi i\, \frac{p_1-p_2}N}\, z \; ,\; w \to e^{2\pi i\, \frac{-p_1+p_2}N}\, w \fstop
\eeqa
Hence, in the notation of Section~\ref{sec:general-orbifolds-general-toric}, we have $k_1=p_1+p_2$, $k_2=p_1-p_2$.

\begin{figure}[!htp]
	\begin{center}
		\includegraphics[width=0.4\textwidth]{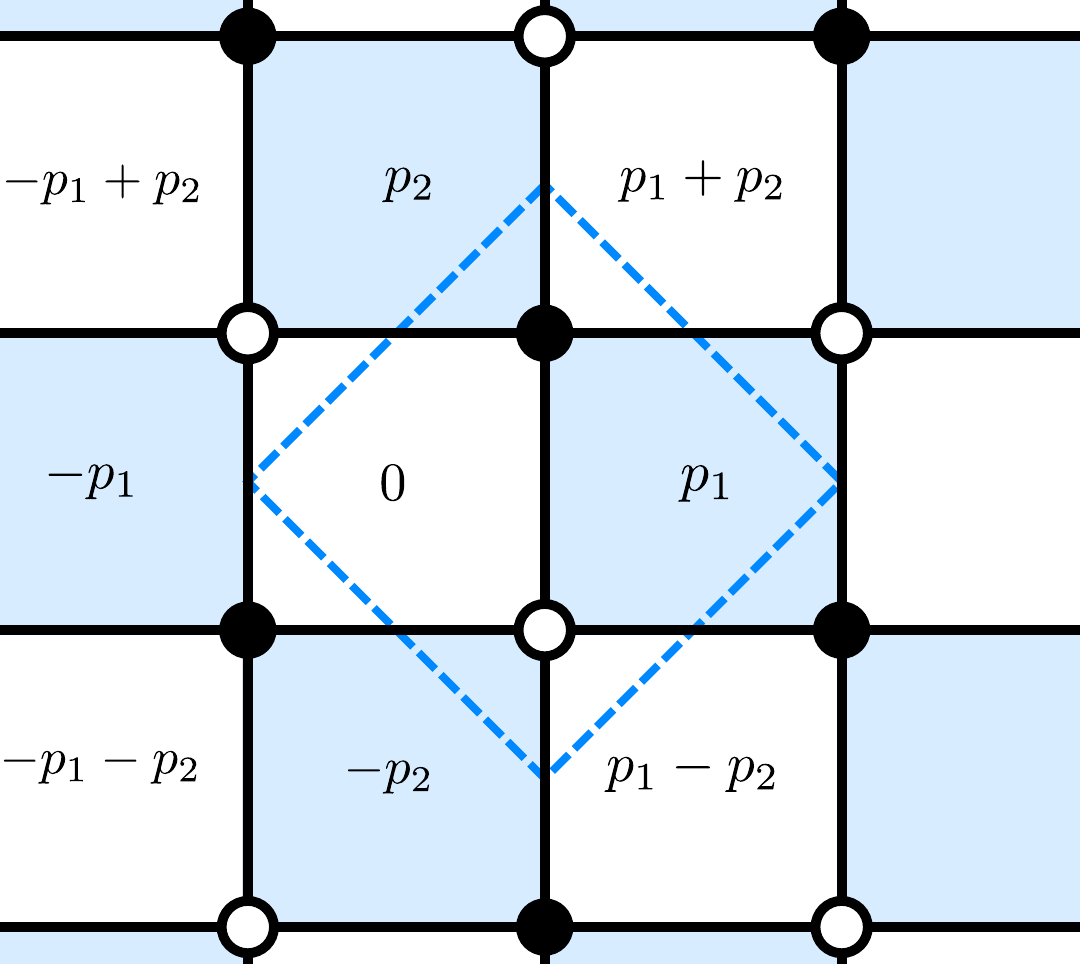}
		\caption{Dimer for general orbifold of the conifold. We show a unit cell of the parent theory with its two faces, and we display different background colors for their images.}
		\label{dimer-infinite-conifold} 
	\end{center}
\end{figure}

The dimer of this theory is shown in Figure \ref{dimer-infinite-conifold}. Note that in this case, there are two faces $F_a$ in the parent theory, and hence two kinds of faces $F_{a,r}$ in the quotient, $r\in\IZ$, shown in different background colors for clarity. The label displayed in the figure corresponds to the index $r$ of the corresponding kind of face. 

The $A$ symmetry acts by $r\to r+1$ as usual. To build the $B$ symmetry, we consider a general linear combination (\ref{q-b}). The anomaly cancellation conditions are
\beqa
n_{a,r+p_1}\, +\, n_{a,r-p_1} \,-\, n_{a,r+p_2}\,-\, n_{a,r-p_2}\,=\, 0\quad {\rm for}\; a=0,1\fstop
\eeqa
The condition for $a=0$ and $a=1$ decouple, and this makes it easy to find solutions. In particular we can take
\beqa
n_{0,r}=r \quad , \quad n_{1,r}=0\fstop
\eeqa

\begin{figure}[!htp]
	\begin{center}
		\includegraphics[width=0.5\textwidth]{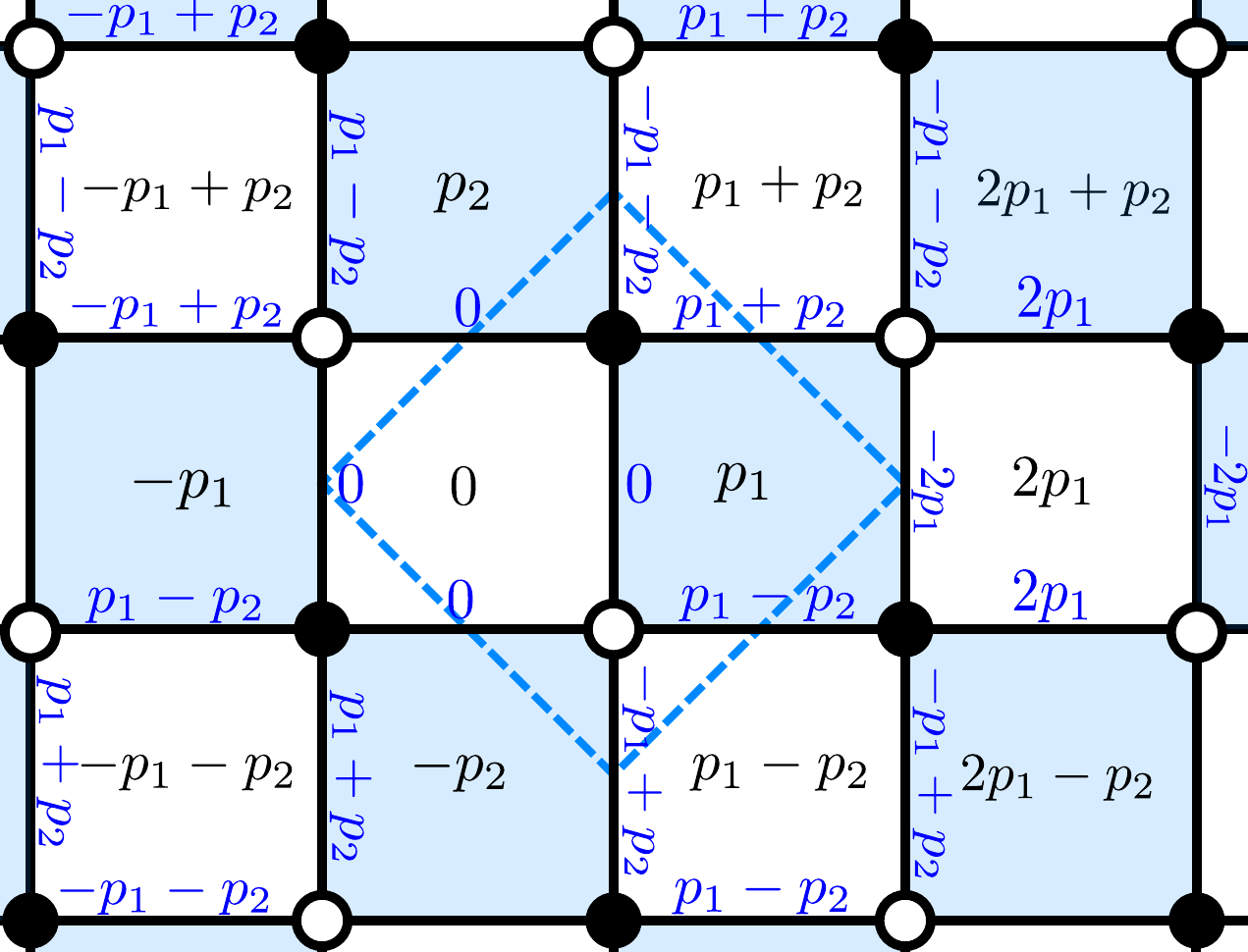}
		\caption{\small $B$-charges for general orbifold of the conifold. We take colored faces to have zero coefficient in the linear combination of $\U(1)$, while the coefficient for white faces is just its label. Hence, charges of edges around a white face are just given by the face label, with a sign corresponding to the bifundamental orientation.}
		\label{charges-infinite-conifold} 
	\end{center}
\end{figure}

The charges obtained are shown in Figure \ref{charges-infinite-conifold}, where the white faces are taken to correspond to $n_{1,r}=0$, and the colored faces to $n_{0,r}=0$. Hence the charges for edges around a face correspond to the face label weighted by the orientation of the bifundamental. It is straightforward to check that the anomalies for an arbitrary face cancel.

The charges $C$ can be read as the jumps in the $B$ charges as one acts with the shifts corresponding to $A$, and read
\beqa
Q_C(A_1)= 1\; ,\; Q_C(A_2)= 1\; ,\; Q_C(B_1)=-1\; ,\; Q_C(B_2)= -1. 
\eeqa
As is clear from these charges, the $C$ symmetry is actually an element of the baryonic $\U(1)$ of the parent theory.

The above results will be re-derived in Section~\ref{sec:general-solution-conifold} from a general procedure.

\section[Discrete Symmetries in Orbifolds of Toric Geometries: General solution]{\texorpdfstring{Discrete Symmetries in Orbifolds of Toric Geometries: \\General solution}{Discrete Symmetries in Orbifolds of Toric Geometries: General solution}}
\label{sec:general-solution}

In this section we provide a systematic recipe to construct the discrete symmetries of general orbifolds of general toric theories, by formulating the problem in the framework of the unit cell ${\cal C}$ of the parent theory. Morally, the problem amounts to solving for the set of charges (eventually, $B$-charges) for the edges/arrows in ${\cal C}$, with a twisted boundary conditions encoding the information of the orbifold action.

%\subsection{Reduction to a twisted 1-form in the unit cell}

We start with a short recap of the main lessons from the previous section. Given a toric theory, we consider the infinite periodic array for its dimer/quiver diagram, and label the copies of the ingredients (faces/nodes, edges/arrows, vertices/plaquettes) in the basic unit cell with an index $r\in\IZ$. Considering a unit cell ${\cal C}$, the orbifold is defined by two integers $(k_1,k_2)$, which specify the jumps $r\to r+k_i$ in the labels of ingredients as one moves from ${\cal C}$ to the adjacent unit cells in the two independent directions. For fixed $(k_1,k_2)$, this defines a family of orbifolds, with an extra parameter $N$ specifying the order of the $\IZ_N$ quotient. 

For a given $N$, there is a discrete Heisenberg group $H_N$ acting as a discrete symmetry of the theory. However, it is useful to consider generators $A$, $B$, $C$, with $AB=CBA$, of the symmetry in general, without explicit reference to $N$. This is done by considering the infinite array of the dimer/quiver diagram as the natural structure on which the symmetry acts. In particular, motion by one unit cell in the two independent directions corresponds to application of $A^{k_i}$, $i=1,2$. Also, the $B$-charges for the edges/arrows $E_{i,r}$ are defined as integer charges $Q_B(E_{i,r})=b_{E_{i,r}}$ for the edges/arrows in the infinite array, satisfying the conditions (\ref{periodic-constraints}) of invariance of the superpotential and anomaly cancellation, exactly and not just mod $N$. Finally, the commutation relations of the Heisenberg group imply that the $C$-charges for $E_{i,r}$ are defined as $r$-independent integers $Q_C(E_{i,r})=c_{E_i}$ in the infinite array, namely satisfying the periodicity of the unit cell ${\cal C}$ of the parent theory (hence, again independent of $N$). From this universal structure, the discrete symmetry generators for a particular choice of $N$ are obtained by restricting to the corresponding unit cell ${\cal C}_N$ and interpreting the $B$- and $C$-charges modulo $N$.

We now show that the infinite set of $B$-charges and equations in the infinite array can be actually reduced to a finite set in the unit cell ${\cal C}$ of the parent theory. This basically follows from the observation that the $B$ charges of two copies of an edge/arrow in the infinite array, $E_{i,r}$ and $E_{i,r+m_1k_1+m_2k_2}$ must be related by
\beqa
Q_B(E_{i,r+m_1k_1+m_2k_2})\,=\, Q_B(E_{i,r}) \, +\, (m_1k_1+m_2k_2)\, Q_C(E_{i})\fstop
\label{qb-periodic}
\eeqa
This moreover allows to reduce the number of constraints from superpotential invariance and anomaly cancellation to just those of the unit cell ${\cal C}$. To show that, we can for instance simply relate the anomaly cancellation constraints corresponding to edges bounding a face $F_{a,r}$ and its copy $F_{a,r+m_1k_1+m_2k_2}$, as follows
\beqa
Q_B({\tilde \partial} F_{a,r+m_1k_1+m_2k_2})\, = \, Q_B({\tilde \partial} F_{a,r})\, +\,(m_1k_1+m_2k_2)\, Q_C({\tilde \partial} F_{a})\fstop
\eeqa
Thus, the anomaly cancellation in the general copy of the unit cell reduces to the cancellation of the anomaly for the $B$- and $C$-charges for faces/nodes in the basic unit cell ${\cal C}$.

\begin{figure}[!htp]
\begin{center}

\includegraphics[scale=0.3]{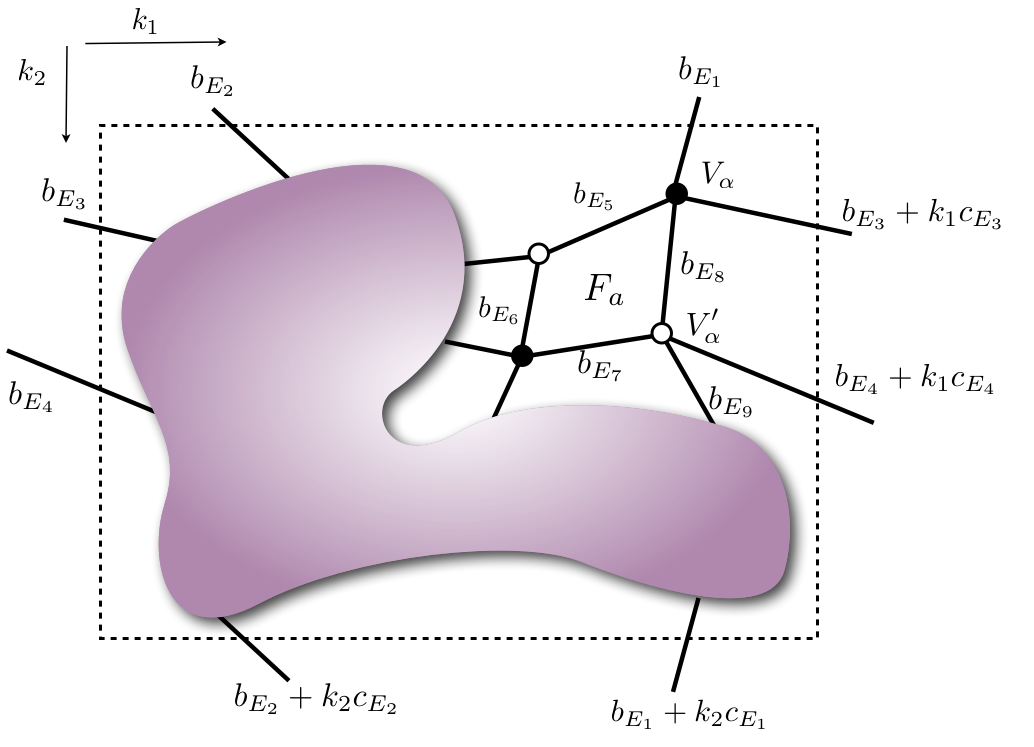}
\caption{$B$-charges in the unit cell of a general toric diagram. We display some of the ingredients, whereas the general structure is suggested by the blob. The structure of the orbifold is encoded in the jumps in $B$-charges in the two periodic directions of the unit cell.}
\label{unit-cell-charges} 

\end{center}
\end{figure}

Let us be a bit more explicit about the structure of $B$- charges. Consider the unit cell of a general theory, as shown in Figure \ref{unit-cell-charges}. We assign general charges $b_{E_i}$ to the edges $E_i$ in  the interior of ${\cal C}$. On the other hand, each edge on the boundary of ${\cal C}$ has a copy at another corresponding point of the boundary; hence, on the boundary we assign charges $b_{E_i}$ to a set of independent edges, and determine the charges of their copies by applying (\ref{qb-periodic}), i.e. $b_{E_i}+\ldots$, where the dots denote a piece depending on the $C$-charge $c_{E_i}$ of the edge, see again Figure \ref{unit-cell-charges}. Hence the general problem has a finite number of unknowns, corresponding to the $b_{E_i}$ and $C$-charges  $c_{E_i}$ of the edges/arrows of the unit cell. 

This number  of variables  would in principle be twice the number $E$ of edges/arrows in ${\cal C}$. However, recall that the $C$-charges have the periodicity of the unit cell ${\cal C}$. Moreover they must be symmetries of the superpotential and cancel mixed anomalies, i.e. regarding the $C$-charges $c_{E_i}$ as defining a 1-form $\gamma$ (via $\gamma(E_i)=c_{E_i}$ in the dimer/quiver in ${\cal C}$, they must satisfy
\beqa
\gamma(\partial V_\alpha)\,=\,\gamma(\partial V_\alpha')\,=\,0\quad ,\quad
\gamma ({\tilde \partial} F_a)=0 \fstop
\label{constraints-general-form}
  \eeqa
 These equations are just (\ref{closed-form-invsupo}) and (\ref{form-anomcancel}). This shows that $C$ is a discrete subgroup of the anomaly-free $U(1)$ symmetries of the theory. This includes the $U(1)^2$ mesonic symmetries $Q_1$, $Q_2$. In addition, there are in general $N_{\bf B}$ baryonic $U(1)$'s which we denote by $Q_{{\bf B}_i}$. Incidentally, we recall that these $U(1)$'s arise from linear relations among the above equations, due to the geometric identities.
The result is that the $C$ charge is a combination of these symmetries
 \beqa
 Q_C\, =\, m_1\, Q_1\, +\, m_2\, Q_2\, +\,\sum_{1=1}^{N_{\bf B}} m_{{\bf B}_i} \, Q_{{\bf B}_i} \fstop
 \label{c-charge-combination}
 \eeqa
 So the actual number of unknowns is given by $E$ from the $B$-charges and $2+N_{\bf B}$ from the coefficients in the above combination for the $C$-charges.
 
 Consider now the superpotential invariance and anomaly cancellation constraints the $B$-charges have to obey. These are given by (\ref{constraints-general-form}) where now $\gamma$ is the 1-form defined by the $B$-charges. Notice however that, since the $B$-charges do not satisfy the periodicities of the unit cell ${\cal C}$, this defines a {\em twisted} 1-form. In order to work with standard forms, we define the 1-form $\gamma$ by $\gamma(E_i)=b_{E_i}$, so the constraints correspond to an inhomogeneous linear set of equations for the $b_{E_i}$, whose associated homogeneous system is precisely (\ref{constraints-general-form}), and the inhomogeneous terms are combinations of the $C$-charges $c_{E_i}$. The number of equations is $F+V$, where $F$ is the number of faces/nodes and $V$ the number of vertices/plaquettes in ${\cal C}$. We may be tempted to consider that this defines a unique solution for the $b$'s in terms of the $c$'s, but additional care is required. Remember that the homogeneous system of equations is not linearly independent, since there are $2+N_{\cal B}$ linear relations arising from the geometric identities. This implies that, for the inhomogeneous system to admit solutions, the inhomogeneous terms must satisfy non-trivial consistency constraints. Namely, evaluating the geometric identities with the (twisted) $B$-charge assignments, the dependence on the $b_{E_i}$ disappears (because they are well-defined in ${\cal C}$ and hence obey the identity automatically), and we obtain certain combinations of the $C$-charges $c_{E_i}$ for some of the edges; these combinations must be zero for the inhomogeneous system to admit solutions. This provides $2+N_{\bf B}$ constraints on the $c_{E_i}'$, which are just enough to fix the $2+N_{\bf B}$ coefficients (\ref{c-charge-combination}) and thus determine the $C$-charges.
 
 We may now take the inhomogeneous system of equations for the $b_{E_i}$ and solve it in terms of the $c_{E_i}$. Since the number of independent equations (again, due to the geometric identities) is $F+V-2-N_{\bf B}$, the solutions for the $b_{E_i}$ are unique up to $(2+N-N_{\bf B})$ free parameters. But this is expected, since the  discrete symmetry $B$-charges can only be defined up to the addition of an arbitrary combination of the $2+N_{\bf B}$ continuous $U(1)$ global symmetries.

\medskip

In our procedure to solve for the $B$-charges we have not used the description in terms of the linear combination (\ref{q-b}), which we exploited in the examples in Sections \ref{sec:example-general-c3}, \ref{sec:orbifold-conifold-app}. Instead, the equations of invariance of the superpotential are written as part of our general linear system and handled simultaneously with the anomaly cancellation conditions. This is because, whereas the twisting of $B$-charges along the periodic directions in the unit cell ${\cal C}$ are easy to understand, it is a priori not clear how the coefficients $n_{a,r}$ change as one moves in these periodic directions. On the other hand, given a solution for the $B$ and $C$-charges, it is easy to go back to the linear combination (\ref{q-b}) and disclose these transformation properties, as follows. 

Consider the unit cell ${\cal C}$ and pick a face/node  $F_{a_0,0}$ in the dimer/quiver, for which we choose $n_{a_0,0}=0$ without loss of generality. Now we may propagate to neighboring faces/nodes by crossing edges / following arrows and obtain the corresponding values of $n_{a,0}$ by adding the $B$ charges of the edges crossed / arrows followed. A particularly interesting case is the behavior when we propagate from a face/node $F_{a,0}$ in ${\cal C}$ to the copy $F_{a,r_1k_1+r_2k_2}$ located in the copy of ${\cal C}$ located in the position $(r_1,r_2)$ with respect to the two basic directions, in the infinite dimer/quiver diagram.
To propagate from the initial to the final face/arrow, we may pick any path, since the result is path independent. For instance, we can pick edges/arrows forming a meson $M_{1,0}$ in the direction of $k_1$ (resp. $M_{2,0}$ in the direction of $k_2$) in ${\cal C}$, and we can follow the sequence of $r_1$ mesons $M_{1,s_1}$ for $s_1=0,\ldots, r_1-1$, to reach $F_{a,r_1k_1}$, and then follow the sequence of $r_2$ mesons $M_{2,r_1k_2+s_2k_2}$ for $s_2=0,\ldots, r_2-1$to reach $F_{a,r_1k_1+r_2k_2}$. Using the $B$- and $C$-charges, we have
\begin{align}
\begin{split}
n_{a,r_1k_1+r_2k_2}- n_{a,0} =& \sum_{s_1=0}^{r_1-1} (\, b_{M_{1}}+k_1s_1c_{M_1}) + \sum_{s_2=0}^{r_2-1} (\, b_{M_{2}}+k_1 r_1c_{M_2}+k_2s_2c_{M_2})\, = \\
= r_1 b_{M_{1}}  + r_2b_{M_{2}}  + & \, r_1r_2k_1c_{M_2}+ \frac{r_1(r_1-1)}2 k_1  c_{M_1} + \frac{r_2(r_2-1)}2 k_2  c_{M_2}\coma 
\label{thenis}
\end{split}
\end{align}
where we have dropped sub-indices for charges in the unit cell ${\cal C}$ at $r=0$.

Note that  mesons carry no baryonic charge, hence the $C$-charges appearing above only have the mesonic contributions. From the above, we can easily understand different patters of growth of the $n$'s with the $r$'s: if the $Q_C$ linear combination (\ref{c-charge-combination}) involves the mesonic $U(1)$'s, the $C$-charges in the above equation are active, and the $n$'s grow quadratically with the $r$'s; if the $Q_C$ linear combination  does not contain the mesonic $U(1)'s$, then the above $C$-charges vanish and the $n$'s grow linearly with the $r$'s. This underlies the different behavior of the $n_i$'s for $\IC^3$ and the conifold, as we see in the examples in the next Section.

\section{Examples: Discrete Symmetries for Infinite Classes of Orbifolds}
\label{sec:examples}

In this Section we illustrate the procedure of the previous Section, by  applying it to systematically construct the discrete symmetries for several infinite classes of orbifolds of different geometries.

\subsection{General orbifolds of $\IC^3$}
\label{sec:general-solution-c3}

\begin{figure}[!htp]
	\begin{center}
		\includegraphics[scale=0.4]{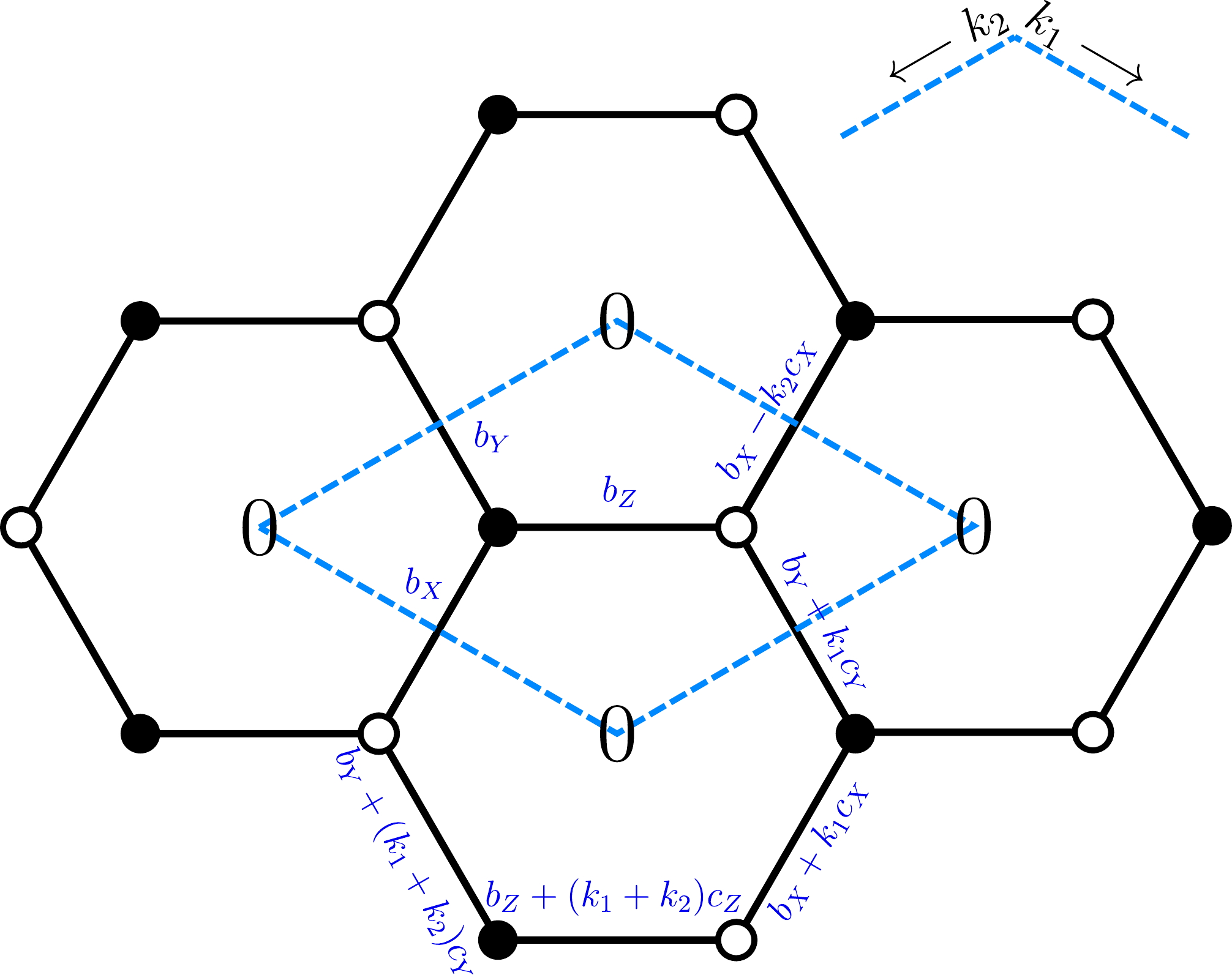}
		\caption{Unit cell in the dimer diagram for $\IC^3$. We display the charge assignments corresponding to the $B$-charges.}
		\label{fig:unit-cell-c3}
	\end{center}
\end{figure}

We consider general orbifolds of the $\IC^3$ theories described in Section~\ref{sec:example-general-c3}. The dimer and unit cell of the parent $\IC^3$ are shown in Figure \ref{fig:unit-cell-c3}. There are no baryonic $\U(1)$'s, and the mesonic charges are
%
%%%%%%%%%%%%%%%%%%%%%%%%%%%%%%%%
%\begin{table}[htb] \footnotesize
\renewcommand{\arraystretch}{1.25}
\begin{center}
	\begin{tabular}{cccc}
		& $Q_{1}$ & $Q_{2}$ & $Q_C$\\
		\hline 
		$X$ & $1$ & $0$ & $m_1$\\
		$Y$ & $0$ & $1$ & $m_2$\\
		$Z$ & $-1$ & $-1$ & $-m_1-m_2$\\
	\end{tabular}
\end{center} 
%\caption{)
%\label{tabpssm} }
%\end{table}
%%%%%%%%%%%%%%%%%%%%%%%%%%%%%%%%%%%%
%
where the last column shows the charges under the combination $Q_C=m_1 Q_{1}+m_2Q_{2}$.

There are two geometric identities in the graph. Denoting $V$ and $V'$ the black and white nodes and $F$ the unique face, they are given by
\begin{align}
\begin{split}
& \partial V - \partial V' =0  \\
& {\tilde \partial} F- \partial V-\partial V'=0\fstop
\label{geometric-c3}
\end{split}
\end{align}

Consider now a general orbifold, and consider the $B$-charge assignment in Figure \ref{fig:unit-cell-c3}.
The conditions of invariance of the superpotential terms in the black and white nodes, and anomaly cancellation are
\begin{align}
\begin{split}
V  & \rightarrow \quad b_X+b_Y+b_Z\,=\,0 \\
V'  & \rightarrow \quad   b_X-k_2 c_X+b_Y+ k_1 c_Y+b_Z \,=\,0 \\
F  & \rightarrow  \quad  2b_X+ k_1c_X+2b_Y+k_1c_Y+(k_1+k_2)c_Y+2b_Z+(k_1+k_2)c_Z\,=\,0\fstop
\label{conditions-b-c3}
\end{split}
\end{align}
To extract the consistency conditions for the charges $c_{E_i}$, we use the geometric combinations (\ref{geometric-c3}), and obtain
\begin{align}
\begin{split}
& -k_2 c_X+k_1 c_Y =\,0  \\
& k_1c_X+(2k_1+k_2)c_Y+(k_1+k_2)c_Z=\, 0\fstop
\end{split}
\end{align}
Expressing the charges in terms of $Q_C=m_1Q_1+m_2Q_2$ in the table above, the equations reduce to
\beqa
-k_2m_1+k_1m_2=0\fstop
\eeqa
Choosing $m_1=k_1$, $m_2=k_2$ to yield integer $C$-charges, we have
\beqa
c_X=k_1\quad ,\quad c_Y=k_2\quad,\quad c_Z=-k_1-k_2\fstop
\eeqa
These $C$-charges ensure that the equations (\ref{conditions-b-c3}) admit a solution for the $B$-charges. The equations reduce to
\beqa
b_X+b_Y+b_Z=0\fstop
\label{unit-b-charges-c3}
\eeqa
As explained above, this determines the $B$-charges up to the action of mesonic $\U(1)^2$. One may choose the latter to set $b_X=b_Y=0$ and then obtain $b_Z=0$. Alternatively, we can recover the general solution in Section~\ref{sec:example-general-c3} by solving (\ref{unit-b-charges-c3}) with the values
\beqa
b_X= -\frac{k_2}{2}+\frac{k_2^{\,2}}{2} \;, \quad 
b_Y= k_1k_2-\frac{k_1}{2}+\frac{k_1^{\,2}}{2} \;, \quad
b_Z= \frac{k_1+k_2}{2}-\frac{(k_1+k_2)^{2}}{2}\fstop
\eeqa

\medskip

\subsection{General orbifolds of the conifold}
\label{sec:general-solution-conifold}

\begin{figure}[!htp]
	\begin{center}
		
		\includegraphics[scale=0.6]{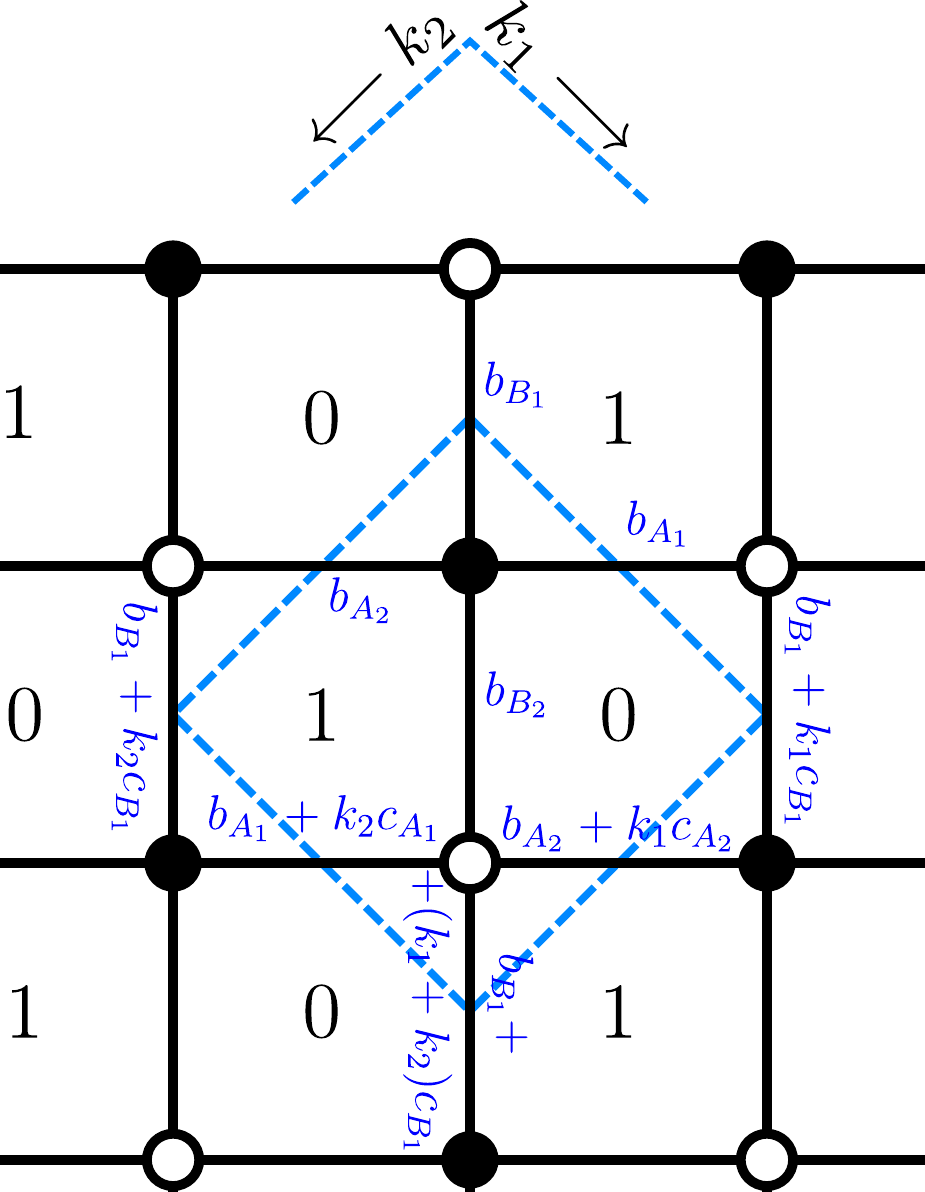}
		\caption{Dimer diagram with a unit cell for the conifold. We display the charge assignments corresponding to the $B$-charges.}
		\label{fig:unit-cell-conifold}
	\end{center}
\end{figure}

Consider the orbifolds of the conifold discussed in Section~\ref{sec:orbifold-conifold-app}. The dimer diagram with a unit cell and ansatz for the B-charge assignment is shown in Figure \ref{fig:unit-cell-conifold}. There are two mesonic $\U(1)$'s and one baryonic $\U(1)$. The latter is associated to the existence of one kind fractional brane, so it corresponds to the overall $\U(1)$ on one of the faces, say face 1. The charges of the different fields under these $\U(1)$'s, and under a general combination $Q_C=m_1 Q_1+m_2Q_2 + m_{\bf B} Q_{\bf B}$, are
%
%%%%%%%%%%%%%%%%%%%%%%%%%%%%%%%%
%\begin{table}[htb] \footnotesize
\renewcommand{\arraystretch}{1.25}
\begin{center}
	\begin{tabular}{ccccc}
		& $Q_{1}$ & $Q_{2}$ & $Q_{\bf B}$ & $Q_C$\\
		\hline 
		$A_1$ & $1$ & $0$ & $1$ & $m_1+m_{\bf B}$\\
		$A_2$ & $-1$ & $0$ & $1$ & $-m_1+m_{\bf B}$\\
		$B_1$ & $0$ & $1$ & $-1$ & $m_2-m_{\bf B}$\\
		$B_2$ & $0$ & $-1$ & $-1$ & $-m_2-m_{\bf B}$\\
	\end{tabular}
\end{center} 
%\caption{ Standard model spectrum and $\U(1)$ charges in the realization in terms of D6-branes with intersection number (\ref{intersm})
%\label{tabpssm} }
%\end{table}
%%%%%%%%%%%%%%%%%%%%%%%%%%%%%%%%%%%%

The geometric identities correspond to the two generic ones, and one associated to the fractional brane. They can be written
\begin{align}
\begin{split}
&\partial V-\partial V'=0\\
& {\tilde \partial} F_1 +{\tilde \partial} F_2-\partial V-\partial V'=0 \\
& {\tilde \partial} F_1 -\partial V=0\fstop
\end{split}
\end{align}
Using the $B$-charge assignments in Figure \ref{fig:unit-cell-conifold}, the constraints from invariance of the superpotential terms at the two nodes, and anomaly cancellation on the two faces, are
\begin{align}
\begin{split}
\partial V\rightarrow & \quad b_{A_1}+b_{B_1}+b_{A_2}+b_{B_2}=0\\
\partial V' \rightarrow & \quad b_{A_1}+k_2c_{A_1}+b_{B_1}+(k_1+k_2)c_{B_1}+ 
b_{A_2}+k_1c_{A_2}+ b_{B_2}=0\\
{\tilde\partial}F_1\rightarrow & \quad b_{A_1}+b_{B_1}+k_1c_{B_1} + b_{A_2}+k_1c_{A_2}+b_{B_2}=0  \\
{\tilde\partial}F_2\rightarrow & \quad b_{A_1}+k_2c_{A_1}+b_{B_1}+k_2c_{B_1} + b_{A_2}+b_{B_2}=0 \fstop
\label{conditions-b-conifold}
\end{split}
\end{align}
Taking combinations of these equations as in the geometric identities, we obtain the consistency conditions for the $C$-charges (there are only two independent ones)
\begin{align}
\begin{split}
& k_2c_{A_1}+(k_1+k_2)c_{B_1}+ k_1c_{A_2}=0 \\
& k_2(c_{A_2}-c_{B_1})=0\fstop
\end{split}
\end{align}
Expressing them in terms of the generator $Q_C=m_1Q_1+m_2Q_2+m_{\bf B}Q_{\bf B}$ as in the table above, the equations imply
\beqa
m_1=m_2=0\coma m_{\bf B} \; {\rm arbitrary}\fstop
\eeqa
Taking the minimal choice to obtain integer charges, we let $m_{\bf B}=1$, and have
\beqa
c_{A_1}=c_{A_2}=1\quad , \quad c_{B_1}=c_{B_2}=-1\fstop
\eeqa
Replacing into (\ref{conditions-b-conifold}) leads to the unique constraint 
\beqa
b_{A_1}+b_{B_1}+b_{A_2}+b_{B_2}=0\fstop
\eeqa
These charges are as usual defined modulo the action of the two mesonic and the baryonic $\U(1)$ symmetries. The simplest solution is to use them to set $b_{A_1}=b_{A_2}=b_{B_1}=0$ and then we get $b_{B_2}=0$. This actually leads to the solution found in Section~\ref{sec:orbifold-conifold-app}.

\subsection{General orbifolds of the $dP_1$ theory}
\label{sec:general-solution-dp1}

\begin{figure}[!htp]
	\begin{center}
		
		\includegraphics[scale=0.35]{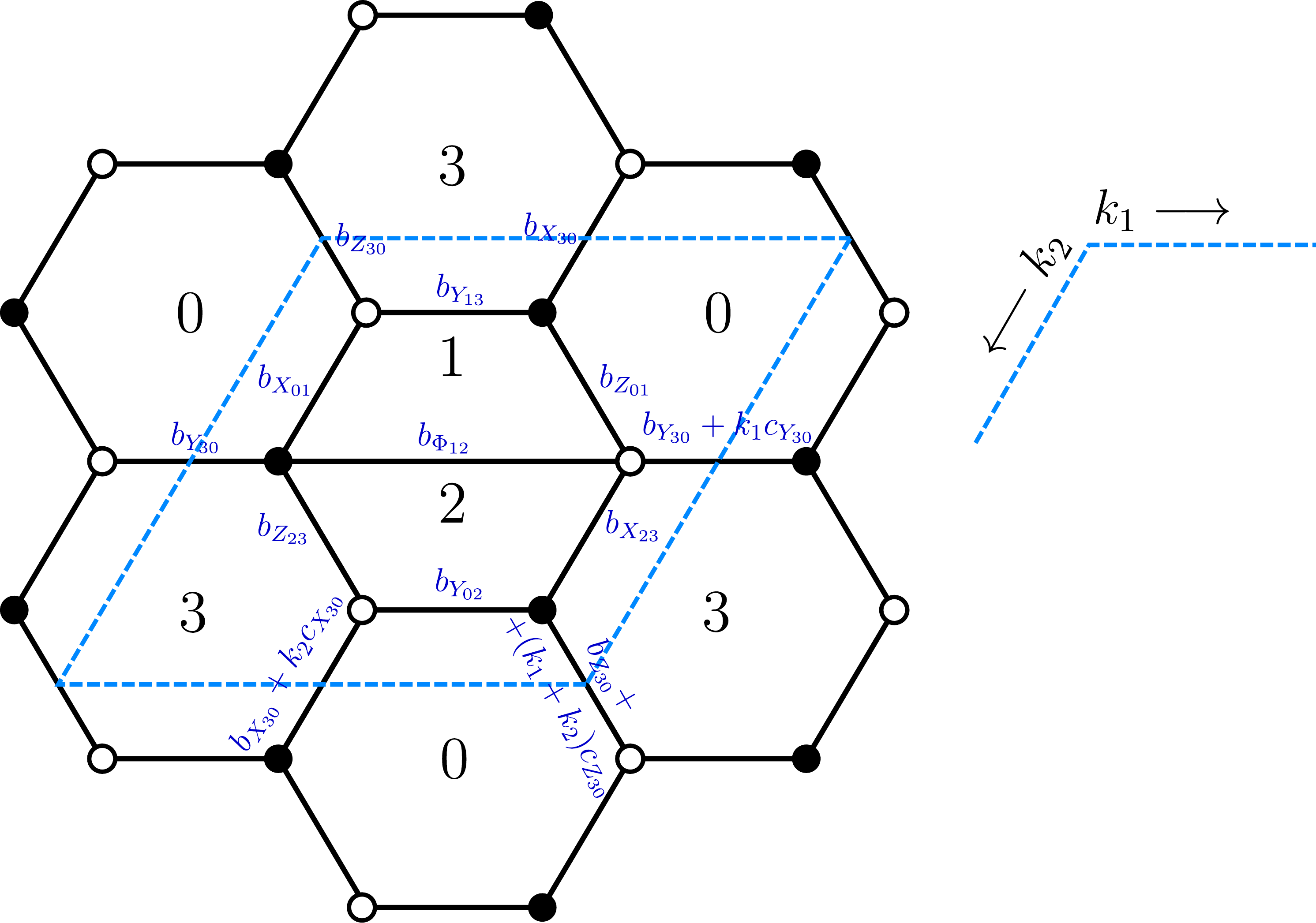}
		\caption{Dimer diagram with a unit cell for the $dP_1$ theory. We display the charge assignments corresponding to the $B$-charges.}
		\label{fig:unit-cell-dp1}
	\end{center}
\end{figure}

To illustrate the power of our method, we construct the discrete symmetries for a new infinite class of theories. They correspond to general orbifolds of the $dP_1$ theory. The dimer diagram with a unit cell and ansatz for the B-charge assignment is shown in Figure \ref{fig:unit-cell-dp1}. There are two mesonic $\U(1)$'s and one baryonic $\U(1)$. The latter is associated to the existence of one kind fractional brane, given by $N_0=1$, $N_1=3$, $N_2=0$, $N_3=2$. The charges of the different fields under these $\U(1)$'s, and under a general combination $Q_C=m_1 Q_1+m_2Q_2 + m_{\bf B} Q_{\bf B}$, are
%
%%%%%%%%%%%%%%%%%%%%%%%%%%%%%%%%
%\begin{table}[htb] \footnotesize
\renewcommand{\arraystretch}{1.25}
\begin{center}
	\begin{tabular}{ccccc}
		& $Q_{1}$ & $Q_2$&$Q_{\bf B}$&$Q_C$\\\hline
		$X_{30}$&$0 $&$1$ &$ 1$ & $m_2+m_{\bf B}$ \\
		$X_{01}$&$-1$ &$0$& $ -2$ & $-m_1-2m_{\bf B}$ \\
		$X_{23}$&$-1$ &$0$& $-2 $& $-m_1-2m_{\bf B}$ \\
		$Y_{30}$&$1$ &$0$&$ 1 $& $m_1+m_{\bf B}$\\
		$Y_{02}$&$1 $&$0$ &$1$ & $m_1+m_{\bf B}$  \\
		$Y_{13}$&$1 $&$0$&$ 1$ & $m_1+m_{\bf B}$  \\
		$Z_{30}$&$0 $&$0 $&$1 $& $m_{\bf B}$  \\
		$Z_{01}$&$-1 $&$-1 $&$-2$ & $-m_1-m_2-2m_{\bf B}$ \\
		$Z_{23}$&$-1 $&$-1 $&$-2 $& $-m_1-m_2-2m_{\bf B}$ \\
		$\Phi_{12}$&$1$ &$1$&$	3$ & $m_1+m_2+3m_{\bf B}$
	\end{tabular}
\end{center} 
%\caption{ Standard model spectrum and $\U(1)$ charges in the realization in terms of D6-branes with intersection number (\ref{intersm})
%\label{tabpssm} }
%\end{table}
%%%%%%%%%%%%%%%%%%%%%%%%%%%%%%%%%%%%

The geometric identities correspond to the two generic ones, and one associated to the fractional brane. They read
\begin{align}
\begin{split}
&\partial V_1+\partial V_2+\partial V_3-\partial V_1'-\partial V_2'-\partial V_3'=0 \\
&{\tilde\partial}F_0+{\tilde\partial}F_1+{\tilde\partial}F_2+{\tilde\partial}F_3-\left(\partial V_1+\partial V_2+\partial V_3\right)-\left(\partial V_1'+\partial V_2'+\partial V_3'\right)=0 \\
&{\tilde\partial}F_0+3{\tilde\partial}F_1+2{\tilde\partial}F_3-3\partial V_1-2\partial V_2-\partial V_3-2\partial V_1'+\partial V_3'=0\fstop
\end{split}
\end{align}
Using the $B$-charge assignments in Figure \ref{fig:unit-cell-dp1}, the constraints from invariance of the superpotential terms at the three nodes, and anomaly cancellation on the four faces, are

\begin{align}
\begin{split}
\partial V_1&\longrightarrow b_{Z_{30}}+b_{Y_{13}}+b_{X_{01}}=0\\
\partial V_2&\longrightarrow k_1 c_{Y_{30}}+b_{Y_{30}}+b_{Z_{01}}+b_{X_{23}}+b_{\Phi_{12}}=0\\
\partial V_3&\longrightarrow k_2 c_{X_{30}}+b_{X_{30}}+b_{Y_{02}}+b_{Z_{23}}=0\\
\partial V_1'&\longrightarrow b_{X_{30}}+b_{Y_{13}}+b_{Z_{01}}=0 \\
\partial V_2'&\longrightarrow \left(k_1+k_2\right) c_{Z_{30}}+b_{Z_{30}}+b_{X_{23}}+b_{Y_{02}}=0 \\
\partial V_3'&\longrightarrow b_{Y_{30}}+b_{Z_{23}}+b_{X_{01}}+b_{\Phi_{12}}=0\\
{\tilde\partial}F_{0}&\longrightarrow b_{Y_{02}}+b_{Z_{30}}+(k_1+k_2)c_{Z_{30}}+b_{X_{01}}+(k_1+k_2)c_{X_{01}}+b_{Y_{30}}+ \\ & + (k_1+k_2)c_{Y_{30}}+b_{Z_{01}}+ k_2c_{Z_{01}}+b_{X_{30}}+k_2c_{X_{30}}=0 \\
{\tilde\partial}F_{1}&\longrightarrow b_{Y_{13}}+b_{Z_{01}}+b_{X_{01}}+b_{\Phi_{12}}=0 \\
{\tilde\partial}F_{2}&\longrightarrow b_{X_{23}}+b_{Y_{02}}+b_{Z_{23}}+b_{\Phi_{12}}=0 \\
{\tilde\partial}F_{3}&\longrightarrow b_{X_{23}}+b_{Y_{30}}+k_1c_{Y_{30}}+b_{Z_{23}}+k_1c_{Z_{23}}+b_{X_{30}}+(k_1+k_2)c_{X_{30}}+b_{Y_{13}} + \\ +& (k_1+k_2)c_{Y_{13}}+ b_{Z_{30}}+(k_1+k_2)c_{Z_{30}}=0\fstop 
\label{eq:dP1BCcont}
\end{split}
\end{align}
The consistency conditions for the $C$-charges are
\beqa
&&k_2 c_{X_{30}}+k_1 c_{Y_{30}}-\left(k_1+k_2\right) c_{Z_{30}}=0\nonumber \\
&&\left(k_1+k_2\right) c_{X_{01}}+\left(k_1+k_2\right) c_{X_{30}}+k_2 c_{Y_{30}}+\left(k_1+k_2\right) c_{Y_{13}}+k_1 c_{Z_{23}}+
%\nonumber\\
%&&
+k_2 c_{Z_{01}}+\left(k_1+k_2\right) c_{Z_{30}}=0\nonumber \\
&&\left(k_1+k_2\right) c_{X_{30}}+\left(k_1+k_2\right) c_{Y_{13}}+k_1 c_{Z_{23}}+2 k_1 c_{Z_{30}}+2 k_2 c_{Z_{30}}=0\fstop
\eeqa
Using the $C-$charges as in the table above, the system reduces to
\begin{align}
\begin{split}
	& k_1m_1+k_2m_2=0 \\
	&2k_1+m_{\bf B}+k_2(m_1+m_2+4m_{\bf B})=0\fstop
	\end{split}
\end{align}
To obtain integer $C-$charges, we choose 
\begin{equation}
	m_{\bf B}=k_2(k_1-k_2) \; ,\; m_1=2k_2(k_1+2k_2)\;,\; m_2=-2k_1(k_1+2k_2)\fstop
\end{equation}
And obtain
\begin{alignat}{2}
%\begin{split}
&c_{X_{30}}=-(k_1+k_2)(2k_1+k_2)\quad\quad &&c_{Y_{13}}=3k_2(k_1+k_2) \nonumber\\ 
&c_{X_{01}}=-2k_2(2k_1+k_2) \quad \quad &&c_{Z_{30}}=k_2(k_1-k_2) \nonumber\\
&c_{X_{23}}=-2k_2(2k_1-k_2) \quad \quad &&c_{Z_{01}}=2\left(k_1^2-k_2^2\right)\\
&c_{Y_{30}}=3k_2(2k_1+k_2) \quad \quad &&c_{Z_{23}}=2\left(k_1^2-k_2^2\right) \nonumber \\
&c_{Y_{02}}=3k_2(k_1+k_2) \quad \quad &&c_{\Phi_{12}}=-(k_1-k_2)(2k_1+k_2)\fstop \nonumber
%\end{split}
\end{alignat}
The solutions for the $B$-charges are
\begin{align}
\begin{split}
b_{X_{01}} =&-b_{Z_{30}}-b_{Y_{30}}+k_2\left(k_1^2-k_2^2\right) \\
b_{X_{23}} =&-b_{Z_{30}}-b_{Y_{30}}-3k_1k_2(k_1+k_2) \\
b_{Y_{02}} =&b_{Y_{30}}+k_2(k_1+k_2)(2k_1+k_2) \\
b_{Y_{13}} =&b_{Y_{30}}+k_2\left(k_1^2-k_2^2\right) \\
b_{Z_{01}} =&-b_{X_{30}}-b_{Y_{30}}-k_2\left(k_1^2-k_2^2\right)\\
b_{Z_{23}} =&-b_{X_{30}}-b_{Y_{30}} \\
b_{\Phi_{12}} =&b_{X_{30}}+b_{Y_{30}}+b_{Z_{30}}+k_2\left(k_1^2-k_2^2\right)\fstop 
\end{split}
\end{align}
where $b_{X_{30}}$, $b_{Y_{30}}$, $b_{Z_{30}}$ are left as undetermined parameters encoding the freedom to shift charges by the global $\U(1)^3$ symmetry.

We hope this example suffices to show the power of our general approach. We provide further examples of orbifolds of the $dP_2$ and $dP_3$ theories in Appendix \ref{sec:more-examples}.

\medskip

\section{Some remarks on the gravity dual}
\label{sec:gravity}

In this Section we sketch some of the main ingredients about the realization of the discrete symmetries in the gravity dual. It was established in \cite{Gukov:1998kn}, that the discrete symmetries in the $\IC^3/\IZ_3$ theory are associated to torsion classes in the 5d horizon $\IS^5/\IZ_3$ of the orbifold theory (see also \cite{Burrington:2006uu} for other geometries), such that objects charged under the generators of the discrete Heisenberg group correspond to branes wrapped on torsion cycles. In a general orbifold, for the 5d horizon $\IX_5=\IS^5/\IZ_N$, the generator of $H_3(\IX_5,\IZ)=\IZ_N$ is a torsion 3-cycle, such that wrapped D5- and NS5-branes produce 5d codimension 2 objects, around which the theory experiences monodromies associated to the $A$ and $B$ generators. The non-abelian nature of the discrete gauge symmetry followed because two torsion 3-cycles intersect over a torsion 1-cycle in $H_1(\IX_5,\IZ)=\IZ_N$; hence when the wrapped NS5- and D5-branes associated to the $A$ and $B$ actions are crossed in 5d, one generates, by the Hanany-Witten effect, \cite{Hanany:1996ie} a D3-brane wrapped on the torsion 1-cycle. This precisely corresponds to the element $C$ in the discrete Heisenberg group. Alternatively, one can characterize the discrete symmetry by the representations formed by (di)baryons, which are realized in the gravity side as D3-branes wrapped in 3-cycles with non-trivial torsion 1-cycles, on which one can turn on $\IZ_N$ valued Wilson lines; this leads to $N$-plets of D3-brane states, on which the discrete Heisenberg group acts faithfully.

We expect a similar mechanism to work in general orbifold theories, and hence are led to looking for such 3-cycles in the horizon geometry of general orbifolds of general toric theories. The 3-cycles on the Sasaki-Einstein 5d horizon of CY threefold singularities have been extensively studied in the context of the holographic description of baryons, and it is well-established that calibrated 3-cycles are in correspondence with non-compact holomorphic 4-cycles in the CY threefold singularity (\cite{Mikhailov:2000ya}, also for instance \cite{Forcella:2007wk,Butti:2007jv,Forcella:2008bb,Forcella:2008eh}). In the toric setup, the non-compact 4-cycles were described in \cite{Franco:2018vqd} (see also \cite{Forcella:2008au}) in terms of pairs of punctures; namely, in the type IIA mirror  geometry, the non-compact 4-cycle becomes a non-compact 3-cycle, which is described as a 1-cycle in the mirror Riemann surface, which comes in through a puncture and goes out through another puncture. In particular, consider the baryonic operator corresponding to antisymmetrizing the indices of a given bifundamental; the corresponding 4-cycle has as mirror a 3-cycle corresponding to two punctures which are mirror to the two zig-zag paths crossing the bifundamental. This leads to a one-to-one correspondence between such baryons and holomorphic 4-cycles in the toric singularity.

\begin{figure}[!htp]
	\centering
	\begin{subfigure}[l]{0.45\textwidth}
		\centering
		\includegraphics[scale=0.25]{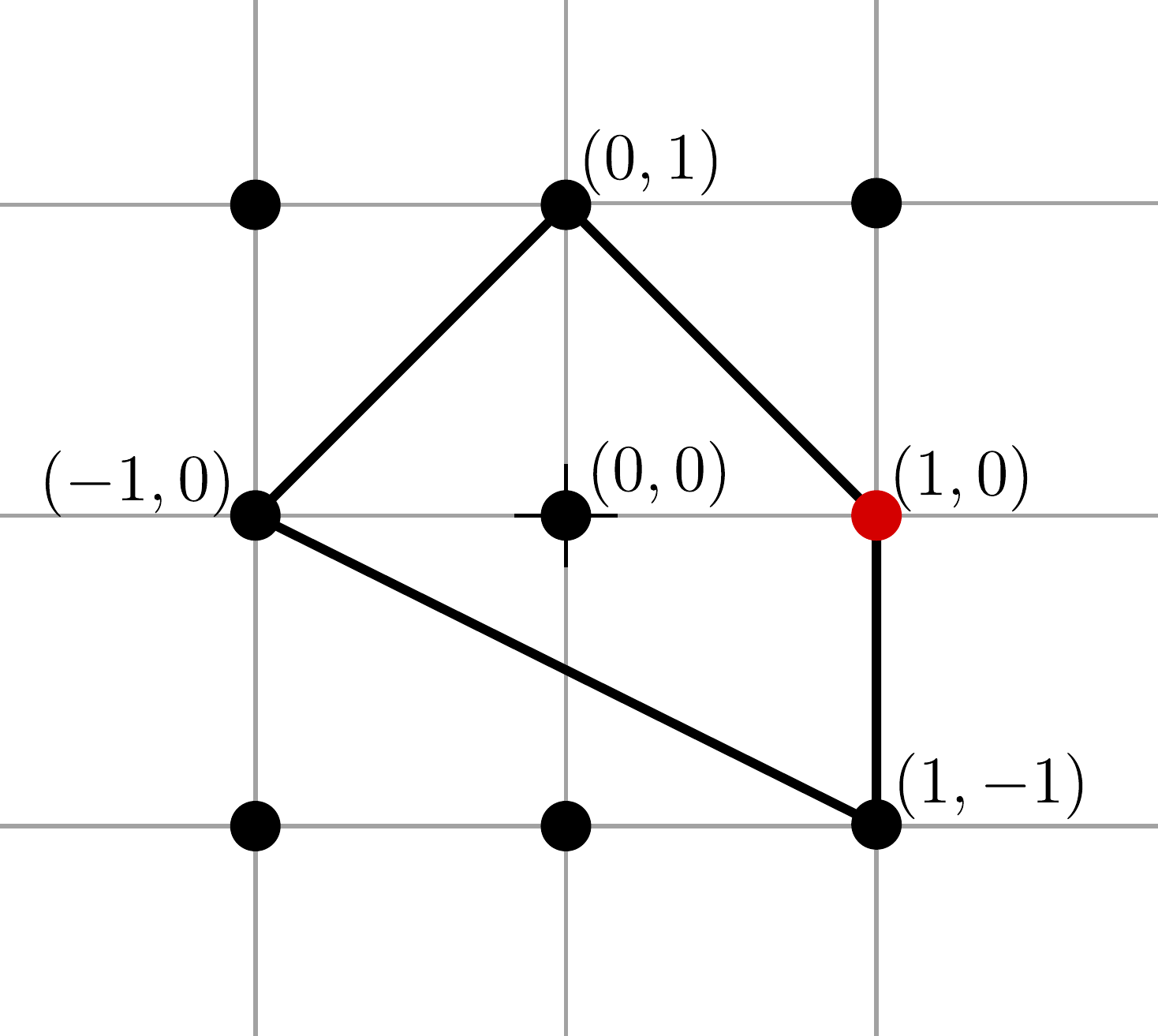}
		\caption{}
		\label{dP1tor2}
	\end{subfigure}\hspace{10mm}
\begin{subfigure}[r]{0.45\textwidth}
	\centering
	\includegraphics[scale=0.15]{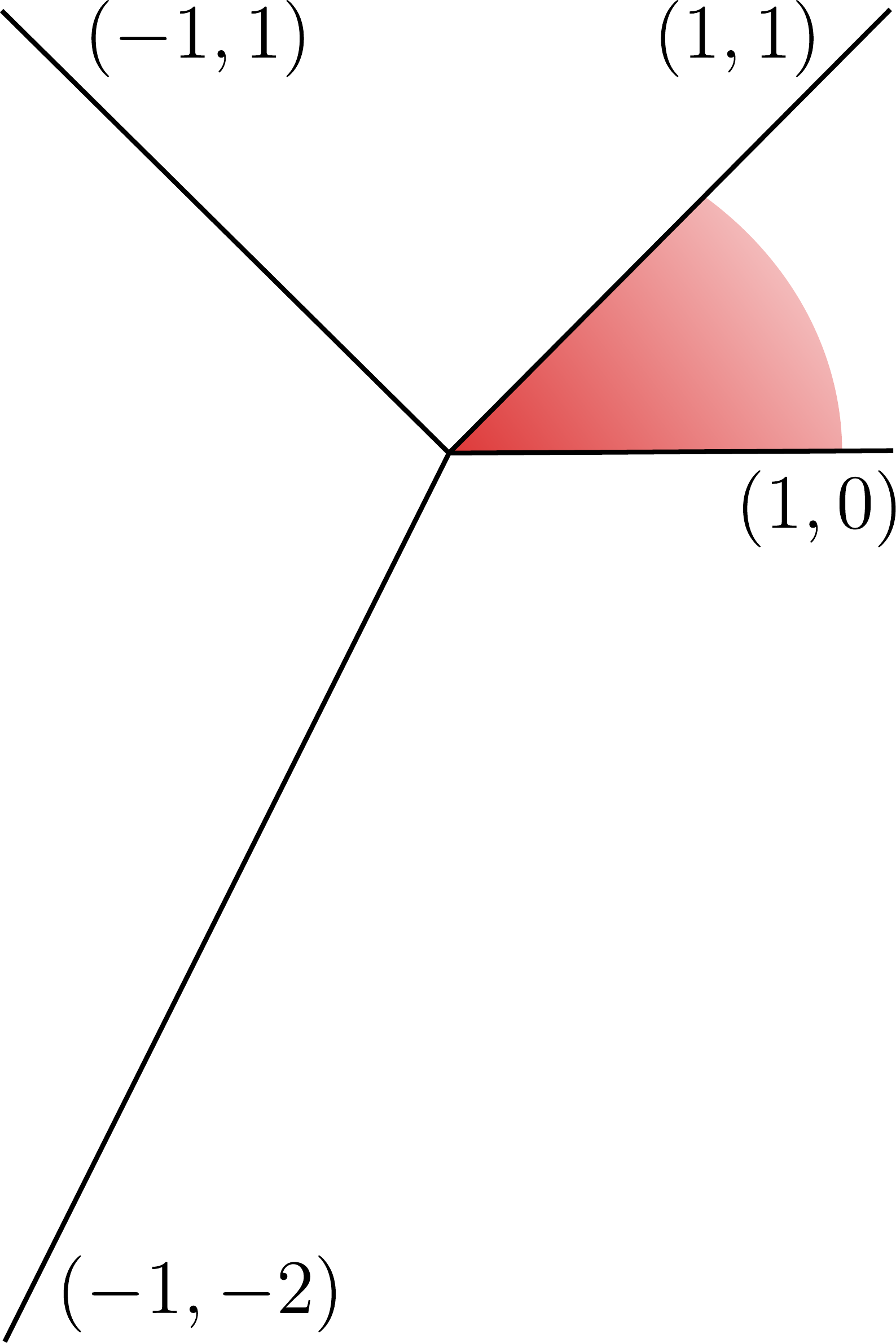}
	\caption{}
	\label{dP1-pq2}
\end{subfigure}\\ \vspace{10mm}
\begin{subfigure}[l]{0.45\textwidth} \vspace{5mm}
\centering
\includegraphics[scale=0.25]{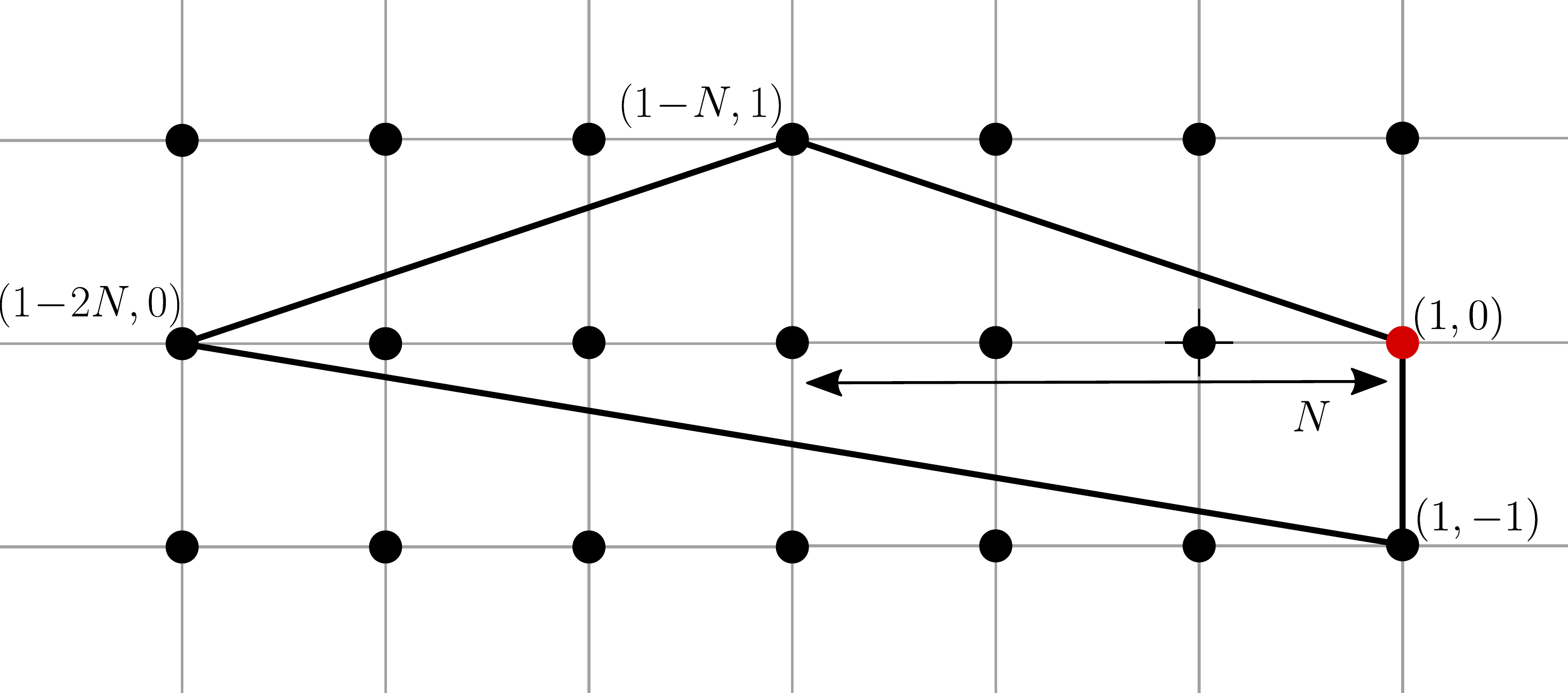}
\vspace{4mm}
\caption{}
\label{dP1torZN}
\end{subfigure} \hspace{10mm}
\begin{subfigure}[l]{0.45\textwidth}
\centering
\includegraphics[scale=0.15]{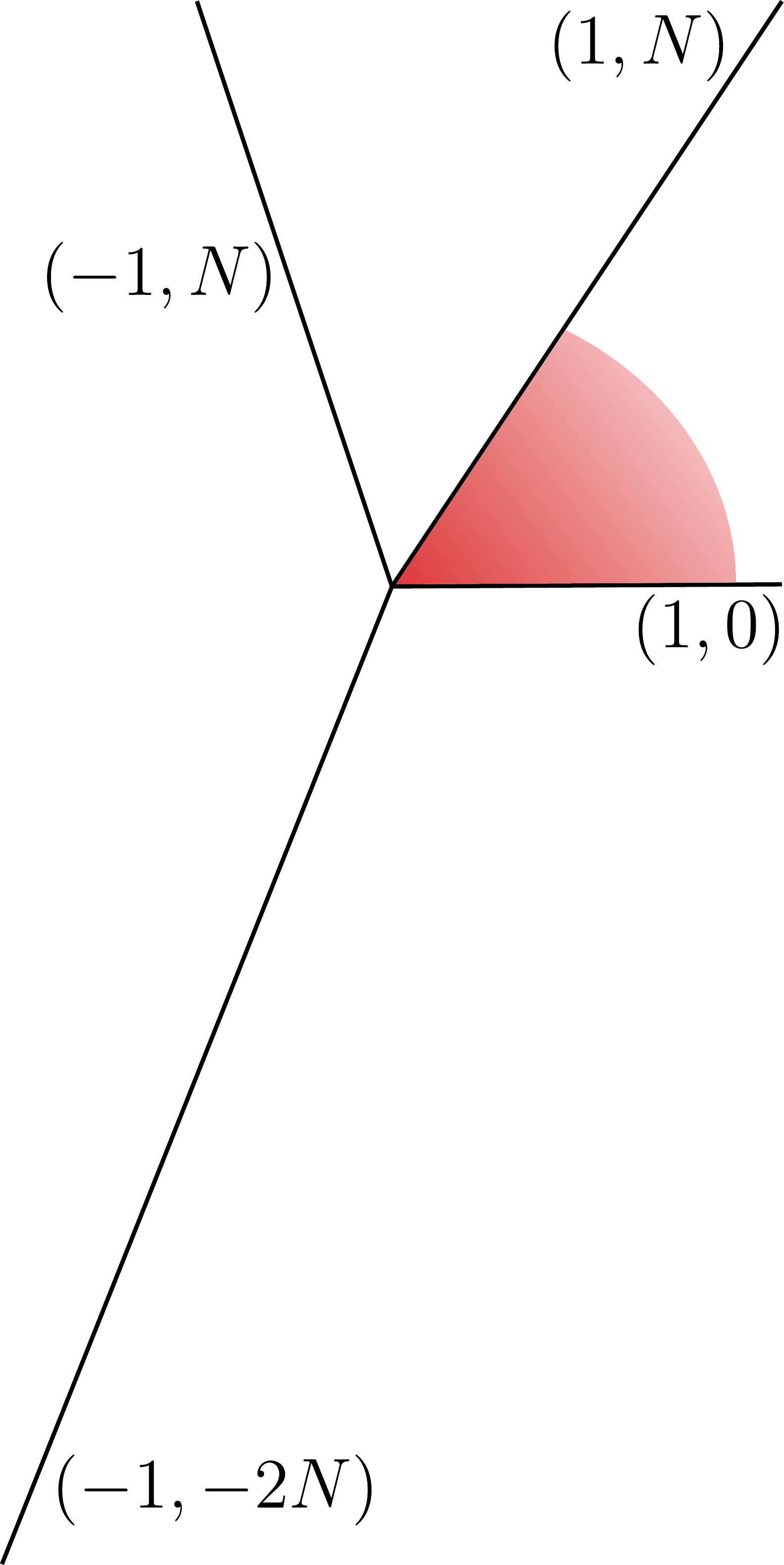}
\caption{}
\label{dP1Zn-pq}
\end{subfigure}
	\caption{Toric and web diagrams of the $dP_1$ theory and its quotient. We have highlighted in red the perfect matching and wedge related to an example of non-compact 4-cycle.} 
	\label{fig:fourcycle} 
\end{figure}

The  description in terms of pairs of punctures is manifest also in the original type IIB picture for the non-compact 4-cycles bounded by adjacent external legs in the web diagram; in this case the non-compact 4-cycle is defined by the equation $p_i=0$ of vanishing of the linear sigma model coordinate corresponding to the perfect matching $p_i$ at the corresponding external point in the toric diagram\footnote{For simplicity, in this section we carry out the discussion for theories with no parallel external legs.}, see Figure \ref{fig:fourcycle}.

When one performs a $\IZ_N$ orbifold of a parent theory, the toric diagram of the original theory is the same as the original one, but in a refined lattice, such that the original is an index-$N$ sub-lattice of the final one.
Now recall that points of the toric diagram correspond to perfect matchings of the dimer; although there is in general not a one-to-one map for general points in the toric diagrams of the parent and quotient theory, there is such one-to-one map for {\em external} points, as follows. Consider an external point $p_i$ of the toric diagram of the parent theory; this corresponds to a perfect matching $p_i$ of the dimer in the parent unit cell ${\cal C}$; we can now obtain a perfect matching of the dimer of the orbifold theory with unit cell ${\cal C}_N$ by simply replicating $p_i$ $N$ times in the $N$ copies of ${\cal C}$ in ${\cal C}_N$. In brief, there is a one-to-one correspondence between external perfect matchings of the toric diagrams of the parent and quotient geometries,  and similarly between external legs of the web diagrams, and hence among 4-cycles, see Figure \ref{fig:fourcycle}. Hence the topology of the 4-cycle in the orbifold is that of the parent 4-cycle, quotiented by the $\IZ_N$ action. At the level of the horizon, the 3-cycle defined by the 4-cycle in the orbifold theory is a quotient of the 3-cycle of the 4-cycle in the parent theory modded out by the $\IZ_N$ action. This  is the origin of the torsion classes, as follows.

An easy to check important feature is that the pairing of $(p,q)$ labels of two external legs (namely, the quantity $p_1 q_2-q_1p_2$ for legs of labels $(p_1,q_1)$, $(p_2,q_2)$)  picks up a factor of $N$ in going from the parent to the orbifold theory. Hence, all the pairings are multiples of $N$ in any $\IZ_N$ orbifold of a general toric singularity. This introduces a subtlety in the relation between 4-cycles and baryonic operators, in the sense that the geometric 4-cycle is actually related to an $N$-plet of baryonic operators. Focusing on the simplest baryonic operators, obtained by antisymmetrizing indices on a given bifundamental, this implies that we have an $N$-plet of bifundamentals; they are just the $N$ copies of the bifundamental of the parent theory in the orbifold theory. These $N$ copies form a representation of the discrete Heisenberg groups, with the $A$ generator acting as a shift and $B$- and $C$-charges as determined in earlier sections. The holographic dual of the baryons associated to these bifundamentals are given by D3-branes wrapped on the 3-cycle with different $\IZ_N$-valued Wilson lines turned on.

 It would be interesting to pursue the gravitational dual description of the Heisenberg group, and in particular to unveil the geometric interpretation of the $B$- and $C$- charges, and their interplay with the mesonic and baryonic $U(1)$'s for general orbifolds of general toric geometries. We leave this question for future work.

\section*{Acknowledgments}
We are pleased to thank S. Franco, L. Ib\'anez and F. Marchesano  for useful discussions. This work is supported through the grants SEV-2016-0597, FPA2015-65480-P and PGC2018-095976-B-C21 from MCIU/AEI/FEDER, UE. AM received funding from ”la Caixa” Foundation (ID 100010434) with fellowship code LCF/BQ/IN18/11660045 and from the European Union’s Horizon 2020 research and innovation programme under the Marie Sklodowska-Curie grant agreement No. 713673. 

\newpage

\appendix

\section{Some topological concepts in dimers and quivers}
\label{sec:cohomology}

In this Appendix we introduce some (co)homological tools for dimer diagrams and their dual periodic quivers.
As in the main text, we use the notation $F_a$ for faces/nodes, $E_i$ for edges/arrows and $V_\alpha$, $V'_\alpha$ for vertices/plaquettes. These ingredients can be regarded as the analogues of simplices in singular homology, hence we consider their formal linear combinations (with negative coefficients corresponding to reversing the orientation these objects carry), which we refer to as 0-, 1- and 2-chains.

On these diagrams we can define $p$-forms as linear maps assigning a (in general, complex) number to every $p$-chain. For instance, in the dimer, we define 2-forms $\sigma$ as assignments of numbers $\sigma(F_a)$ to the faces $F_a$, and similarly 1-forms $\lambda(E_i)$ and 0-forms $f(V_\alpha)$, $f(V'_\alpha)$. The assignments defining 2-, 1- and 0-forms in the dimer, when regarded in the quiver, define 0-, 1- and 2-forms. This can be regarded as a duality (in a constructions known as quad-edge in computational physics), although we will not exploit it at present.

In the quiver there is a very natural realization of (co)homology. The boundary $\partial V$ of a plaquette $V$ (similarly for $V'$) is the 1-chain given by the sum of the arrows surrounding it; the boundary $\partial E$ of an edge $E$ is the formal difference of the nodes at its tail and its head $\partial E=t(E)-h(E)$. Clearly $\partial^2\equiv 0$, and we can define a homology. At the level of forms, we introduce an exterior derivative $d$ as follows.  For a 0-form $f$, we define $df$ as the 1-form given by
\beqa
df(E)=f(\partial E)=f(h(E))-f(t(E))\coma
\eeqa
where $h(E)$, $t(E)$ denote the node at the head and tail of the arrow $E$.
Similarly, for a 1-form $\lambda$, we define the 2-form $d\lambda$ by $d\lambda(F)=\lambda(\partial F)$. Finally, for a 2-form $\sigma$ we define $d\sigma\equiv 0$. One clearly has $d^2\equiv 0$ so this defines a cohomology. By defining integration by evaluation, $d$ and $\partial$ obey Stokes' theorem.
The above homology and cohomology are realizations of those of the underlying surface on which the quiver is embedded, in our case the 2-torus (or, as we occasionally focus on the infinite cover, $\IR^2$).

In the main text, the assignments of (continuous or discrete) charges to bifundamentals are used to define 1-forms $\lambda$ on the quiver, and the conditions of invariance of the superpotential amount to closedness, $d\lambda=0$. Non-trivial cohomology classes correspond to the mesonic $\U(1)$ symmetries, while exact forms $\lambda=df$ correspond to $\U(1)$ baryonic symmetries in the 2-torus, or related to the discrete $B$-symmetry in $\IR^2$.

\medskip

In the dimer, the notion of boundary and exterior derivative convenient for us includes a subtlety. We define the boundary ${\tilde \partial} F$ of a face $F$ as the sum of the edges bounding it, with their natural orientation (i.e. from black to white vertices). This differs from the geometric intuition, where the boundary involves the same edges but with a weight $\pm$ determined by the incidence relation between the edge and the face (i.e. the chirality of the bifundamental). We use a tilde to emphasize this difference. We define the boundary ${\tilde \partial} E$ of and edge $E$ as the difference between the corresponding black and white vertices, namely ${\tilde \partial} E=b(E)-w(E)$. Correspondingly, we define the exterior derivative ${\tilde d}$ as follows. For a 0-form $f$, we define ${\tilde d} f$ by 
\beqa
{\tilde d}f(E)=f({\tilde \partial}E)=f(b(E))-f(w(E))\fstop
\eeqa
Similarly, for a 1-form $\lambda$ we define ${\tilde{d}}\lambda(F)=\lambda({\tilde\partial}F)$. Defining integration by evaluation, this obeys Stokes' theorem. However, in general ${\tilde d}^2\neq 0$, ${\tilde\partial}^2\neq 0$; there is a well defined cohomology only if we restrict to 0-forms $f$ which satisfy that for any face $F$, the value of $f$ on the sum of black nodes equals its value on the sum of white nodes (and one may define homology in a similar restricted sense). Since these restrictions render these tools less natural, in the main text we simply use ${\tilde \partial}$ as a notational device. In terms of it, if the charge assignments under continuous or discrete symmetries are used to define a 1-form $\lambda$, the anomaly cancellation conditions read ${\tilde d}\lambda=0$.

\section{More infinite classes of orbifolds}
\label{sec:more-examples}

\subsection{General orbifolds of the $dP_2$ theory}
\label{sec:general-solution-dp2}

\begin{figure}[!htp]
	\centering
	\includegraphics[scale=0.4]{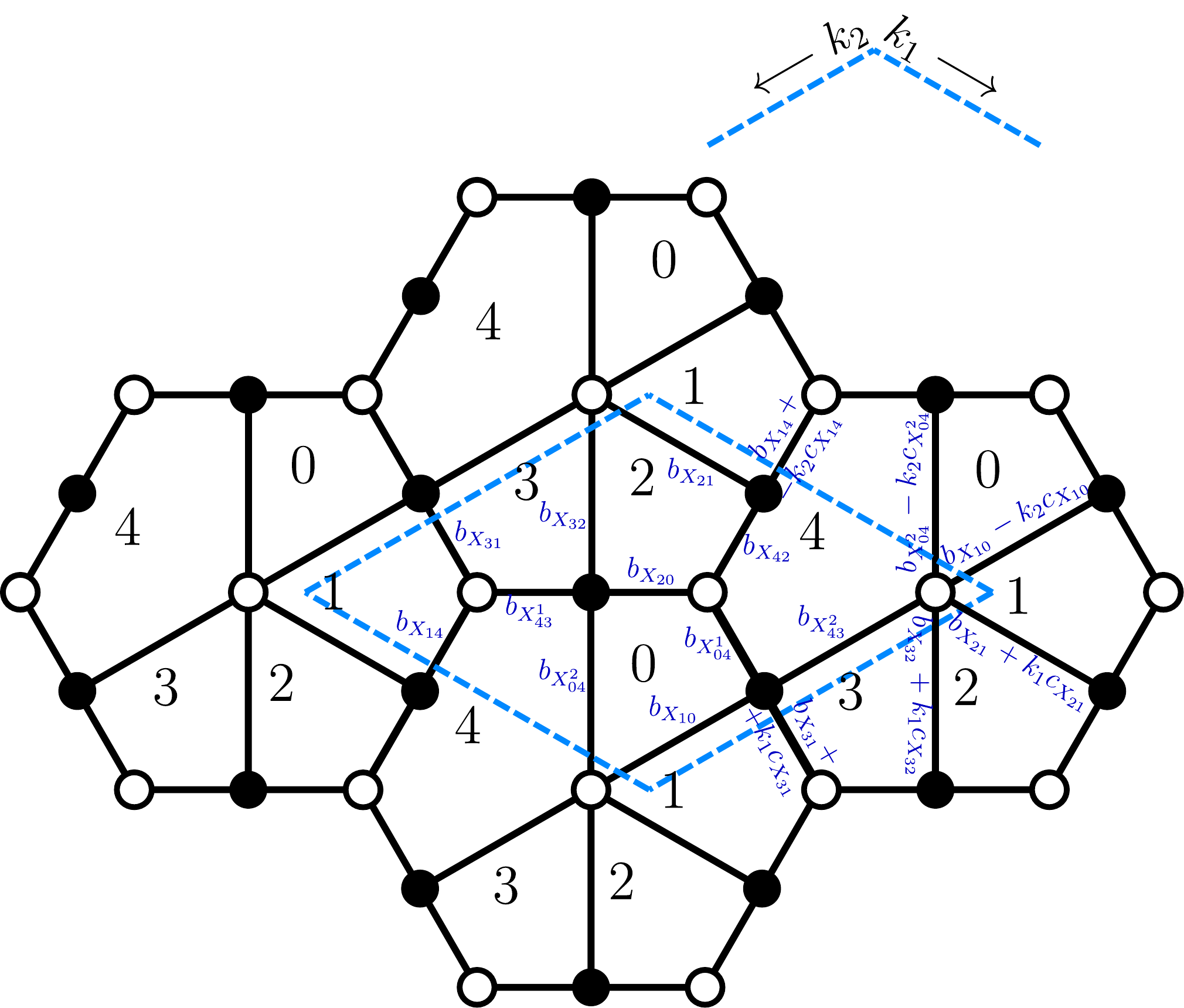}
	\caption{Dimer diagram with a unit cell for the $dP_2$ theory. We display the charge assignments corresponding to the $B$-charges.}
	\label{fig:unit-cell-dp2}
\end{figure}

We now construct the discrete symmetries for the general orbifolds of the $dP_2$ theory. The dimer diagram with a unit cell and the $B-$charge assignment is shown in Figure~\ref{fig:unit-cell-dp2}. There are two mesonic and baryonic $\U(1)$'s. The charges of the different fields under these $\U(1)$'s, and under a general combination $Q_C=m_1Q_1+m_2Q_2+m_{\bf B_{1}}Q_{\bf B_1}+m_{\bf B_2} Q_{\bf B_2}$, are

\begin{center}
	\begin{small}
		\renewcommand*{\arraystretch}{0.7}
	\begin{tabular}{c|ccccc}
	& $Q_1$&$ Q_2$&$Q_{\bf B_1}$&$Q_{\bf B_2}$&$Q_C$\\\hline
	$	X_{10}    $&$  0 $&$-1/2 $&$ 3$&$-1 $&$ -1/2m_2+3m_{\bf B_1}-m_{\bf B_2}$\\
	$	X_{21}    $&$  0 $&$ 0 $&$ -4$&$ 0 $&$ -4m_{\bf B_1}$\\
	$	X_{32}    $&$  -1/2 $&$ 0 $&$ 3 $&$ 1$&$ -1/2m_1+3m_{\bf B_1}+m_{\bf B_2}$ \\
	$	X_{43}^1 $&$ 0 $&$-1/2 $&$ -1$&$-1$&$-1/2m_2-m_{\bf B_1}-m_{\bf B_2}$\\
	$	X_{43}^2 $&$0$&$1/2$&$-1 $&$-1$&$1/2m_2-m_{\bf B_1}-m_{\bf B_2}$\\
	$	X_{04}^1$&$-1/2$&$0$&$-1 $&$1$&$-1/2m_1-m_{\bf B_1}+m_{\bf B_2}$\\
	$	X_{04}^2$&$1/2$&$0$&$-1 $&$1$&$1/2m_1-m_{\bf B_1}+m_{\bf B_2}$\\
	$	X_{20}    $&$ 0$&$1/2$&$ -1 $&$-1$&$1/2m_2-m_{\bf B_1}-m_{\bf B_2}$\\
	$		X_{31}    $&$ 1/2$&$0$&$-1$&$1$&$1/2m_1-m_{\bf B_1}+m_{\bf B_2}$\\
	$	X_{14}    $&$ -1/2$&$1/2$&$ 2$&$0$&$-1/2m_1+1/2m_2+2m_{\bf B_1}$\\
	$	X_{42}     $&$ 1/2$&$ -1/2 $&$2$&$0$&$1/2m_1-1/2m_2+2m_{\bf B_1}$
\end{tabular}
	\end{small}
\end{center}

On $dP_2$ there are two kinds of fractional branes given by: 
\begin{enumerate}
	\item $N_0=0$, $N_1=2$, , $N_2=0$, $N_3=1$, $N_4=1$;
	\item  $N_0=1$, $N_1=0$, $N_2=0$, $N_3=1$, $N_4=0$.
\end{enumerate}
The geometric identities correspond to the two generic ones, and two associated to fractional branes. They read 

\beqa
&&\partial V_1+\partial V_2+\partial V_3-\partial V'_1-\partial V'_2-\partial V'_3=0 \nonumber\\
&&{\tilde\partial}F_0+{\tilde\partial}F_1+{\tilde\partial}F_2+{\tilde\partial}F_3+{\tilde\partial}F_4-\left(\partial V_1+\partial V_2+\partial V_3\right)-\left(\partial V'_1+\partial V'_2+\partial V'_3\right)=0\nonumber\\
&&2(2{\tilde\partial}F_1+{\tilde\partial}F_3+{\tilde\partial}F_4)+\partial V_1-\partial V_2-\partial V_3-5\partial V'_1-\partial V'_2-3\partial V'_3=0\nonumber\\
&&{\tilde\partial}F_0+{\tilde\partial}F_3+\partial V_2-\partial V'_1-\partial V'_2-\partial V'_3=0 \fstop
\eeqa
Using the $B-$charge assignments in Figure~\ref{fig:unit-cell-dp2}, the constraints from invariance of the superpotential terms at the six nodes, and anomaly cancellation on the five faces, are

\beqa
\partial V_1&\longrightarrow&b_{X_{32}}+b_{X_{43}^1}+b_{X_{04}^2}+b_{X_{20}}=0 \nonumber\\
\partial V_2&\longrightarrow& -k_2 c_{X_{14}}+b_{X_{14}}+b_{X_{21}}+b_{X_{42}}=0\nonumber\\ 
\partial V_3&\longrightarrow& k_1 c_{X_{31}}+b_{X_{31}}+b_{X_{43}^2}+b_{X_{10}}+b_{X_{04}^1}=0\nonumber\\ 
\partial V'_1&\longrightarrow&b_{X_{31}}+b_{X_{14}}+b_{X_{43}^1}=0\nonumber\\ 
\partial V'_2&\longrightarrow&b_{X_{20}}+b_{X_{42}}+b_{X_{04}^1}=0\nonumber\\ 
\partial V'_3&\longrightarrow&b_{X_{32}}+k_1 c_{X_{32}}+b_{X_{21}}+k_1 c_{X_{21}}+b_{X_{10}}-k_2c_{X_{10}}+ b_{X_{04}^2}-k_2c_{X_{04}^2}+b_{X_{43}^2}=0\nonumber\\
{\tilde\partial}F_0&\longrightarrow& b_{X_{10}}+b_{X_{04}^2}+b_{X_{20}}+b_{X_{04}^1}=0\\ 
{\tilde\partial}F_1&\longrightarrow&k_1 c_{X_{31}}+b_{X_{31}}+k_1 c_{X_{14}}+b_{X_{14}}+\left(k_1+k_2\right) c_{X_{21}}+b_{X_{21}}+b_{X_{10}}=0\nonumber\\
{\tilde\partial}F_2&\longrightarrow&  b_{X_{32}}+b_{X_{21}}+b_{X_{20}}+b_{X_{42}}=0\nonumber\\ 
{\tilde\partial}F_3&\longrightarrow&  k_1 c_{X_{31}}+b_{X_{31}}+k_1 c_{X_{32}}+b_{X_{32}}+k_1 c_{X_{43}^1}+b_{X_{43}^1}+b_{X_{43}^2}=0\nonumber\\ 
{\tilde\partial}F_4&\longrightarrow&  -k_2 c_{X_{14}}+b_{X_{14}}-k_2 c_{X_{04}^2}-k_2 c_{X_{43}^1}+b_{X_{43}^1}+b_{X_{43}^2}+b_{X_{04}^2}+b_{X_{42}}+b_{X_{04}^1}=0\fstop \nonumber
\label{eq:dP2BCcont}
\eeqa
Using the $C-$charges as in the table above, the system reduces to

\begin{align}
\begin{split}
& k_2 \left(m_1-m_2\right)+k_1 m_1=0  \\
& -2 k_2 \left(4 m_{\bf B_{1}}-m_1+m_2\right)+\frac{1}{2} k_1 \left(-8 m_{\bf B_{1}}+4 m_{\bf B_{2}}+3 m_1+m_2\right)=0 \\
& \frac{1}{2} k_1 \left(4 m_{\bf B_{1}}+m_1-m_2\right)-k_2 \left(m_2-m_1\right)=0 \fstop 
\end{split}
\end{align}
To obtain integer $C-$charges, we choose

\begin{equation}
	m_{\bf B_{1}}=\frac{k_1}{4}(k_1+2k_2)\coma m_1=k_1k_2 \coma m_2=k_1(k_1+k_2) \coma m_{\bf B_{2}}=\frac{1}{4}\left(k_1^2+8k_1k_2+8k_2^2\right)\fstop
\end{equation}
And we obtain

\begin{alignat}{2}
&		c_{	X_{42}    } =2k_2(k_1+k_2) \quad \quad \quad &&c_{	X_{54}^2} =-2k_2(k_1+k_2) \nonumber \\
&		c_{	X_{43}    }=(k_1+2k_2)(k_1+k_2)   &&c_{	X_{15}^1}=k_2(k_1+2k_2) \nonumber \\
&		c_{	X_{25}   }  =k_1(k_1+k_2) && c_{	X_{15}^2}=2k_2(k_1+k_2) \nonumber \\
&		c_{	X_{32}  }   =- k_1(k_1+2k_2)&& c_{	X_{31}  }   =-2k_2(k_1+k_2) \\
&		c_{	X_{21}  }   =-k_2(k_1+2k_2)&& c_{	X_{53}    }=k_1k_2 \nonumber\\
&		c_{	X_{54}^1 } =-(k_1+2k_2)(k_1+k_2)  \fstop \nonumber
\end{alignat}
The solutions for the $B-$charges are
\begin{align}
\begin{split}
		b_{	X_{21}  }  & = -b_{X_{42}}-b_{X_{25}}-b_{X_{32}}\\
		b_{	X_{54}^1 }& = -b_{X_{42}}-b_{X_{25}}\\
		b_{	X_{54}^2} &= -b_{X_{43}}+b_{X_{25}}-2k_1k_2(k_1+k_2)\\
		b_{	X_{15}^1}&= b_{X_{43}}+b_{X_{32}}\\
		b_{	X_{15}^2}&= b_{X_{42}}+k_1k_2(k_1+k_2)\\
		b_{	X_{31}  }  & = -b_{X_{43}}+b_{X_{25}}-k_1k_2(k_1+k_2)\\
		b_{	X_{53}    } &= -b_{X_{25}}-b_{X_{32}}+k_1k_2(k_1+k_2)\coma 
\end{split}
\end{align}
where $b_{X_{42}}$, $b_{X_{43}}$, $b_{X_{25}}$, $b_{X_{32}}$ are left as undetermined parameters encoding the freedom to shift charges by the global $\U(1)^4$ symmetry.

\subsection{General orbifolds of the $dP_3$ theory}
\label{sec:general-solution-dp3}

\begin{figure}[!htp]
	\centering
	\includegraphics[scale=0.4]{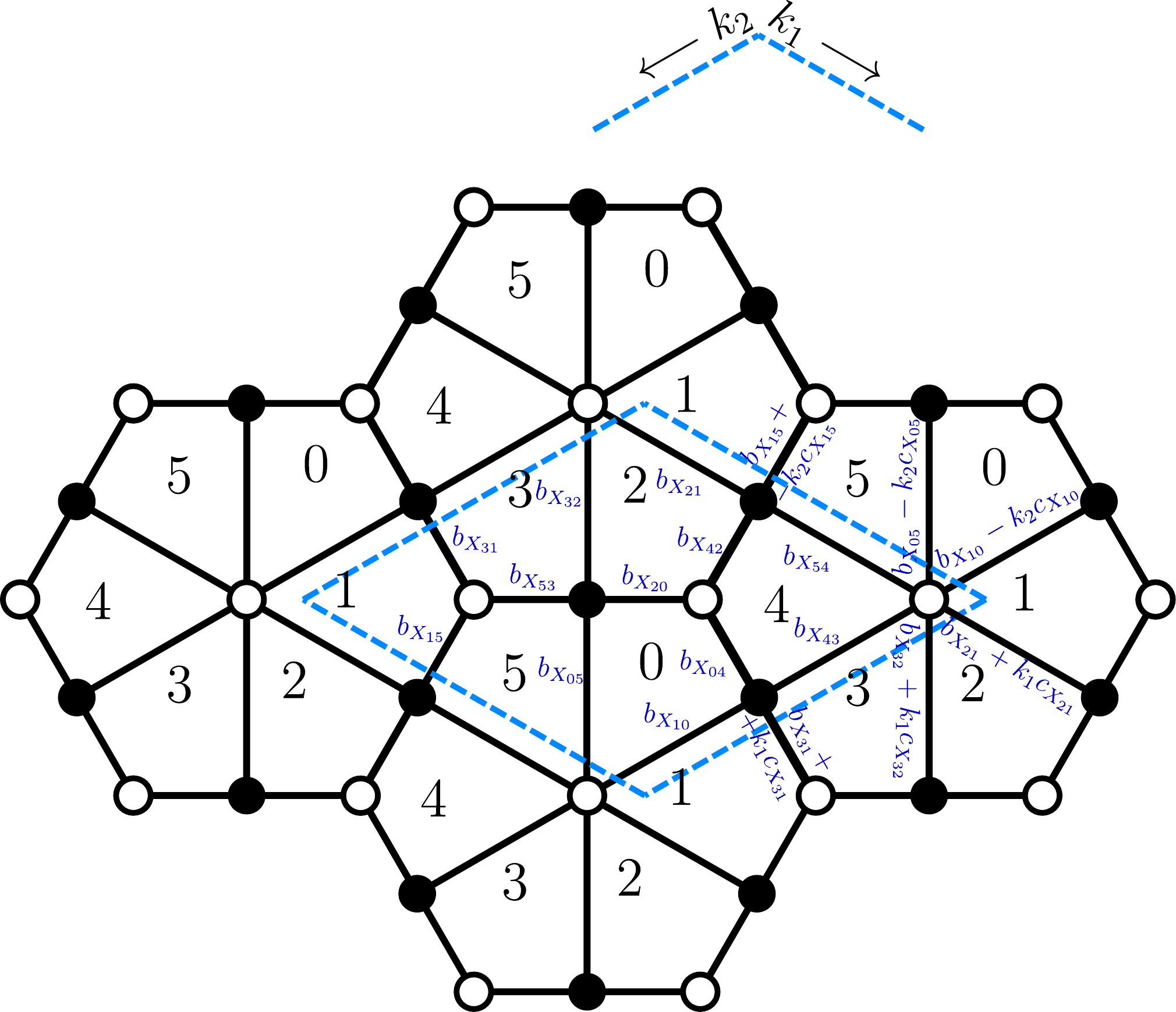}
	\caption{Dimer diagram with a unit cell for the $dP_3$ theory. We display the charge assignments corresponding to the $B$-charges.}
	\label{fig:unit-cell-dp3}
\end{figure}

We now construct the discrete symmetries for the general orbifolds of the $dP_3$ theory. The dimer diagram with a unit cell and the $B-$charge assignment is shown in Figure~\ref{fig:unit-cell-dp3}. There are two mesonic $\U(1)$'s and three baryonic $\U(1)$'s. The charges of the different fields under these $\U(1)$'s, and under a general combination $Q_C=m_1Q_1+m_2Q_2+m_{\bf B_{1}}Q_{\bf B_1}+m_{\bf B_2} Q_{\bf B_2}+m_{\bf B_3}Q_{\bf B_3}$, are

\begin{center}
	\begin{small}
		\renewcommand*{\arraystretch}{0.7}
		\begin{tabular}{c|cccccc}
			&$ Q_{1} $&$ Q_2$&$Q_{\bf B_1}$&$Q_{\bf B_2}$&$Q_{\bf B_3}$&$Q_C$\\\hline
			$X_{10} $&$    -1   $&$    0    $&$    1 $&$ 0 $&$ -1 $&$ -m_1+m_{\bf B_{1}}-m_{\bf B_{3}}  $\\
			$		X_{54} $&$    0    $&$    0    $&$	-1 $&$ 1 $&$ -1 $&$ -m_{\bf B_{1}}+m_{\bf B_{2}}-m_{\bf B_{3}}  $\\
			$	X_{32} $&$    0     $&$   -1   $&$	0 $&$ -1 $&$ -1 $&$ -m_2-m_{\bf B_{2}}-m_{\bf B_{3}} $\\
			$	X_{43} $&$    -1     $&$   1   $&$	1 $&$ 0 $&$ 1 $&$ -m_1+m_2+m_3+m_{\bf B_{3}}  $\\
			$		X_{21} $&$    1     $&$    0   $&$	-1 $&$ 1 $&$ 1 $&$ m_1-m_{\bf B_{1}}+m_{\bf B_{2}}+m_{\bf B_{3}} $\\
			$	X_{05} $&$    1     $&$    0   $&$	0 $&$ -1 $&$ 1 $&$ m_1-m_{\bf B_{2}}+m_{\bf B_{3}}  $\\
			$		X_{31} $&$    1     $&$    0   $&$	-1 $&$ 0 $&$ 0 $&$ m_1-m_{\bf B_{1}}  $\\
			$		X_{04} $&$   1      $&$   -1   $&$	-1 $&$ 0 $&$ 0 $&$ m_1-m_2-m_{\bf B_{1}}  $\\
			$		X_{15} $&$   -1      $&$  0    $&$	1 $&$ -1 $&$ 0 $&$ -m_1+m_{\bf B_{1}}-m_{\bf B_{2}}  $\\
			$		X_{42} $&$    0     $&$   0    $&$	1 $&$ -1 $&$ 0 $&$ m_{\bf B_{1}}-m_{\bf B_{2}}  $\\
			$		X_{53} $&$    0     $&$   0    $&$	0 $&$ 1 $&$ 0 $&$ m_{\bf B_{2}}  $\\
			$		X_{20} $&$  -1       $&$  1   $&$	   0 $&$ 1 $&$ 0 $&$ -m_{1}+m_2+m_{\bf B_{2}} $
		\end{tabular}
	\end{small}
\end{center}

On $dP_3$, instead, there are three kinds of fractional branes given by: 
\begin{enumerate}
	\item $N_0=0\coma N_1=1\coma N_2=0\coma N_3=0 \coma N_4=1 \coma N_5=0$;
	\item $N_0=0\coma N_1=0\coma N_2=1\coma N_3=0 \coma N_4=0 \coma N_5=1$;
	\item $N_0=1\coma N_1=0\coma N_2=1\coma N_3=0 \coma N_4=1 \coma N_5=0$.
\end{enumerate}
The geometric identities correspond to the two generic ones, and three associated to fractional branes. They read 

\beqa
&&\partial V_1+\partial V_2+\partial V_3-\partial V'_1-\partial V'_2-\partial V'_3=0\nonumber\\
&&{\tilde\partial}F_0+{\tilde\partial}F_1+{\tilde\partial}F_2+{\tilde\partial}F_3+{\tilde\partial}F_4+{\tilde\partial}F_5-\left(\partial V_1+\partial V_2+\partial V_3\right)-\left(\partial V'_1+\partial V'_2+\partial V'_3\right)=0\nonumber\\
&&{\tilde\partial}F_1+{\tilde\partial}F_4-\partial V_2-\partial V_3=0\nonumber\\
&&{\tilde\partial}F_2+{\tilde\partial}F_5-\partial V_1-\partial V_2=0\nonumber\\
&&{\tilde\partial}F_0+{\tilde\partial}F_2+{\tilde\partial}F_4-2\partial V'_2-\partial V'_3=0\fstop
\eeqa
Using the $B-$charge assignments in Figure~\ref{fig:unit-cell-dp3}, the constraints from invariance of the superpotential terms at the six nodes, and anomaly cancellation on the six faces, are

\beqa
\partial V_1&\longrightarrow&b_{X_{32}}+b_{X_{20}}+b_{X_{05}}+b_{X_{53}}=0 \nonumber \\
\partial V_2&\longrightarrow& -k_2 c_{X_{15}}+b_{X_{15}}+b_{X_{54}}+b_{X_{42}}+b_{X_{21}}=0 \nonumber \\ 
\partial V_3&\longrightarrow& k_1 c_{X_{31}}+b_{X_{31}}+b_{X_{10}}+b_{X_{04}}+b_{X_{43}}=0 \nonumber \\ 
\partial V'_1&\longrightarrow&b_{X_{31}}+b_{X_{15}}+b_{X_{53}}=0 \nonumber \\ 
\partial V'_2&\longrightarrow&b_{X_{20}}+b_{X_{04}}+b_{X_{42}}=0 \nonumber \\ 
\partial V'_3&\longrightarrow&b_{X_{32}}+k_1 c_{X_{32}}+b_{X_{21}}+k_1 c_{X_{21}}+b_{X_{10}}-k_2c_{X_{10}}+ b_{X_{05}}-k_2c_{X_{05}}+\nonumber\\
&&+b_{X_{54}}+b_{X_{43}}=0 \nonumber \\
{\tilde\partial}F_0&\longrightarrow& b_{X_{10}}+b_{X_{04}}+b_{X_{20}}+b_{X_{05}}=0\\ 
{\tilde\partial}F_1&\longrightarrow&k_1 c_{X_{31}}+b_{X_{31}}+k_1 c_{X_{15}}+b_{X_{15}}+\left(k_1+k_2\right) c_{X_{21}}+b_{X_{21}}+b_{X_{10}}=0 \nonumber \\
{\tilde\partial}F_2&\longrightarrow&  b_{X_{32}}+b_{X_{21}}+b_{X_{20}}+b_{X_{42}}=0 \nonumber \\ 
{\tilde\partial}F_3&\longrightarrow&  k_1 c_{X_{31}}+b_{X_{31}}+k_1 c_{X_{32}}+b_{X_{32}}+k_1 c_{X_{53}}+b_{X_{53}}+b_{X_{43}}=0 \nonumber \\ 
{\tilde\partial}F_4&\longrightarrow&  b_{X_{42}}+b_{X_{04}}+b_{X_{43}}+b_{X_{54}}=0 \nonumber \\
{\tilde\partial}F_5&\longrightarrow& b_{X_{15}}-k_2c_{X_{15}}+b_{X_{54}}+b_{X_{05}}-k_2c_{X_{05}}+b_{X_{53}}+k_{2}c_{X_{53}}=0 \fstop\nonumber 
\label{eq:dP3BCcont}
\eeqa
Using the $C-$charges as in the table above, the system reduces to

\begin{align}
\begin{split}
&k_2 m_1+k_1 m_2=0 \\
&\left(k_1+k_2\right) m_{\bf B_3}=0 \\
&k_2 \left(m_{\bf B_3}+m_1\right)=0\\
&k_1 \left(m_{\bf B_1}-m_1+m_2\right)-k_2 \left(m_{\bf B_2}-m_{\bf B_1}\right)=0\fstop
\end{split}
\end{align}
To obtain integer $C-$charges, we choose

\begin{equation}
	m_{\bf B_1}=-k_2\coma m_1=0 \coma m_2=0 \coma m_{\bf B_2}=-k_2-k_1 \e m_{\bf B_3}=0\fstop
\end{equation}
And we obtain

\begin{alignat}{2}
	&	c_{		X_{42} } =k_2  \quad \quad && c_{		X_{54} } =-k_2\nonumber\\
	&	c_{		X_{43} } =k_1+k_2 \quad \quad  && c_{		X_{16} } =k_1+k_2\nonumber \\
	&	c_{		X_{26} } =k_1 && c_{		X_{15} } =k_2\nonumber\\
	&	c_{		X_{32} } =-k_1 && c_{		X_{53} } =k_1\\
	&	c_{		X_{65} } =-k_1 && c_{		X_{64} } =-k_1+k_2 \nonumber\\
	&	c_{	    X_{21} } =-k_2 && c_{		X_{31} } =k_1+k_2\fstop\nonumber
\end{alignat}
The solutions for the $B-$charges are
\begin{align}
\begin{split}
		b_{	    X_{21} }& = -b_{X_{42}}-b_{X_{26}}-b_{X_{32}} \\
		b_{		X_{54} }& = -b_{X_{43}}+b_{X_{26}}-k_1k_2 \\
		b_{		X_{16} }& = b_{X_{42}}-b_{X_{65}}+k_1k_2 \\
		b_{		X_{15} }& = b_{X_{43}}+b_{X_{32}} \\
		b_{		X_{53} }&= -b_{X_{26}}-b_{X_{32}}-b_{X_{65}}+k_1k_2\\
		b_{		X_{64} }&= -b_{X_{42}}-b_{X_{26}} \\
		b_{		X_{31} }&= -b_{X_{43}}+b_{X_{26}}+b_{X_{65}}-k_1k_2\coma 
\end{split}
\end{align}
where $b_{X_{42}}$, $b_{X_{43}}$, $b_{X_{26}}$, $b_{X_{32}}$, $b_{X_{65}}$ are left as undetermined parameters encoding the freedom to shift charges by the global $\U(1)^5$ symmetry.

\bibliographystyle{JHEP}
\bibliography{mybib}

\end{document}